\newcommand\maxSearchArea{16293}    
\newcommand\flagZeroArea{12590}     
\newcommand\cosmoSearchArea{10317}  
\newcommand\massLimFull{5.3}        
\newcommand\massLimCosmo{5.0}       
\newcommand\massLimFullArnaud{3.0}  
\newcommand\DESOverlap{4391}
\newcommand\HSCOverlap{1078}
\newcommand\KiDSOverlap{802}
\newcommand\eRASSDeOverlap{7044}

\documentclass[iop,twocolumn,twocolappendix,numberedappendix]{openjournal}
\usepackage[T1]{fontenc}
\usepackage{hyperref} 
\hypersetup{colorlinks=true,linkcolor=blue,citecolor=blue,filecolor=blue,urlcolor=blue}
\usepackage{longtable}
\usepackage{orcidlink}
\usepackage{lineno}

\mathchardef\mhyphen="2D

\defcitealias{Hasselfield_2013}{H13}
\defcitealias{Hilton_2018}{H18}
\defcitealias{Hilton_2021}{H21}
\defcitealias{Arnaud_2010}{A10}
\defcitealias{Naess_2020}{N20}

\newcommand{\yt}{{\tilde{y}_0}}
\newcommand{\yti}{{\tilde{y}_{0i}}}
\newcommand{\snr}{{q}} 
\newcommand{\snrt}{{\tilde{q}}} 

\newcommand{\cosmo}{Legacy }

\begin{document}

\newcommand\totalCandidates{21558}
\newcommand\totalFlagZeroCandidates{15752}
\newcommand\totalFlagZeroConfirmed{8943}
\newcommand\totalConfirmed{10040}
\newcommand\totalCosmoSample{3747}
\newcommand\totalFixedSNRFiveCandidates{7786}
\newcommand\totalFixedSNRSixCandidates{4754}
\newcommand\totalNew{6854}
\newcommand\totalHighZ{1180}
\newcommand\totalHighestZ{124}
\newcommand\totalSpecRedshifts{4133}
\newcommand\percentageSpecRedshifts{41.2}
\newcommand\numWISEBCGz{59}
\newcommand\numProjected{128}
\newcommand\minRedshift{0.03}
\newcommand\medianRedshift{0.58}
\newcommand\medianRedshiftNew{0.58}
\newcommand\maxRedshift{2.00}
\newcommand\minMass{1.3}
\newcommand\medianMass{2.8}
\newcommand\maxMass{22.9}

\begin{nolinenumbers}
\vspace*{-\headsep}\vspace*{\headheight}
\footnotesize \hfill  FERMILAB-PUB-25-0527-PPD\\
\vspace*{-\headsep}\vspace*{\headheight}
\footnotesize \hfill DES-2025-0892
\end{nolinenumbers}

\title{\vspace{-0.8cm}The Atacama Cosmology Telescope: DR6 Sunyaev-Zel'dovich Selected Galaxy Clusters Catalog\vspace{-1.5cm}}

\author{M.~Aguena$^{1,2}$\orcidlink{0000-0001-5679-6747}}
\author{S.~Aiola$^{3,4}$\orcidlink{0000-0002-1035-1854}}
\author{S.~Allam$^{5}$\orcidlink{0000-0002-7069-7857}}
\author{F.~Andrade-Oliveira$^{6}$\orcidlink{0000-0003-0171-6900}}
\author{D.~Bacon$^{7}$}
\author{N.~Bahcall$^{8}$\orcidlink{0000-0002-8226-9825}}
\author{N.~Battaglia$^{9}$\orcidlink{0000-0001-5846-0411}}
\author{E.~S.~Battistelli$^{10}$\orcidlink{0000-0001-5210-7625}}
\author{S.~Bocquet$^{11}$\orcidlink{0000-0002-4900-805X}}
\author{B.~Bolliet$^{12,13}$\orcidlink{0000-0003-4922-7401}}
% ...
\collaboration{ACT-DES-HSC Collaboration}

\email{$^*$E-mail: matt.hilton@wits.ac.za}

\shorttitle{ACT: DR6 SZ-Selected Galaxy Clusters}
\shortauthors{ACT-DES-HSC Collaboration}



\begin{abstract}

We present the results of a search for galaxy clusters in the
Atacama Cosmology Telescope (ACT) Data Release 6 (DR6) microwave sky maps
covering \maxSearchArea{} square degrees in three frequency bands, 
using data obtained over the lifetime of the project (2008--2022). We report
redshifts and mass estimates for \totalConfirmed{} clusters detected via
their Sunyaev-Zel’dovich (SZ) effect with signal-to-noise greater than 4
at a 2.4 arcminute filter scale. The catalog includes \totalHighZ{} clusters
at redshifts greater than 1, and \totalHighestZ{} clusters at redshifts
greater than 1.5. 
Using a relation between cluster SZ signal and mass that is consistent with
recent weak-lensing measurements, we estimate that clusters detected with
signal-to-noise greater than 5 form a sample which is 90\% complete for
clusters with masses greater than $5 \times 10^{14}\,M_{\sun}$ 
(measured within a spherical volume with mean density 500 times the critical
density). El Gordo, a cluster found in an initial ACT survey of 755
square degrees, remains the most extreme cluster in mass and redshift; we find
no cluster with a mass and redshift combination high enough to falsify the standard $\Lambda$CDM
cosmology with Gaussian initial perturbations. We make public a variety of data
products, including the full cluster candidate list, noise maps, and sky masks,
along with our software for cluster detection and instructions for reproducing
our cluster catalogs from the public ACT maps. 

\end{abstract}


\section{Introduction} 
\label{sec:intro}

Over the past two decades, the development of increasingly sensitive millimeter wavelength instruments, capable of surveying large areas of the sky with arcminute resolution, has progressed significantly. These advancements have enabled the detection of massive galaxy clusters spanning the last 10\,Gyr of cosmic history by exploiting the thermal Sunyaev-Zel’dovich effect \citep[SZ;][]{SZ_1970, SZ_1972}.

The SZ effect is the inverse
Compton scattering of cosmic microwave background (CMB) photons by
electrons in the hot gaseous atmospheres of galaxy clusters 
\citep[see reviews by][]{Birkinshaw_1999, Carlstrom_2002,
Mroczkowski_2019}. The SZ signal, which is in principle unaffected
by the distance to the observer, provides a measure of the integrated
thermal pressure along the line of sight. This correlates with the
total mass of clusters, allowing SZ surveys to construct approximately
mass-limited cluster samples with unlimited redshift reach.
One application of such catalogs is to constrain cosmological
parameters using the evolution of cluster number counts, with the
current state-of-the-art represented by the \citet{Bocquet_2024}
analysis of 1005 SZ-selected clusters detected by the 
South Pole Telescope (SPT).

Thousands of clusters, reaching up to $z = 2$, have now been detected using blind SZ surveys by ACT 
\citep{Marriage_2011, Hasselfield_2013, Hilton_2018, Hilton_2021},
\textit{Planck} \citep{Planck_XXIX_2013, PlanckPSZ2_2016, Zubeldia_2024},
and SPT \citep{Staniszewski_2009, Williamson_2011, Bleem_2015, 
Bleem_2020, Huang_2020, Bleem_2024, KleinSPT_2024, Archipley_2025}. The majority of these have been found
using ACT data \citep{Hilton_2021, Klein_2024}, due to a combination
of the sensitivity of the AdvACT receiver \citep{Henderson_2016, Choi_2018},
and the fact that ACT observed one-third of the extragalactic sky
\citep{Naess_2020}. The deepest observations, detecting the lowest mass
SZ-selected clusters to date ($M_{\rm 500c} \approx 1.5 \times 10^{14}$\,$M_{\sun}$), have been performed by SPT \citep{Bleem_2024, Kornoelje_2025}.

\begin{figure}
\includegraphics[width=\columnwidth]{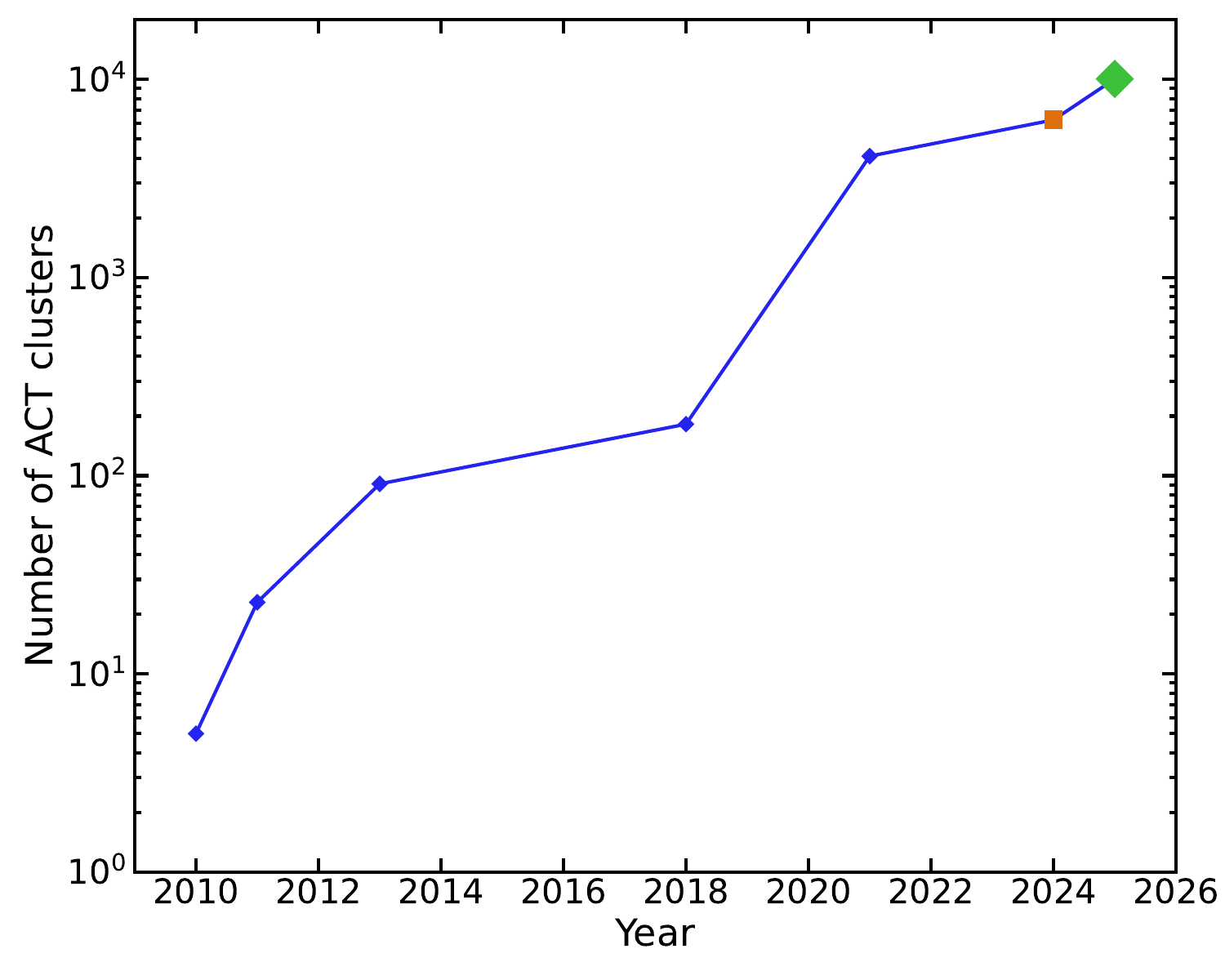}
\caption{The number of ACT-detected SZ-selected clusters reported over
the lifetime of the project. 
References for the blue points: 
First detections \citep{Hincks_2010}; 
ACT DR1 \citep{Marriage_2011, Menanteau_2010};
ACT DR2 \citep{Hasselfield_2013, Menanteau_2013};
ACT DR3 \citep{Hilton_2018};
ACT DR5 \citep{Hilton_2021}.
The orange square marks the ACT-DR5 MCMF catalog \citep{Klein_2024}, which is based on an independent re-run of the \textsc{Nemo} software on the ACT DR5 maps, followed by optical confirmation using the DESI Legacy Imaging Surveys DR10 \citep{Dey_2019}.
The green diamond marks the \totalConfirmed{} clusters reported in the current work, based on the ACT DR6 maps,
which is $>1.5$ times larger than catalogs produced using ACT DR5 data.}
\label{fig:numClustersVsTime}
\end{figure}

\begin{figure*}
\includegraphics[width=\textwidth]{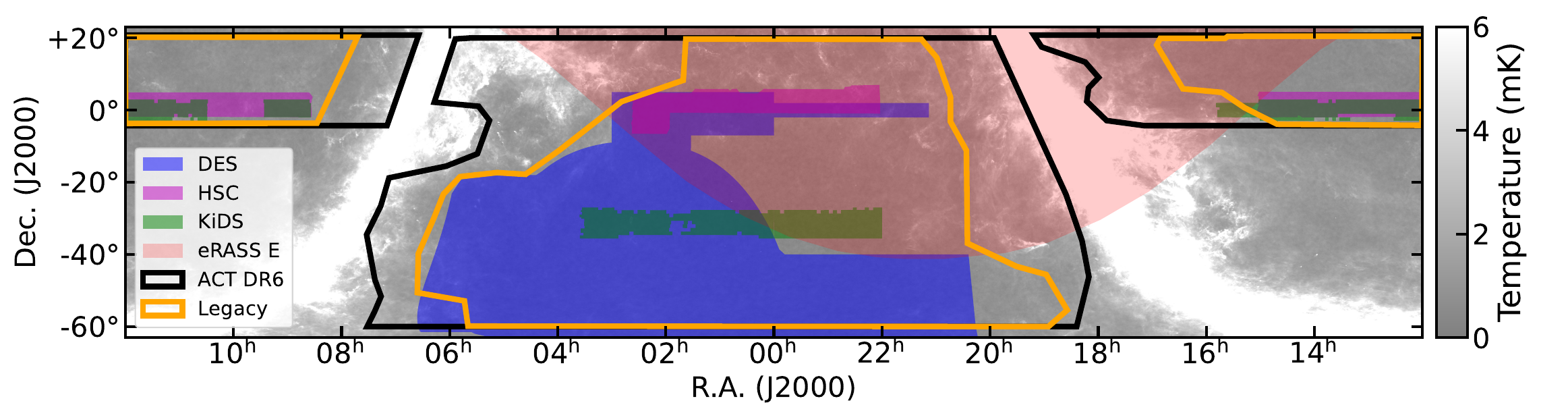}
\includegraphics[width=\textwidth]{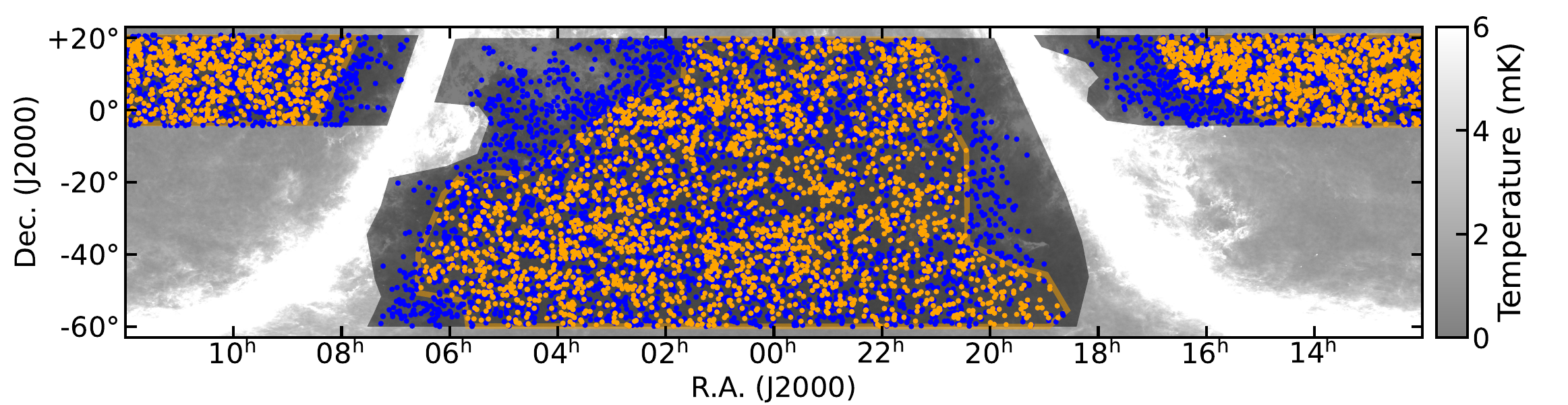}
\caption{Top panel: The full ACT DR6 cluster search region (black outline),
covering \maxSearchArea{}\,deg$^2$ before masking. The orange outline shows the 
\cosmoSearchArea{}\,deg$^2$ \cosmo footprint (after masking and flagging)
used to define the `Legacy sample' (see Section~\ref{sec:catalog}). 
The footprints of deep and wide optical surveys are highlighted, with the overlapping area with ACT DR6 (after masking and flagging) given in square degrees: 
DES (blue; \citealt{DESDR2_2021}; $\DESOverlap~\deg^2$); 
HSC (magenta; \citealt{Aihara_2022}; $\HSCOverlap~\deg^2$); 
and KiDS (green; \citealt{Wright_2024}; $\KiDSOverlap~\deg^2$). 
The red shaded region indicates the
region of sky covered by the Russian half of the eROSITA survey; the rest of the ACT
search area has $\eRASSDeOverlap~\deg^2$ of overlap with the German half of the eROSITA sky \citep{Meloni_2024}. 
Bottom panel: Blue points mark the locations of \totalConfirmed{} optically confirmed
SZ-detected clusters reported in this paper. The `Legacy sample' of
\totalCosmoSample{} clusters detected at signal-to-noise $\snrt > 5.5$ are marked 
with orange points. The full ACT search area is shown by the dark gray shaded region. 
In both panels, the \textit{Planck} 353\,GHz map is shown in the background,
to highlight regions of sky that have more thermal emission from dust.
}
\label{fig:dustOverlay}
\end{figure*}

In this paper we report on a search for SZ clusters in the ACT Data
Release 6 (DR6) maps \citep{Naess_2025}, using 
data collected from 2008 to 2022. 
Figure~\ref{fig:numClustersVsTime} 
illustrates how the number of SZ clusters detected by ACT has grown with
each data release. Together with this paper, we release
a number of data products related to the cluster 
search.\footnote{All data products are available from LAMBDA:
\url{https://lambda.gsfc.nasa.gov/product/act/actadv_prod_table.html}}
This includes a catalog of \totalConfirmed{} clusters with mass and redshift
estimates detected across the maximum \maxSearchArea{}\,deg$^2$ search area
(see Table~\ref{tab:FITSTableColumns} for a description of the contents and a list of symbols used in this paper);
a complete, high signal-to-noise subsample of \totalCosmoSample{}
clusters over a \cosmoSearchArea{}\,deg$^2$ area that we refer to as the
`Legacy sample', which is suggested for cosmological analyses; the full list of \totalCandidates{} cluster candidates (i.e., including objects without redshift estimates and optical confirmation); and various files related to the cluster selection (masks, flagged areas,
random catalogs, and noise properties as a function of sky position). 
The top panel of Figure~\ref{fig:dustOverlay} shows the location of the
ACT DR6 cluster search area compared with the footprints of several other
surveys on the sky. 
The bottom panel of Figure~\ref{fig:dustOverlay} shows the location of the
optically confirmed SZ-detected clusters reported in this work,
highlighting in particular the Legacy sample. 

This paper is organized as follows. In Section~\ref{sec:data} we describe
the ACT maps, detection of cluster candidates, and mass completeness
estimates, highlighting differences with respect to the previous analysis
based on ACT DR5. In Section~\ref{sec:optical} we explain the process
of obtaining redshift estimates for the cluster candidates.
In Section~\ref{sec:catalog}, we describe the ACT DR6 cluster search data
products, and present the properties of the cluster catalog.
We discuss how the DR6 sample compares to the ACT DR5 and SRG/eROSITA \citep{Sunyaev_2021} All-Sky Survey \citep[eRASS1 Western Galactic hemisphere;][]{Bulbul_2024} cluster catalogs in Section~\ref{sec:discussion}.
We summarize our findings in Section~\ref{sec:summary}.

We assume a flat $\Lambda$CDM cosmology with $\Omega_{\rm m}=0.3$, $\Omega_\Lambda=0.7$, and $H_0=70$~km~s$^{-1}$~Mpc$^{-1}$ 
throughout. We report cluster mass estimates ($M_{\rm \Delta c}$) within a spherical radius that encloses an average 
density equal to $\Delta$ times the critical density at the cluster redshift ($R_{\rm \Delta c}$), where $\Delta = 200$ or $500$. We replace the subscript $\rm c$ with subscript $\rm m$ to indicate a mass estimate with respect to $\Delta$ times the mean density of the Universe at the cluster redshift.
Note that the mass calibration adopted in this paper
differs from that used in previous ACT cluster catalog papers (see Section~\ref{sec:masses} for details).

\section{ACT Observations and SZ Cluster Detection} 
\label{sec:data}

\subsection{Observations and Maps} 
\label{sec:maps}

ACT observed the millimeter sky from Chile with arcminute resolution
from 2007
\footnote{The telescope saw first light in 2007; in this paper we use data from 2008 onwards.}
to 2022, using three generations of receivers:
the Millimeter Bolometer Array Camera \citep[MBAC;][2007--2011]{Swetz_2011};
ACTPol \citep[][2013--2016]{Thornton_2016}; and AdvACT \citep[][2016--2022]{Henderson_2016, Choi_2018}.
In this work we use maps covering three frequency bands named
f090, f150, and f220, where the number indicates the approximate center of
the observed frequency range in GHz. 
These have angular resolutions of
$2.1\arcmin$, $1.4\arcmin$, and 1.0$\arcmin$ respectively.
To maximize our sensitivity to the SZ signal, we use the deepest
available ACT data, which are the co-added maps from ACT DR6 that
include all usable daytime and nighttime data \citep{Naess_2025}. The method used
to create these co-added maps is described in detail in \citet{Naess_2020}.
We use the DR6 release ACT+Planck day+night coadds with ACT DR4 data included,
which therefore includes observations from 2008 to 2022. These
cover approximately 19000\,deg$^2$.

\subsection{Cluster Finding} 
\label{sec:nemo}

We use the \textsc{Nemo}\footnote{\url{https://github.com/simonsobs/nemo/}
; see \url{https://nemo-sz.readthedocs.io} for documentation, including
a tutorial on how to reproduce the cluster search data products 
described in this paper.}
package for SZ cluster detection and much of the subsequent analysis
presented in this work. Earlier versions of \textsc{Nemo} were used for
the ACT DR3 \citep{Hilton_2018} and ACT DR5 \citep[][H21 hereafter]{Hilton_2021} cluster
searches. Cluster candidates are identified by using
a multi-frequency matched filter \citep[e.g.,][]{Melin_2006, Williamson_2011},
\begin{equation}
\label{eq:MFMF}
\psi(k_x, k_y, \nu_i) = A \sum_j \textbf{N}^{-1}_{ij}(k_x, k_y) f_{\rm SZ}(\nu_j) S (k_x, k_y, \nu_j) \, ,
\end{equation}
where $\psi$ is the filter, ($k_x$, $k_y$) denote the spatial frequencies in the horizontal and vertical directions
in the maps, $\textbf{N}$ is the noise covariance between the maps at different frequencies $\nu$, $S$ is a beam-convolved signal template, and $A$ is a normalization factor. When applied to a set of maps containing a beam-convolved cluster signal (in temperature units), $A$ is chosen such
that the matched filter returns the central Comptonization parameter (see Section~\ref{sec:masses} below), and $f_{\rm SZ}$ is the non-relativistic form for the spectral dependence of the SZ
effect,
\begin{equation}
\label{eq:fSZ}
f_{\rm SZ} = x \frac{e^x+1}{e^x-1} - 4 \, ,
\end{equation}
where $x = h \nu / k_B T_{\rm CMB}$ (the relativistic SZ suppresses the SZ signal for very high temperatures; we apply a correction for this when inferring cluster masses, as noted in Section~\ref{sec:masses} below). As in previous ACT cluster searches,
we use the map itself to form the noise covariance $\textbf{N}$, as the maps are dominated by the CMB on large scales, and white noise on small scales, rather than by the thermal SZ signal. 

When filtering the map across multiple angular scales during the cluster search,
we adopt the maximum S/N across all filter scales for each candidate as the
`optimal' S/N detection, for which we use the symbol $\snr$
throughout this paper. However, as in previous ACT work, to mitigate 
inter-filter noise bias \citep[][H13 hereafter; see also the discussion of optimization bias in Appendix~\ref{ap:optBiasSims}]{Hasselfield_2013},
we use the S/N measured at a single reference filter scale 
($\theta_{\rm 500c} = 2.4 \arcmin$ as in H21; 
$\theta_{\rm 500c} = R_{\rm 500c}/D_{\rm A}$, where $D_{\rm A}$
is the angular diameter distance), for which we use the symbol $\snrt$. 
Similarly, we use the symbol $\yt$ to refer
to the central Comptonization parameter, which is our adopted measurement
of the SZ signal (see Section~\ref{sec:masses}).

Below we describe the improvements made for the DR6 cluster search compared to
the ACT DR5 analysis.

\subsubsection{Cluster Search Area and Tiling}
\label{sec:tiling}
We use a more extended cluster search area mask compared to DR5, extending
to lower Galactic latitudes. This is shown as the black outline in the top
panel of Figure~\ref{fig:dustOverlay}, and covers a sky area of
\maxSearchArea{}\,deg$^2$. This allows us to identify cluster candidates in
dusty regions that may have been missed by previous surveys, although in
practice we flag such areas when constructing statistical cluster
samples for cosmological analyses (see Section~\ref{sec:flagging}).

\begin{figure*}
\includegraphics[width=\textwidth]{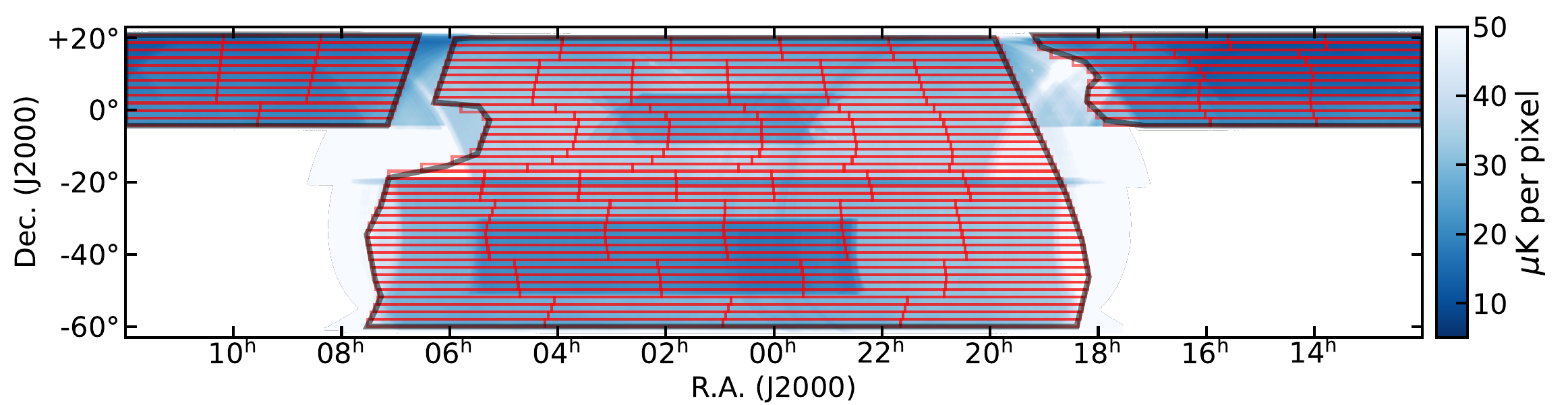}
\caption{A map illustrating the tiling geometry chosen for the ACT DR6 cluster
search. The red boxes indicate the locations of tiles, of nominal size
25\,deg\,$\times 2$\,deg, that are processed independently by the cluster finder
(see Section~\ref{sec:tiling}). The
full \maxSearchArea{}\,deg$^2$ cluster search area (before masking and flagging)
is shown as the black outline. The map in the background shows the variation
in the white noise level as a function of position in the ACT+Planck
day+night f090 map.
}
\label{fig:tiling}
\end{figure*}

For efficiency, \textsc{Nemo} breaks up the ACT maps into tiles, which
are processed separately (each with its own, different matched filter)
and subsequently combined (see H21). 
Figure~\ref{fig:tiling} shows the tiling arrangement chosen for the DR6
analysis. We have chosen to use tiles that nominally cover
25\,deg\,$\times 2$\,deg (right ascension $\times$ declination). These
are thinner than the 10\,deg\,$\times 5$\,deg tiles used in H21. This
change was made due to a concern that the flat sky approximation,
combined with the variable pixel scale due to the plate
carr\'{e}e projection (\texttt{CAR} in the terminology of Flexible Image Transport System (FITS) world 
coordinate systems; \citealt{Calabretta_2002}), may introduce a small bias
in cluster detection efficiency at high declination. Simulations have
shown that this is not a problem for the 25\,deg\,$\times 2$\,deg tiling
used in DR6 (see Appendix~\ref{ap:optBiasSims}, which presents the results of applying \textsc{Nemo} to simulated maps).
As in H21, noise differences within a tile are handled by making noise estimates within cells in the filtered maps (here we use cells that are nominally 80$\arcmin{}$ on a side, with the
exact size depending on the tile geometry).

\subsubsection{Multi-pass Filtering and Object Detection}
\label{sec:multipass}

For the DR6 analysis, we make use of \textsc{Nemo}'s new multi-pass filtering
and object detection feature, which was developed subsequent to the ACT DR5 release. This allows both point sources and clusters to
be detected in a single run of the software. 
This multi-pass approach mitigates some signal loss which would occur if the cluster signal
is not subtracted from the maps when estimating $\textbf{N}$ from the maps themselves. \citet{Zubeldia_2023} describe
a similar approach to the one used here, applied to \textit{Planck} data.
The procedure used for this work is as follows:

\begin{enumerate}
\item{\textit{Source detection run 1.} Detect ${\rm S/N} > 50$ point sources by applying a matched filter with the signal 
template set to the appropriate ACT beam, for each of the f090, f150, and f220 maps
individually (i.e., no spectral templates are used here).}

\item{\textit{Source detection run 2.} 
Detect ${\rm S/N} > 5$ point sources by applying a matched filter with the signal 
template set to the appropriate ACT beam, for each of the f090, f150, and f220 maps
individually. Locations with sources detected in Step 1 are
masked and in-filled with a median-filtered ($10\arcmin$-scale) version of the map, when
estimating $\textbf{N}$ in Equation~(\ref{eq:MFMF}).}

\item{\textit{Initial cluster detection run.} Detect $\snr > 3$ clusters by applying a multi-frequency matched filter to the
f090, f150, and f220 maps. Here, the Universal Pressure Profile 
\citep[UPP;][]{Arnaud_2010} is used for the signal template, after convolution
with the appropriate ACT beam. Detection is performed over 16 different angular
scales, corresponding to UPP models with $M_{\rm 500c} \in \{ (1, 2, 4, 8) \times 10^{14}$\,$M_\sun \}$ and $z \in \{0.2, 0.4, 0.8, 1.2\}$ (these are the same as in
H21). Locations with sources detected in Step 1 are
masked and in-filled as described in Step 2. Sources detected
in Step 2 are subtracted from the maps, before estimating $\textbf{N}$ in Equation~(\ref{eq:MFMF}).}

\item{\textit{Final cluster detection run.} Detect $\snrt > 4$ clusters by applying a 
multi-frequency matched filter to the
f090, f150, and f220 maps, using the same filter scales as in Step 3.
Before estimating $\textbf{N}$ in Equation~(\ref{eq:MFMF}), sources detected in Steps 1
and 2 are handled as described in Step 3. Cluster candidates detected in Step 3 are
also subtracted before estimating $\textbf{N}$, using the signal template with the maximum S/N for each detected
cluster.  We find that further passes are not necessary (i.e., do not result in
the detection of additional cluster candidates).
}

\end{enumerate}

We have also made some changes to how object detection is handled and how
multi-scale catalogs are combined compared to H21. We now apply a 3$\sigma$ threshold
when constructing the object segmentation map
\footnote{A segmentation map identifies which pixels in an image (the filtered map in our case) belong to which object in a catalog.}
and measuring object properties
(position, $\snr$, $\yt$, etc.), and then cut the catalog to the chosen $\snrt >4$
limit in the final step. This avoids incompleteness near the chosen $\snrt > 4$ limit,
as noted by \citet{Klein_2024}, who re-ran \textsc{Nemo} to reproduce the H21 cluster
candidate list.
\footnote{The \texttt{thresholdSigma} parameter in \textsc{Nemo} sets the S/N threshold used for constructing the segmentation map from which objects are identified. The segmentation map includes only pixels for which the S/N value is $>$ \texttt{thresholdSigma}. So, if \texttt{thresholdSigma} is set to the same value as the desired S/N cut of the output catalog (as was done for the ACT DR5 release), the resulting catalog is incomplete at the S/N cut, because the sub-pixel interpolation routine cannot include pixels with S/N values just below the threshold. More details can be found in the \textsc{Nemo} documentation.}
We also fixed a bug related to the minimum number of pixels required
for object detection (in H21, the minimum number of pixels required for a detection
was accidentally set to 2, rather than the intended 1 pixel, which means that the 
H21 catalog was purer than might have been expected at $q \approx 4$). Finally,
object positions in DR6 are reported from the reference 2.4$\arcmin$ filter scale
map (see Section~\ref{sec:masses} below), rather than from the highest S/N object
detection across all filter scales (in practice this change has little impact, but it was made for the purposes of minimizing optimization bias and calculating the correction for this consistently; see Appendix~\ref{ap:optBiasSims}).
Fig.~\ref{fig:SNRDist} shows the signal-to-noise ($\snrt$) distribution close to the detection limit for the full candidate list.

\begin{figure}
\includegraphics[width=\columnwidth]{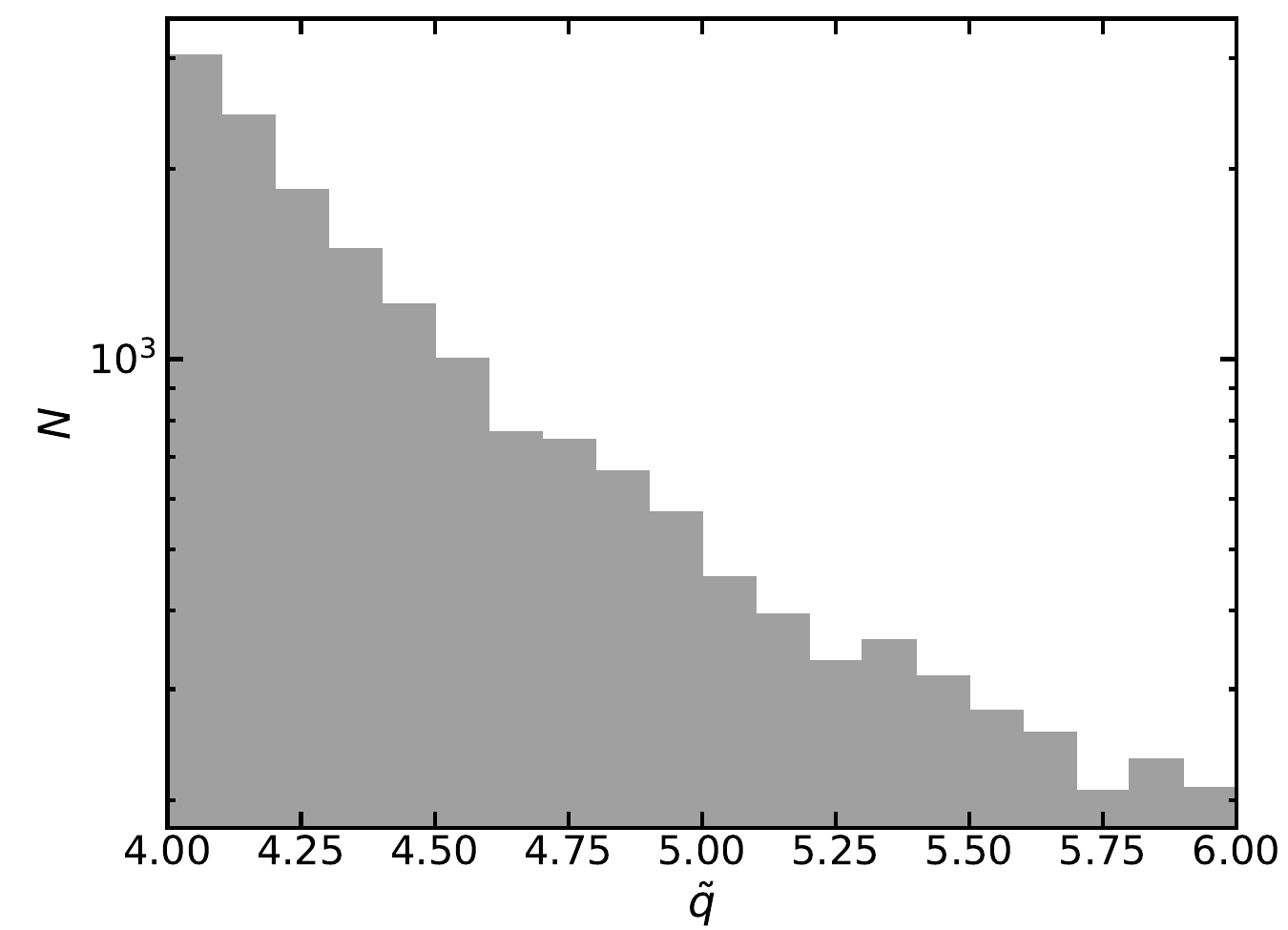}
\caption{Distribution of signal-to-noise ($\snrt$) values close to the detection limit for the full cluster candidate list.
}
\label{fig:SNRDist}
\end{figure}

\subsubsection{Flagging}
\label{sec:flagging}

\begin{figure*}
\includegraphics[width=\textwidth]{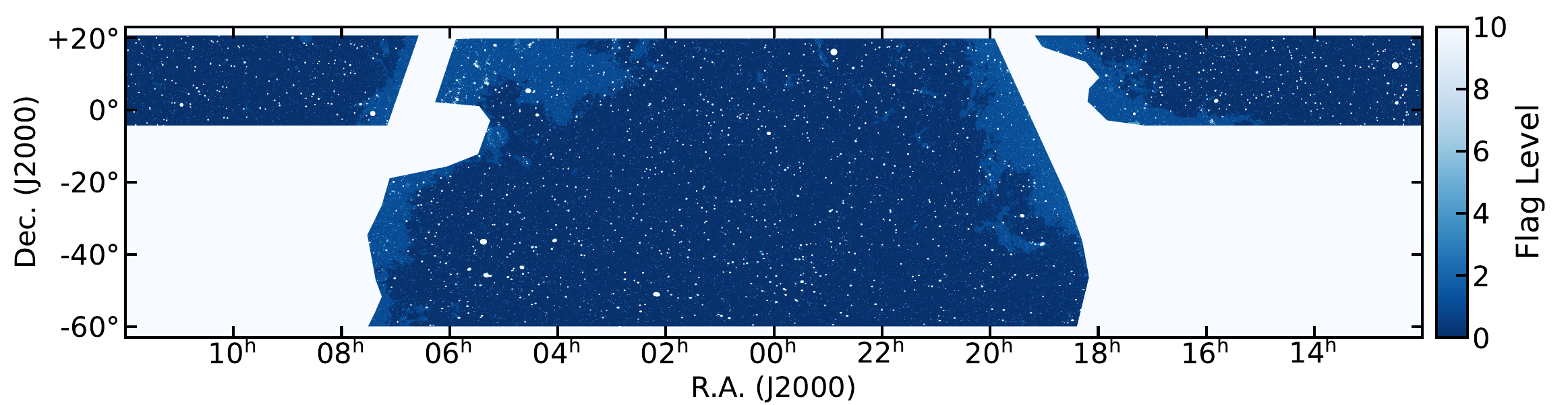}
\caption{Flag values across the cluster search region, used to identify
objects located in potentially problematic regions (see Section~\ref{sec:flagging}).
Regions where \texttt{flags = 0} are taken to be `clean'; most of the \texttt{flags = 1} region corresponds to the dust mask; and regions with larger flag values contain point sources (e.g. AGNs and dusty galaxies) detected by ACT. 
The full \maxSearchArea{}\,deg$^2$ cluster search area (before masking and flagging)
is shown as the black outline. The total `clean' area within this covers
\flagZeroArea{}\,deg$^2$.
}
\label{fig:flagMask}
\end{figure*}

In H21, we took the approach of masking point sources, zeroing such regions of the 
filtered map. For the DR6 analysis, we make use of \textsc{Nemo}'s new flagging
features instead. We produce a separate flag map (see Figure~\ref{fig:flagMask})
that records pixels in the map that are potentially affected by any of the following
issues:

\begin{enumerate}
\item{
\textit{Point source subtraction.} 
Map pixels where a point source (found in Steps 1-2 in Section~\ref{sec:multipass})
was subtracted are flagged. These are
considered independently across the f090, f150, and f220 maps, so a point source
detected in all three frequencies will be flagged multiple times in the flag mask. The main motivation for subtracting and flagging, rather than masking, was to attempt
to recover clusters affected by central point sources, but this was not successful (see Section~\ref{sec:sourceContamination}).
Objects affected by point source subtraction
are indicated with a non-zero value for
\texttt{finderFlag} in the cluster catalog
(see Table~\ref{tab:FITSTableColumns}).
}

\item{
\textit{Dusty regions.}
We identify dusty regions by thresholding the \textit{Planck} 353\,GHz map above
2\,mK (this is more conservative than the 4\,mK threshold used in H21).
This threshold was chosen based on visual inspection of cluster candidates in dusty
regions, where positive residuals, identified as spurious detections, were found
to be correlated with dust features visible in the f220 map.
Large scale dust features, mostly at low Galactic latitude, can be seen in Figure~\ref{fig:flagMask}.
Objects that are located in dusty regions
are indicated with a non-zero value for
\texttt{dustFlag} in the cluster catalog
(see Table~\ref{tab:FITSTableColumns}).
}

\item{
\textit{Extended objects.}
Objects in the local Universe with large angular size, such as nearby galaxies
and planetary nebulae, can also cause spurious detections in the filtered maps.
We therefore flag circular regions, of radius 1.2 times the total elliptical
aperture radius, around objects listed in the
2 Micron All Sky Survey (2MASS)
Large Galaxy Atlas
\citep{Jarrett_2003}. We also flag circular regions around all objects with
radius $>1\arcmin$ in the New General Catalog
\citep[NGC2000 version;][]{Sinnott_1988}.
Objects located within regions containing
extended foreground objects 
are indicated with a non-zero value for
\texttt{extObjFlag} in the cluster catalog
(see Table~\ref{tab:FITSTableColumns}).
}

\item{
\textit{Bright stars.}
It is difficult to optically confirm and measure photometric redshifts for cluster
candidates that are near bright stars. We use $W1$-band (3.4\,$\mu$m wavelength) 
magnitude measurements in the AllWISE catalog \citep[produced by the Wide-field Infrared Survey Explorer (WISE) mission;][]{Wright_2010, Mainzer_2011} 
to identify such regions. We mask circular regions of radius 
$3.5\arcmin$ ($3 < W1 < 7$\,mag); $7\arcmin$ ($-1 < W1 < 3$\, mag);
and $10\arcmin$ ($W1 < -1$\,mag).
Objects located close to bright stars
are indicated with a non-zero value for
\texttt{starFlag} in the cluster catalog
(see Table~\ref{tab:FITSTableColumns}).
}
\end{enumerate}

There are also two, separate flag classes that are not recorded in the flag mask:

\begin{enumerate}
\item{
\textit{Ring artifacts around high S/N clusters}. 
Clusters with large SZ signals (e.g., El Gordo, \citealt{Menanteau_2012},
detected at ${\rm S/N} = 82$ in DR6) can induce ringing artifacts in the filtered maps,
some of which have S/N above the object detection threshold. We identify candidate
spurious ring features by creating a segmentation map that identifies objects made
from a minimum of 30 connected pixels with ${\rm S/N} > 0.5$. When constructing the
cluster catalog, we then check if an object is found in this candidate ring mask.
If the object is then found to be located within 12$\arcmin$ of another cluster
candidate detected with ${\rm S/N} > 20$, the object is flagged as a ring artifact
(\texttt{ringFlag = 1} in the cluster candidate list). For convenience, we increment
the \texttt{flags} value of the object in the cluster catalog as well.
}

\item{
\textit{Objects split across tile boundaries.}
Some high S/N cluster candidates detected at the edge of a tile may be detected in
more than one tile. 
Objects which are located within $2.5\arcmin$ of another candidate found
in a different tile have the \texttt{tileBoundarySplit} flag set in the cluster
candidate list produced by \textsc{Nemo}. These are then
visually inspected in the filtered maps, and the candidate corresponding to the part of the cluster that
`leaked' across the tile boundary (generally the candidate with the lower S/N, corresponding to the edge rather than center of a cluster) is rejected when the final cluster catalog is
assembled. This process could be automated or avoided entirely with some modifications to
\textsc{Nemo}, but this is left to future work as this situation occurs rarely in the ACT DR6
catalog (only 47 out of \totalCandidates{} candidates have \texttt{tileBoundarySplit = 1}).
}
\end{enumerate}

Pixels (and in turn cluster candidates) may be flagged multiple times using the above criteria. The \texttt{flags} value in the catalog contains the sum across all the above categories, except for \texttt{tileBoundarySplit} (see Table~\ref{tab:FITSTableColumns}). Hence, a conservative, clean catalog can be selected by choosing only objects for which 
\texttt{flags = 0}. This corresponds to
\totalFlagZeroCandidates{} out
of \totalCandidates{} candidates detected
with $\snrt > 4$. We focus on this subsample of objects for the results presented
in this work, but the full list of candidates, regardless of \texttt{flags} level,
is included in the data products released with this paper.

\subsection{False Positive Rate}
\label{sec:signalFree}

We estimate the expected number of false positive detections in the cluster
candidate list by running \textsc{Nemo} on signal-free simulations, using a similar
method to that described in Section~\ref{sec:multipass} above.
We create simulated f090, f150, and f220 maps that contain CMB, $1/f$ noise, and
white noise that follows the inverse variance maps produced by the map maker
\citep{Naess_2025}. We scale the white noise level up by a factor of 1.16 across
all the simulated frequency maps, which results in the output noise map estimated by
\textsc{Nemo} having the same median level (to within 2\%) of the noise map
obtained from running on the real data. 
\footnote{Rescaling the white noise level
has no effect on the number of false positive detections as a function of S/N $\snrt$ (i.e., this makes no difference to the results shown in Figure~\ref{fig:signalFree}). We apply this factor only for consistency with the end-to-end simulations described in Appendix~\ref{ap:optBiasSims}.}
We do this because the simulations do not contain point sources,
or other astrophysical components (e.g., dust, Cosmic Infrared Background), resulting in an underestimate
of the noise level seen in the real data. 
As the simulations do not contain point sources, 
we only run
Steps 3--4 of the detection algorithm described in Section~\ref{sec:multipass}.
Nevertheless, this exercise gives an idea of the expected false positive cluster
candidate detection rate due to noise fluctuations, and is consistent
with the fraction of optically confirmed clusters detected as a function of
S/N in the real data, as shown in Section~\ref{sec:purity}.

\begin{figure}
\includegraphics[width=\columnwidth]{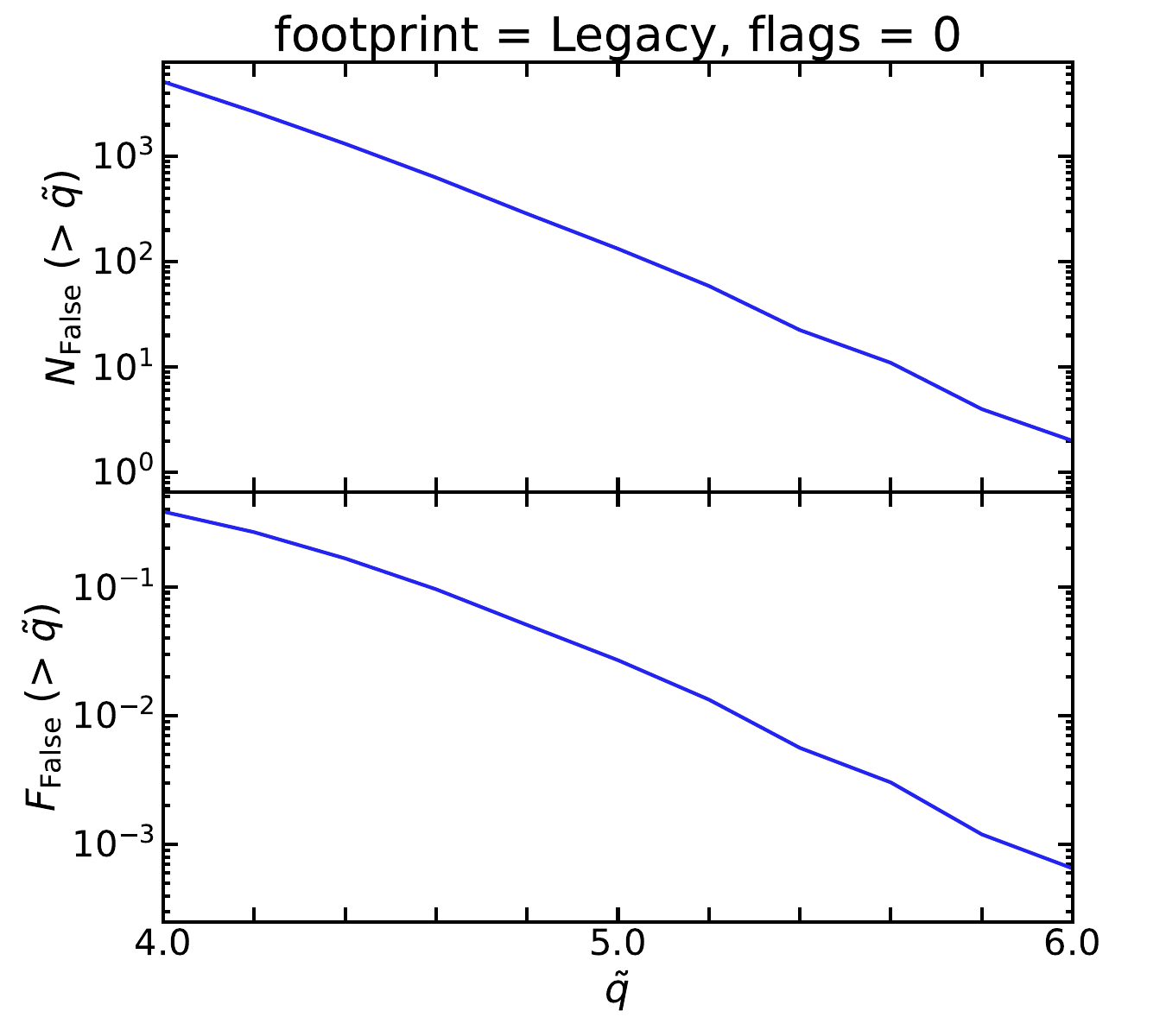}
\caption{An estimate of the false positive cluster candidate detection
rate in the \texttt{flags = 0} area of the \cosmoSearchArea\,deg$^2$
\cosmo footprint. The top panel shows the cumulative number of
detections as a function of signal-to-noise $\snrt$ in signal-free
simulations (averaged over 10 realizations).
The bottom panel presents the expected
false positive fraction of the real cluster candidate list as a
function of $\snrt$ cut.}
\label{fig:signalFree}
\end{figure}

Figure~\ref{fig:signalFree} shows the results for the \texttt{flags = 0} area
within the \cosmo footprint, which are similar to those reported in H21 for this
exercise. This shows that we expect only 11 false detections due to noise
fluctuations in the \cosmo footprint for the $\snrt > 5.5$ cut used to define
the `Legacy sample'. This is a lower limit as the simulations used do not
include all possible sources of contamination (e.g., artifacts related to point sources, dust, etc.).

\subsection{Position Recovery}
As in H21, we use source injection simulations to assess the accuracy of position
recovery for cluster candidates. Note that this only accounts for the effect of noise
fluctuations in the map, and not any other astrophysical causes (e.g., cluster
mergers). We paint UPP model clusters into the real ACT frequency maps with
uniformly distributed amplitudes and sizes selected from
$\theta_{500c} (\arcmin{}) \in \{7.8, 4.2, 2.4, 1.5\}$. We recover a total
of $\approx 10^{5}$ detections with $\snrt > 4$ over these injection runs. To
quantify the accuracy of position recovery, we use the Rayleigh distribution,
\begin{equation}
\label{eq:rayleigh}
P(\theta, \sigma) = \frac{\theta}{\sigma^2} \exp{\left(-\theta^2 / 2 \sigma^2 \right)}\,,
\end{equation}
where $\theta$ is the angle between the true and recovered cluster position, 
and $\sigma$ is the scale parameter for the distribution. 

\begin{figure}
\includegraphics[width=\columnwidth]{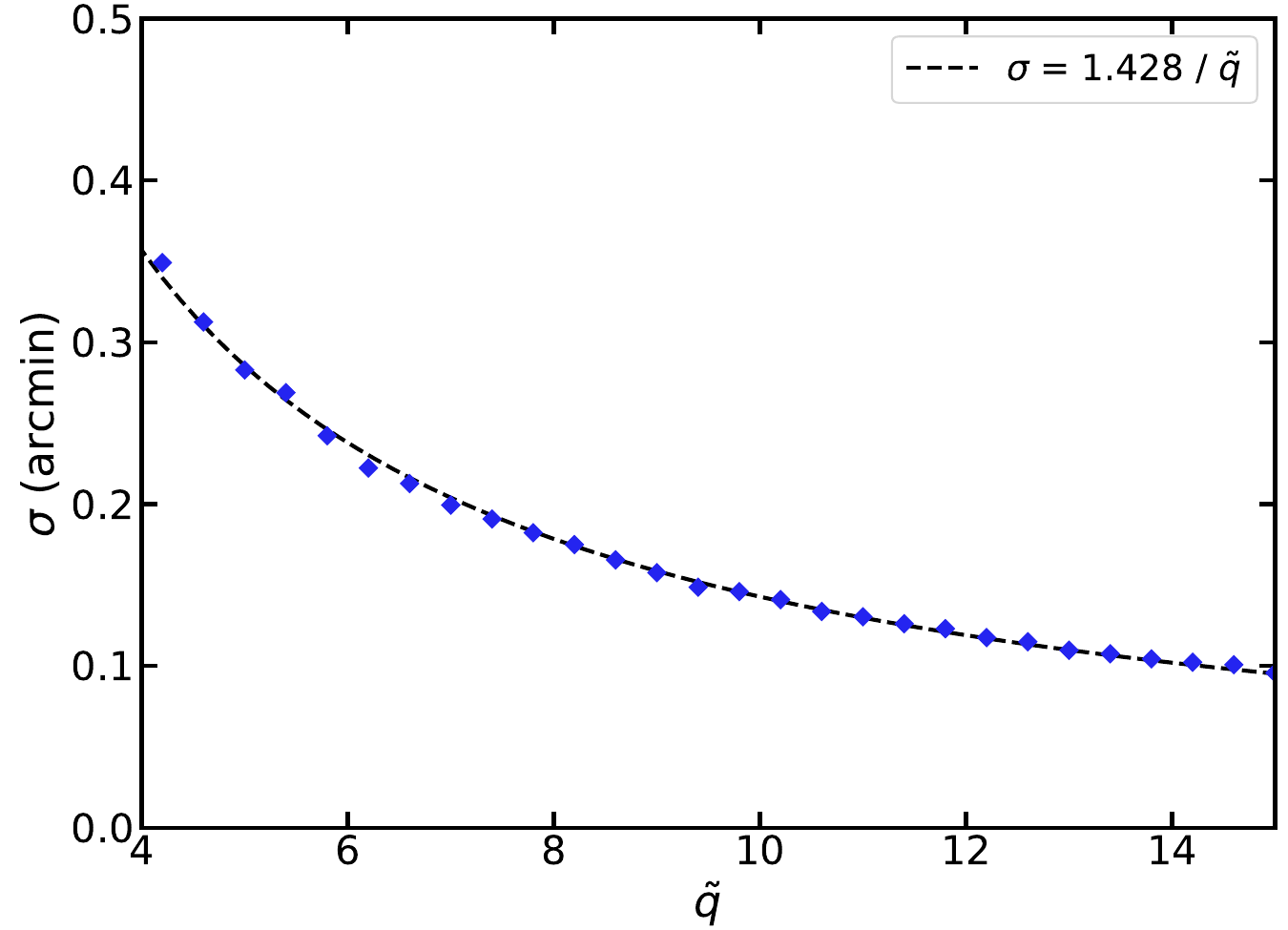}
\caption{Results of fitting the distribution of recovered cluster position
offsets obtained from source injection simulations with the Rayleigh distribution
(Equation~\ref{eq:rayleigh}), as a function of signal-to-noise $\snrt$. 
A single parameter model is a good description for how $\sigma$ changes with
$\snrt$.}
\label{fig:rayleigh}
\end{figure}

Figure~\ref{fig:rayleigh} shows the result of fitting Equation~(\ref{eq:rayleigh}) to
recovered position offsets, binned by $\snrt$, to find $\sigma$. We find that
a good fit to $\sigma$ as a function of $\snrt$ is given by 
$\sigma = 1.428 / \snrt$, with $\sigma$ specified in arcminutes.
We make use of this model in cross-matching ACT cluster candidates against other
cluster catalogs in Section~\ref{sec:crossMatching} below.

\subsection{Mass Estimates} 
\label{sec:masses}

We estimate cluster masses with the same methodology used for previous ACT cluster
catalogs. The motivation for the method is described in H13.
We assume that the cluster central Comptonization parameter $\yt$ is related to
mass via
\begin{equation}
\tilde{y}_{0} = 10^{A_0} E(z)^2 \left( \frac{M_{\rm 500c}}{M_{\rm pivot}} \right)^{1+B_0} Q(\theta_{\rm 500c}) f_{\rm rel} (M_{\rm 500c}, z) \, ,
\label{eq:y0}
\end{equation}
where $10^{A_0}$ is a normalization that sets the overall mass scale, 
$B_0 = 0.08$,  $M_{\rm pivot} = 3 \times 10^{14}$\,$M_{\sun}$, 
$Q(\theta_{\rm 500c})$ is the filter mismatch function (this accounts for
the angular size difference between clusters with different masses and 
redshifts to the reference $2.4 \arcmin$ filter scale used for measuring $\yt$, 
see H13), and $f_{\rm rel}$ is a relativistic correction (implemented as
described in H21, using the formulae of \citealt{Itoh_1998}).
$E(z)$ describes the evolution of the Hubble parameter in units of $H_{\rm 0}$ with redshift, i.e., 
$E(z) = (\Omega_{\rm m}(1+z)^3 + \Omega_{\Lambda})^{1/2}$.

In a departure from previous ACT cluster catalog papers, we adopt 
$10^{A_0} = 3.0 \times 10^{-5}$ as the normalization for the scaling relation,
rather than the value derived from \citet[][based on X-ray
observations of low redshift clusters; $10^{A_0} = 4.95 \times 10^{-5}$, see H13]{Arnaud_2010}, that was used for
both previous ACT and \textit{Planck} cluster catalogs. All figures in this paper
that use the
symbol $M_{\rm 500c}$ assume $10^{A_0} = 3.0 \times 10^{-5}$. 
For comparison with previous work, we still report mass estimates that use the \citet{Arnaud_2010} mass calibration in the cluster catalog (see Section~\ref{sec:catalog}), for which we use the symbol
$M_{\rm 500c}^{\rm A10}$.
The $10^{A_0} = 3.0 \times 10^{-5}$ normalization chosen here, which is $\approx 60$\% of the previous value, gives masses
that are consistent with stacked weak-lensing analyses of the ACT DR5 cluster sample,
for which $M_{\rm 500c}^{\rm A10} / M_{\rm 500c}^{\rm WL} = 1 - b = 0.65 \pm 0.05$
(\citealt{Robertson_2024}, based on Kilo-Degree Survey [KiDS] data; which is consistent with results
found by \citealt{Shirasaki_2024} using Hyper Suprime-Cam [HSC] data), where the superscript WL indicates the
average stacked weak-lensing mass.

To obtain mass estimates from the SZ signal using Equation~(\ref{eq:y0}),
correcting for the Eddington bias \citep{Eddington_1913} due to the steepness of the halo mass function
(HMF) and $\sigma_{\rm int} = 0.2$ log normal intrinsic scatter (based
on the results of numerical simulations, see H13),
we use the same method as in previous ACT cluster papers. We compute the posterior probability
\begin{equation}
P( M_{\rm 500c} | \tilde{y}_{0}, z) \propto P ( \tilde{y}_{0} | M_{\rm 500c}, z) P(M_{\rm 500c} | z) \, ,
\label{eq:PM500}
\end{equation}
where $P(M_{\rm 500c} | z)$ is the HMF at redshift $z$, for which we use the implementation of the \citet{Tinker_2008} HMF in the Core Cosmology Library v3.1.2 \citep[\texttt{CCL};\footnote{\url{https://github.com/LSSTDESC/CCL}}][]{Chisari_2019}. 
We assume $\sigma_8 = 0.80$ for HMF calculations throughout this work.
The uncertainties we quote on our mass estimates account for measurement errors
and the effect of intrinsic scatter, but otherwise do not include uncertainties
on the scaling relation parameters themselves. We also provide a set of mass
estimates in the cluster catalog (see Section~\ref{sec:catalog}) that do not
include the Eddington bias correction, allowing them to be compared against, 
e.g., the mass estimates in the PSZ2 catalog \citep{PlanckPSZ2_2016}.

\subsection{Completeness} 
\label{sec:completeness}
We estimate the completeness of the cluster sample as a function of mass by
using a semi-analytic method that has been validated using end-to-end simulations, accounting for the optimization bias introduced by the matched filter (see Appendix~\ref{ap:optBiasSims} below). The overall approach 
is similar to that used by \citet{Planck2015_XXIV}. This accounts for the completeness due to the noise properties of the map, but does not account for the impact of sources correlated with clusters (see Section~\ref{sec:sourceContamination} below for a brief discussion of this issue).
 
We calculate the area-weighted average completeness $\chi$ 
above a given signal-to-noise cut 
$\snrt_{\rm cut}$ on the ($M_{\rm 500c}$, $z$) grid over $i$ patches in the map (with
$i$ different noise levels) using
\begin{equation}
\label{eq:completeness}
\chi = \int d{\rm ln}\yt \, P(\yt|M_{\rm 500c},z) \sum_i \chi_i(\snrt) W_i   \, ,
\end{equation}
where
\begin{equation}
P(\yt|M_{\rm 500c},z) = \frac{1}{\sqrt{2\pi} \sigma_{\rm int} } 
\exp \left[ - \left(
\frac{\ln \yt - \ln \yt^{\rm corr} }{\sqrt{2} \sigma_{\rm int}} 
\right)^2 \right]
\, ,
\end{equation}
and
\begin{equation}
\chi_i(\snrt) = \frac{1}{2}
\left[
1 - {\rm {erf}}
\left(
\frac{{\snrt_{\rm cut} - \snrt_i}}  {\sqrt{2}}
\right)
\right]
\,.
\end{equation}
Here, $\yt$ is taken to be the true value obtained through Equation~(\ref{eq:y0}),
neglecting the relativistic correction (as this is not significant for clusters near our chosen signal-to-noise cut),
and $\yt^{\rm corr}$ is this value with optimization bias applied, using the model
described in Appendix~\ref{ap:optBiasSims}.
The area weight for patch $i$ is represented
by $W_i$, and $\snrt_i = \yt / \delta \yti$, where $\delta \yti$ is the noise level in
patch $i$.

\begin{figure}
\includegraphics[width=\columnwidth]{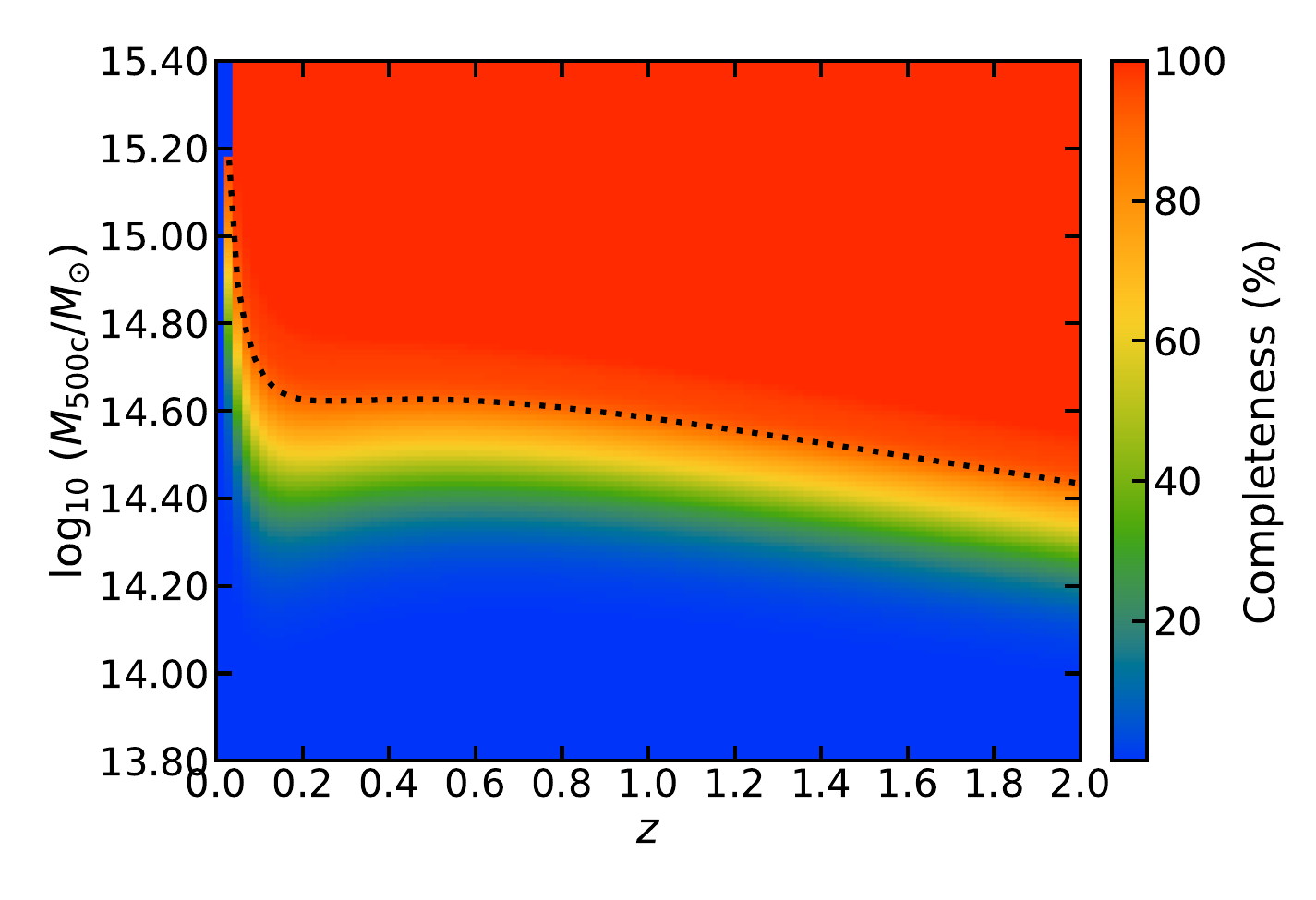}
\caption{Completeness in the ($M_{\rm 500c}$, $z$) plane for a signal-to-noise cut $\snrt_{\rm cut} = 5$, 
for the \texttt{flags = 0} search area, calculated as described in
Section~\ref{sec:completeness}. The dotted black line indicates the 
90\% completeness limit.}
\label{fig:MzComp}
\end{figure}

\begin{figure*}
\includegraphics[width=\textwidth]{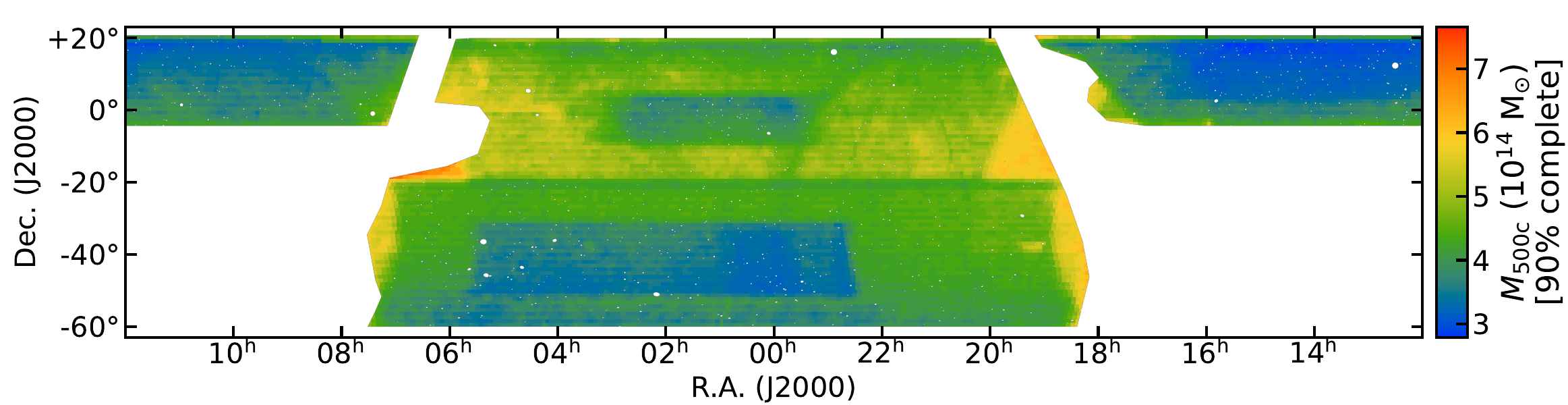}
\caption{Mass limit map (90\% completeness) for clusters with $\snrt_{\rm cut} = 5$, evaluated
at $z = 0.5$ as described in Section~\ref{sec:completeness}. The spatial variation in
the mass limit is largely a reflection of the ACT survey strategy, which evolved over
the lifetime of the project.}
\label{fig:massLimMap}
\end{figure*}

\begin{figure*}
\includegraphics[width=\columnwidth]{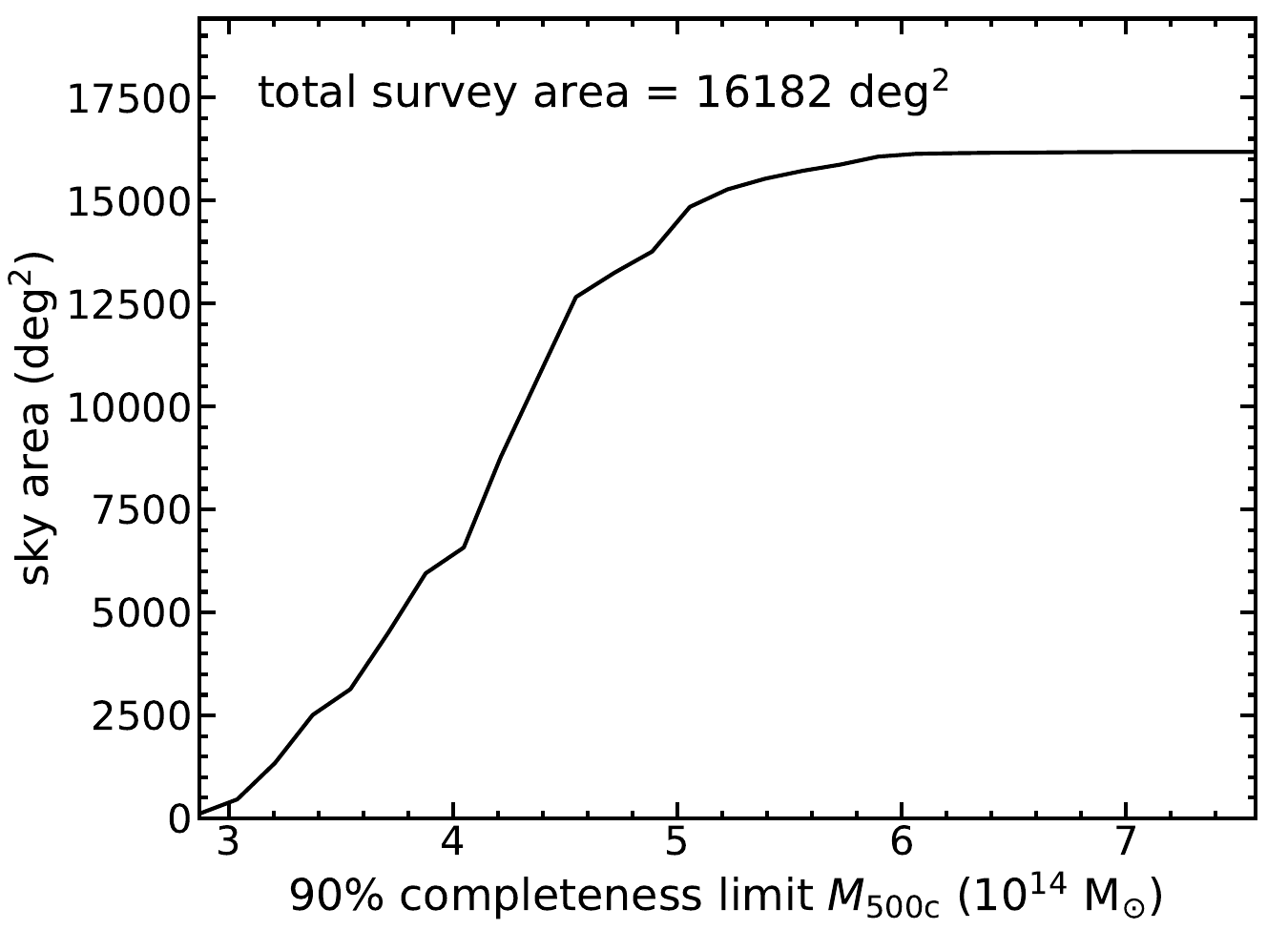}
\includegraphics[width=\columnwidth]{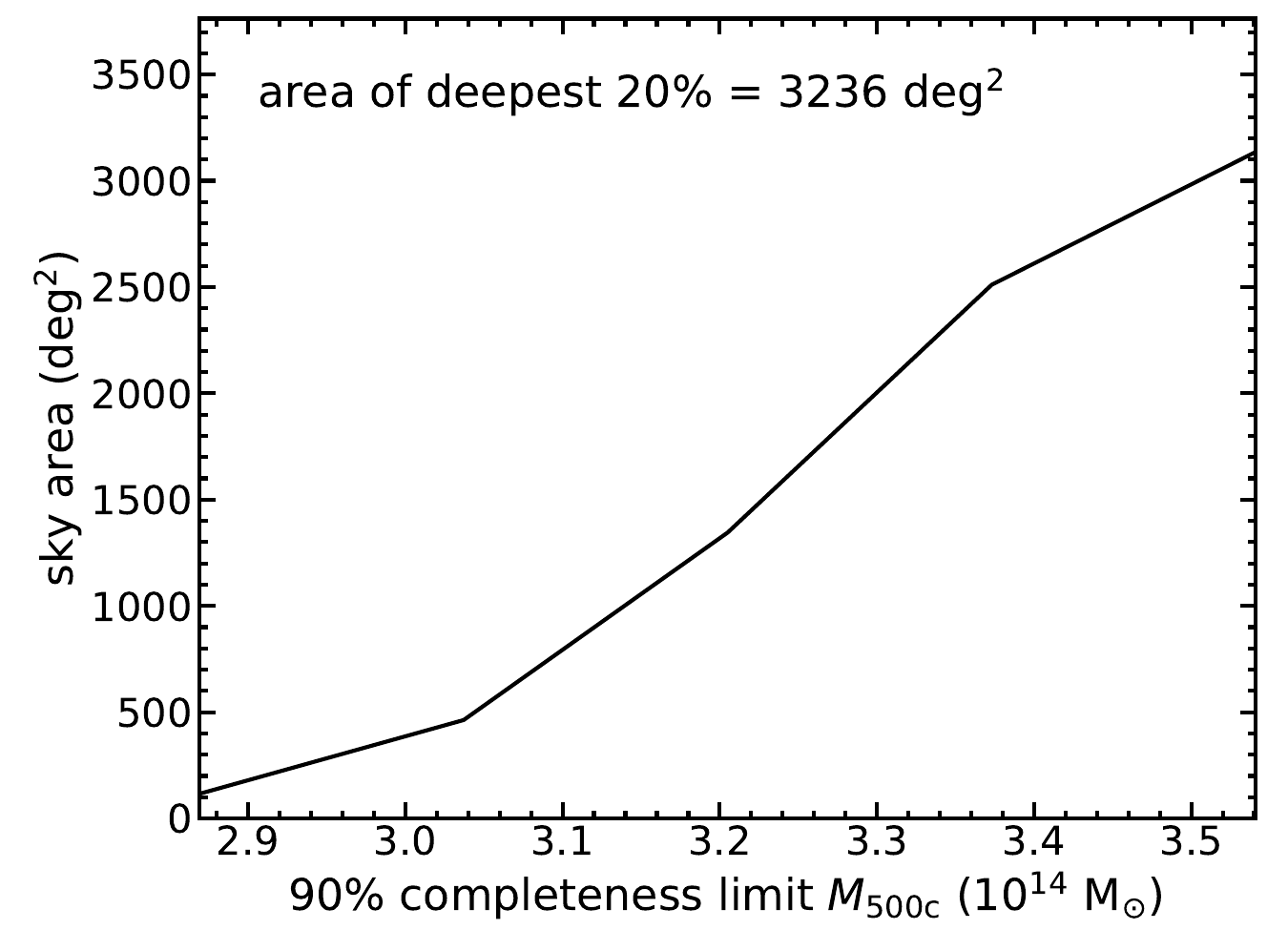}
\caption{
Sky area for which the signal-to-noise $\snrt = 5$ cluster sample is at least 90\% complete above the given mass limit 
for clusters with $z = 0.5$ 
(see Figure~\ref{fig:MzComp} for an illustration of the mild $z$-dependence over most of the redshift range).
The left panel shows this for the full survey area, while the right panel shows this for the deepest 20\% of survey area.}
\label{fig:cumulativeMassLim}
\end{figure*}

Figure~\ref{fig:MzComp} shows the area-weighted average completeness in the 
($M_{\rm 500c}$, $z$) plane for $\snrt > 5$ over the \flagZeroArea{}\,deg$^2$ 
\texttt{flags = 0} search area, as calculated using Equation~(\ref{eq:completeness}),
assuming the scaling relation described by Equation~(\ref{eq:y0}), the optimization bias model described in Appendix~\ref{ap:optBiasSims}, log-normal intrinsic scatter, and Gaussian noise. 
At $z = 0.5$,
the 90\% mass completeness limit is 
$M_{\rm 500c} > \massLimFull \times 10^{14}$\,$M_{\sun}$. For the \cosmo footprint,
the limit is slightly lower, $M_{\rm 500c} > \massLimCosmo \times 10^{14}$\,$M_{\sun}$.
The mass limits within the DES, HSC, and KiDS footprints are similar.
These mass limits are higher than those quoted in H21 due to the lower value of
$10^{A_0}$ used in this work. The equivalent mass limit using the
\citet{Arnaud_2010} mass calibration is 
$M_{\rm 500c}^{\rm A10} > \massLimFullArnaud \times 10^{14}$\,$M_{\sun}$ for DR6, compared to $M_{\rm 500c}^{\rm A10} > 3.8 \times 10^{14}$\,$M_{\sun}$
for DR5, as reported in H21. 

The mass limit varies spatially, driven by the evolution of the ACT observing strategy
from 2008 to 2022, as shown by the map in Figure~\ref{fig:massLimMap}. 
We plot the cumulative area as function of mass limit derived from this map in Figure~\ref{fig:cumulativeMassLim}. The deepest 20\% of the full search area, 
covering 3237\,deg$^2$, is sensitive to clusters 
with lower limits to the mass lying in the range
$2.9 < M_{\rm 500c} / 10^{14}\,M_{\sun} < 3.5$.

\section{Optical/IR Follow-up and Redshifts}
\label{sec:optical}

In this Section we describe the process of assigning redshifts to ACT cluster candidates,
using the heterogeneous optical/IR data currently available over the large ACT cluster
search area. The overall procedure is similar to that used in H21; here we aim to
highlight differences with respect to the previous analysis, which come primarily
from new redshift sources that were not available at the time of the ACT DR5 release.
We may update the catalog of confirmed clusters with redshifts and mass estimates
after publication as newer optical/IR data become available. If so, a set of DR6 cluster
search data products with a new version number will be posted on LAMBDA.

\subsection{Redshift Assignment}
\label{sec:crossMatching}
We assign redshifts to most of the cluster candidates through automated cross-matching
against a set of cluster catalogs that we deem to be reliable, using a method that is
more conservative than used in H21. In the previous analysis, we matched clusters
using a combination of the ACT positional uncertainty, with an additional 0.5\,Mpc
projected distance cross-matching radius that was intended to account for
`astrophysical' uncertainties; i.e., miscentring in optical/IR catalogs
\citep[e.g.,][]{Zhang_2019, Ding_2025}. While cross-matching using a large projected
radius will result in a more complete sample, it carries the risk of matching
low-richness clusters in optical catalogs with noise fluctuations in the ACT map
(see Section~\ref{sec:signalFree}). Three such cases were identified by \citet{Ding_2025}
using the combination of ACT DR5 and HSC data.

We start by considering the Rayleigh distribution model
fit shown in
Figure~\ref{fig:rayleigh}. A fixed value of $\sigma = 0.357 \arcmin$ corresponds to $\snrt = 4$, which is the minimum S/N that we consider in our sample.
The 99.7 percentile of this Rayleigh distribution corresponds to a cross-match distance 
of $1.22 \arcmin$, which is equivalent to cross-matching within a projected distance
between 0.24--0.61\,Mpc over the redshift range $0.2 <  z < 2$.
For comparison, 95\% of cluster centers are found within a 0.5\,Mpc projected cross-match distance according to the \citet{Ding_2025} model (which includes a miscentered component, and has been cleaned of non-astrophysical sources of miscentring), while 99.7\% are found within 1\,Mpc.
We therefore choose to cross-match using the projected distance corresponding to $1.22 \arcmin$ at the cluster redshift, if this
is larger than 0.5\,Mpc (where the ACT positional accuracy may
be the dominant cause of miscentring), or otherwise use
a projected cross-match radius of 0.5\,Mpc. 
This choice is a compromise; it may result in up to 5\% of
clusters being missed by automated cross-matching, but it
avoids the possibility of adding a number of spurious cross-matches (particularly at low redshift) that would result from using a 1\,Mpc cross-match radius.

We perform automated cross-matching against cluster catalogs produced by the following teams:
CAMIRA \citep[][here we use the CAMIRA23B catalog; see Oguri et al., in prep.]{Oguri_2014, Oguri_2018};
eRASS1 \citep[][]{Bulbul_2024};
MaDCoWS \citep[][]{Gonzalez_2019, ThongkhamI_2024, ThongkhamII_2024};
MCMF \citep[][]{Klein_2023, Klein_2024};
redMaPPer \citep[][here we use the v0.8.5 DES Y6 catalog]{Rykoff_2014, Rykoff_2016};
SPIDERS \citep[][]{Clerc_2020, Kirkpatrick_2021};
SPT \citep[]{Bleem_2015, Bleem_2020, Bleem_2024};
WaZP \citep[][here we use the v5.0.12.6801 DES Y6 catalog]{Aguena_2021, Benoist_2025};
and
\citet{WH_2015}.

Table~\ref{tab:crossMatches} lists the total number of ACT DR6 cluster candidates cross-matched against
cluster catalogs drawn from the DES and HSC surveys, and
the chance association probability estimated from matching against ACT DR6 random catalogs.
We choose to only consider the DES and HSC cluster catalogs here because estimating the chance association probability requires accurate knowledge of the area of intersection between the ACT cluster search area and the respective optical survey (this is not readily available in a convenient format for all the optical cluster catalogs that we cross-match against).
We find that there is a 1-6\% probability of matching an ACT DR6 cluster candidate against a cluster in these optical cluster catalogs by chance. For comparison, the
less conservative automated cross-matching procedure
used in constructing the ACT DR5 cluster catalog resulted in a 5\% chance of ACT DR5 cluster candidates being randomly associated with redMaPPer or CAMIRA clusters (see H21).
The chance association probability depends upon the details of how the optical cluster catalog was constructed and the cuts applied to the catalog.

\begin{deluxetable*}{ccccc}
\small
\tablecaption{Summary of the results of automated cross matching against
              DES and HSC optical cluster catalogs that intersect with the ACT cluster search area
              (see Section~\ref{sec:crossMatching}).
\label{tab:crossMatches}
}
\tablehead{
\colhead{Catalog}           &
\colhead{Optical Survey}    &
\colhead{Cut Applied}       &
\colhead{Number of Matches} &
\colhead{Chance Association Probability (\%)}
}
\startdata
redMaPPer & DES Y6           & $\lambda > 20$     & 2154 & 1.0\\
WaZP      & DES Y6           & WaZP SNR~$ > 4$    & 3204 & 6.4\\
CAMIRA    & HSC Wide S23B    & $N_{\rm mem} > 16$ & 753  & 1.7\\
\enddata
\tablecomments{The Cut Applied column refers to a constraint placed on the input optical cluster catalog, which
               is either in terms of S/N reported in that catalog (WaZP), or richness ($\lambda$ or $N_{\rm mem}$).}
\end{deluxetable*}

We also estimate photo-$z$s using the \textsc{zCluster}\footnote{v0.5.1;
\url{https://github.com/ACTCollaboration/zCluster}} package,
as described in H21, the only change being that here we use updated photometry from
DR10.1 of the DESI Legacy Imaging Surveys \citep{Dey_2019}.
We ran \textsc{zCluster} on all \texttt{flags = 0} candidates, 
and all $\snrt > 5$ candidates with any \texttt{flags} value.
We visually inspect the available optical/IR imaging before assigning a \textsc{zCluster}
redshift estimate to a candidate. As in H21, in the \texttt{warnings} field of the
DR6 cluster catalog we indicate 
objects where the \textsc{zCluster} optical density contrast statistic has a low value 
($\delta < 3$), although this does not necessarily indicate that the redshift is 
unreliable.

\begin{figure}
\includegraphics[width=\columnwidth]{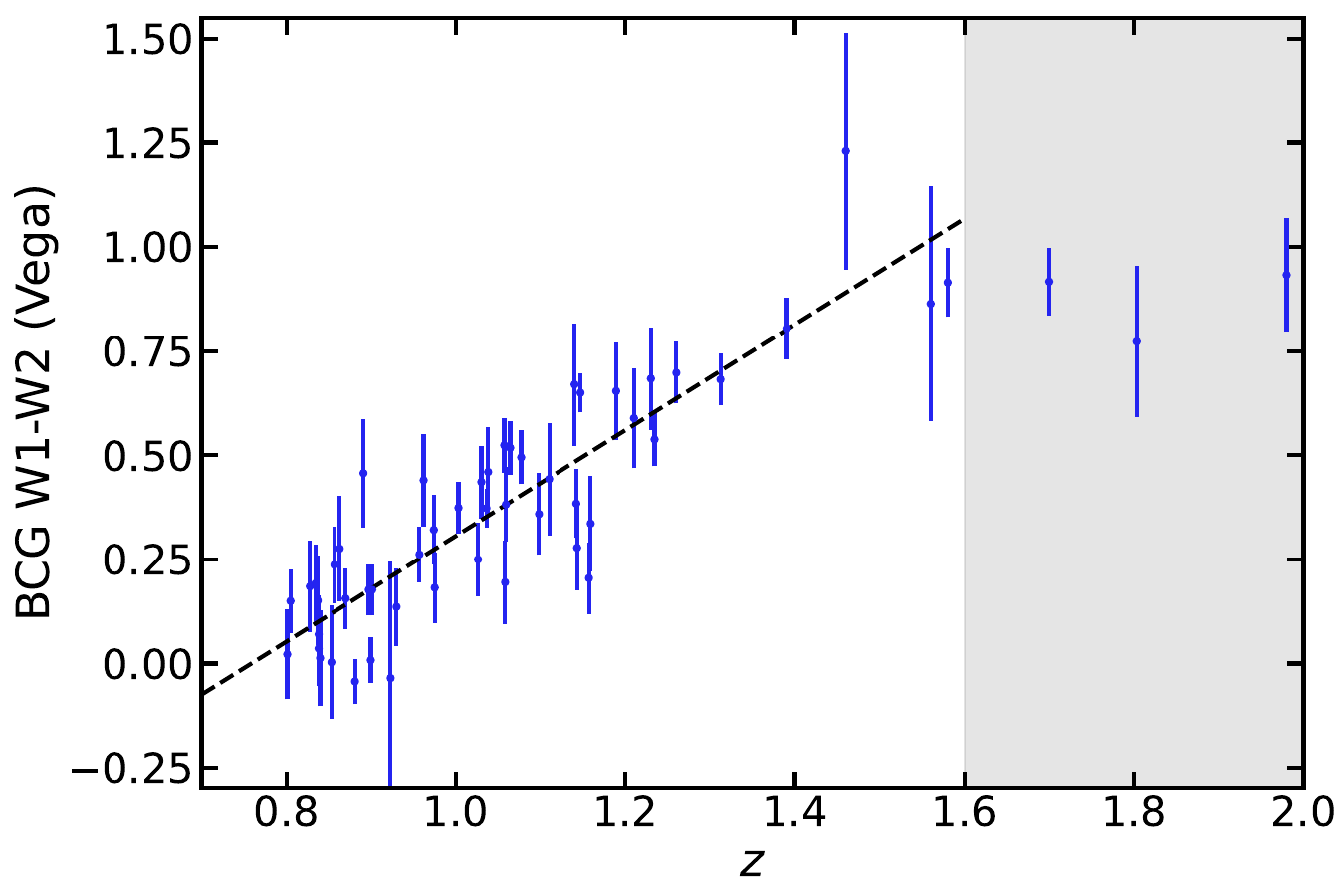}
\caption{Relationship between WISE $W1-W2$ color and $z$, derived from BCGs in clusters
with spectroscopic redshifts (see Section~\ref{sec:crossMatching}). 
This arises because of the `stellar bump' feature at 1.6\,$\micron$ in galaxy spectral energy distributions \citep[e.g.,][]{Papovich_2008}.
The dashed line
shows a weighted least squares fit to the data (Equation~\ref{eq:WISEBCGs}) over the range $0.8 < z < 1.6$. This was used to
obtain redshift estimates for \numWISEBCGz{} clusters with no other redshift source
available.}
\label{fig:WISEBCGs}
\end{figure}

After the above procedures were completed, we visually inspected WISE images of
$\snrt > 5.5$ candidates for which no redshift estimate was available. 
For those deemed to be high-redshift
clusters, we estimated their redshifts by using an empirical relationship between the 
$W1-W2$ color of the object identified as the brightest central galaxy (BCG),
\begin{equation}
W1-W2 = (1.52 \pm 0.08) \left( \frac{z}{1.2} \right) - (0.96 \pm 0.07)\,,
\label{eq:WISEBCGs}
\end{equation}
(see Figure~\ref{fig:WISEBCGs}).
This was obtained from a weighted least squares fit to WISE colors of 52 
$0.8 < z < 1.6$ BCGs in clusters with spectroscopic redshifts taken
from \citet{Stott_2010, Lidman_2013, Fassbender_2011, Fassbender_2011b, Santos_2011},
in addition to clusters with spectroscopic redshifts and BCG positions recorded in
the ACT cluster catalog. Redshifts were assigned to \numWISEBCGz{} clusters
using this procedure, by inverting Equation~(\ref{eq:WISEBCGs}),
with redshift uncertainty $\Delta z \approx 0.1$ resulting from the uncertainty
in the fit parameters. 

In addition to all of the above sources, we also used a number of redshifts from
miscellaneous sources in the literature. Table~\ref{tab:RedshiftSources} provides
a full breakdown of all the redshift sources used in constructing the catalog.

The final catalog contains a single redshift estimate per candidate, using what
we judged to be the best available redshift at the time of publication. We adopt spectroscopic redshifts 
where possible. 
Photometric redshifts assigned via automated cross-matching used the following order of preference: 
(1) redMaPPer in DES Y6; (2) CAMIRA S23B; (3) SPT; (4) MaDCoWS2 DR2; (5) WaZP in DES Y6; (6) eRASS1; (7) MCMF; (8) redMaPPer in SDSS; (9) \citet{WH_2015}. To some extent this prioritizes cluster catalogs that used deeper optical data and/or incorporate WISE IR photometry.

Note that for many objects in the catalog where multiple photometric redshifts are available, the different redshift sources are usually consistent, so in practice the order of assignment for photo-$z$s does not have much impact on the resulting ACT DR6 cluster catalog (though it does impact the number of redshifts drawn from each source as listed in Table~\ref{tab:RedshiftSources}).
However, we have identified \numProjected{} possible projected systems (i.e., candidates with plausible matches to multiple clusters along the line of sight),
with many of these being identified after visual inspection of
candidates with discrepant redshifts ($\Delta z > 0.5$) from multiple sources. 
These objects
are indicated in the \texttt{warnings} field of the cluster catalog. 
Only one of these projected systems has so far been identified as a clearly blended detection in the SZ filtered map
(ACT-CL\,\,J0335.1$-$4036, as first noted in H21).

\begin{deluxetable*}{cccl}
\tablecaption{Breakdown of redshift sources used in the ACT DR6 cluster catalog. The labels given in
the Source column correspond to those used in the \texttt{redshiftSource} column in the 
\texttt{FITS} Table format cluster catalog (see Table~\ref{tab:FITSTableColumns}).\label{tab:RedshiftSources}}

\tablehead{
\colhead{Source}       &
\colhead{Number} &
\colhead{Fraction (\%)} &
\colhead{Reference(s)} \\
}
\startdata
MaDCoWS & 2575 & 25.6 & \begin{minipage}[t]{90mm}\citet{Gonzalez_2019}; \citet{ThongkhamI_2024}; \citet{ThongkhamII_2024}\end{minipage}\\
redMaPPer & 2019 & 20.1 & \begin{minipage}[t]{90mm}\citet{Rykoff_2014}; \citet{Rykoff_2016}\end{minipage}\\
MCMF & 1403 & 14.0 & \begin{minipage}[t]{90mm}\citet{Klein_2023}; \citet{KleinSPT_2024}; \citet{Klein_2024}\end{minipage}\\
PublicSpec & 1223 & 12.2 & \begin{minipage}[t]{90mm}Includes: 2dFLens, OzDES, SDSS, VIPERS, SPIDERS, DESI;  \citet{2016MNRAS.462.4240B}; \citet{2017MNRAS.472..273C}; \citet{2020ApJS..249....3A}; \citet{2018AandA...609A..84S}; \citet{Clerc_2020}; \citet{Kirkpatrick_2021}; \citet{DESIDR1_2025}\end{minipage}\\
ACT & 1049 & 10.4 & \begin{minipage}[t]{90mm}Includes: \textsc{zCluster} redshifts from this work;  \citet{2013ApJ...765...67M}; \citet{2016MNRAS.461..248S}; \citet{Hilton_2018}; \citet{Hilton_2021}\end{minipage}\\
WaZP & 477 & 4.8 & \begin{minipage}[t]{90mm}\citet{Aguena_2021}; \citet{Benoist_2025}\end{minipage}\\
SPT & 436 & 4.3 & \begin{minipage}[t]{90mm}\citet{Bocquet_2019}; \citet{Bleem_2020}; \citet{Bleem_2024}; \citet{Kornoelje_2025}\end{minipage}\\
WH & 295 & 2.9 & \begin{minipage}[t]{90mm}\citet{WHL_2012}; \citet{WH_2015}; \citet{2022MNRAS.513.3946W}; \citet{2024ApJS..272...39W}\end{minipage}\\
eRASS1 & 285 & 2.8 & \begin{minipage}[t]{90mm}\citet{Bulbul_2024}\end{minipage}\\
CAMIRA & 128 & 1.3 & \begin{minipage}[t]{90mm}\citet{Oguri_2018}\end{minipage}\\
Lit & 91 & 0.9 & \begin{minipage}[t]{90mm}See table notes\end{minipage}\\
WISEBCGz & 59 & 0.6 & \begin{minipage}[t]{90mm}This work\end{minipage}\\
\hline
Total spectroscopic & 4133 & 41.2 & \nodata \\
Total photometric & 5907 & 58.8 & \nodata \\
\enddata
\tablecomments{Sources for literature redshifts: \citet{1991ApJS...77..363S}; \citet{1996ApJ...470..172S}; \citet{1997ApJ...479...70S}; \citet{1999ApJ...514..148D}; \citet{1999ApJS..125...35S}; \citet{2000ApJS..126..209R}; \citet{2000ApJS..129..435B}; \citet{2002ApJS..140..239C}; \citet{2003ApJ...594..154M}; \citet{2004AandA...423...75V}; \citet{2004AandA...425..367B}; \citet{Allen_2004}; \citet{2005ApJ...623L..85M}; \citet{2007ApJS..172..561B}; \citet{2008ApJ...677L..89G}; \citet{2008ApJ...682..821C}; \citet{2009AJ....137.2981G}; \citet{2009AJ....137.4795C}; \citet{2011ApJ...737...74G}; \citet{2011MNRAS.410.1797S}; \citet{2012AandA...538A..35C}; \citet{2014AandA...564A..17N}; \citet{2014ApJ...788...51N}; \citet{2014ApJ...792...76B}; \citet{2014ApJS..213...25S}; \citet{2015AandA...581A..14P}; \citet{2015AJ....150...91P}; \citet{PlanckPSZ2_2016}; \citet{Adami_2018}; \citet{2020Natur.577...39W}; \citet{Ansarinejad_2023}}
\end{deluxetable*}

\begin{figure*}
\centering
\includegraphics[width=59mm]{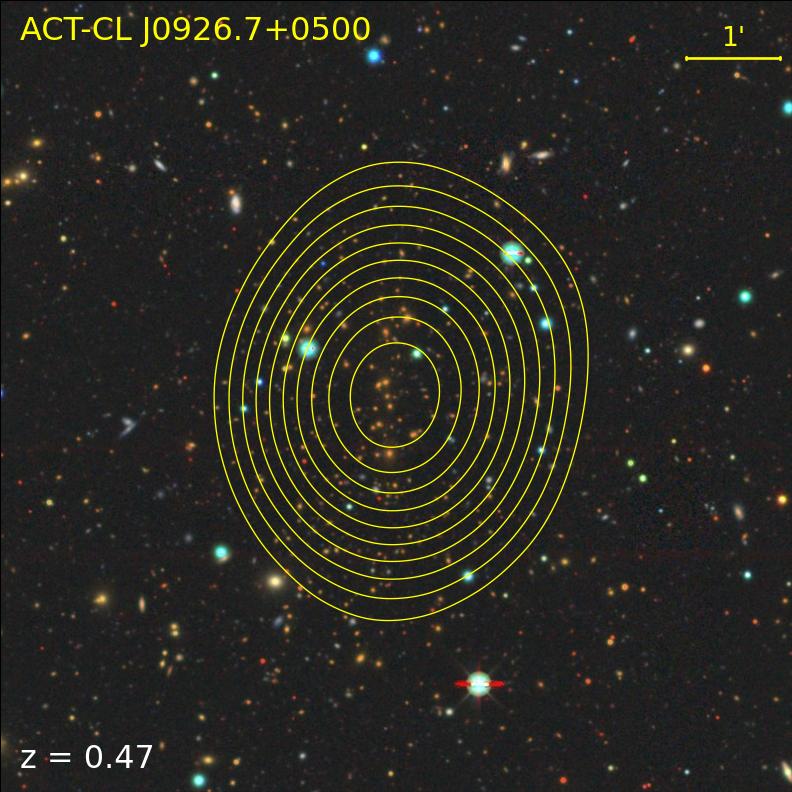}
\includegraphics[width=59mm]{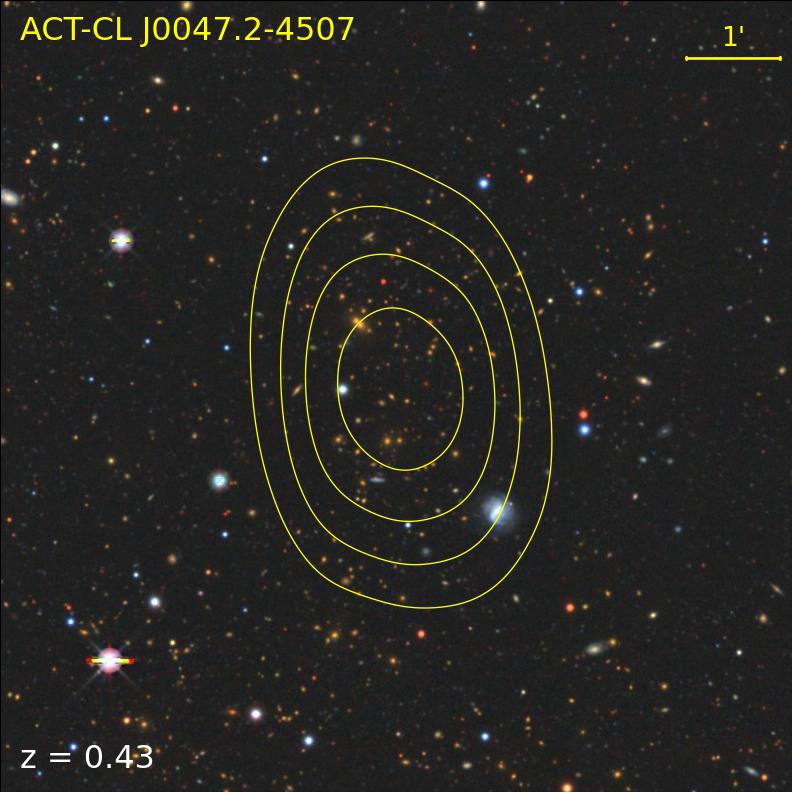}
\includegraphics[width=59mm]{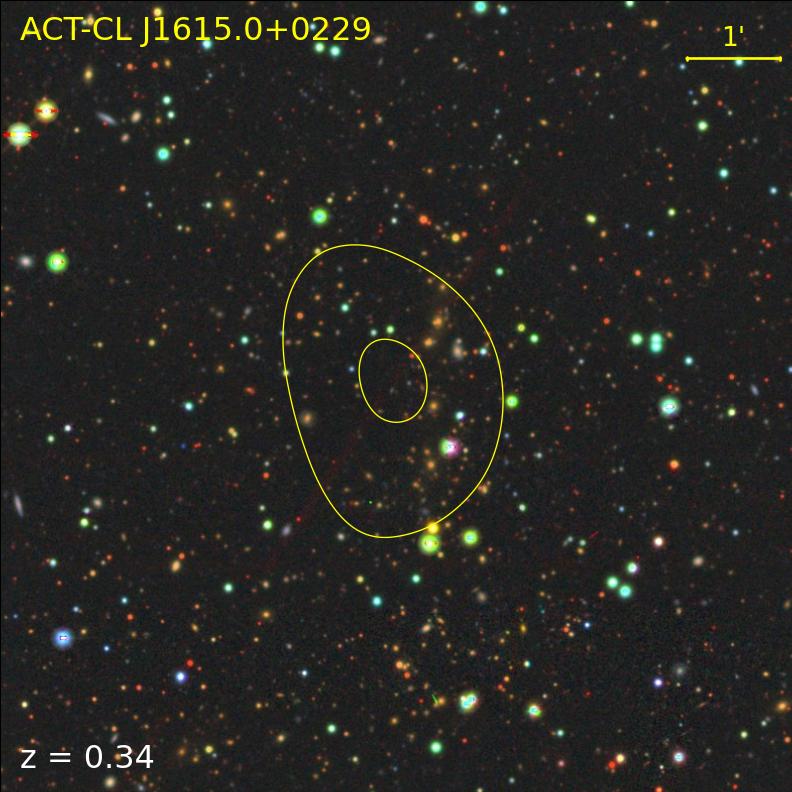}
\includegraphics[width=59mm]{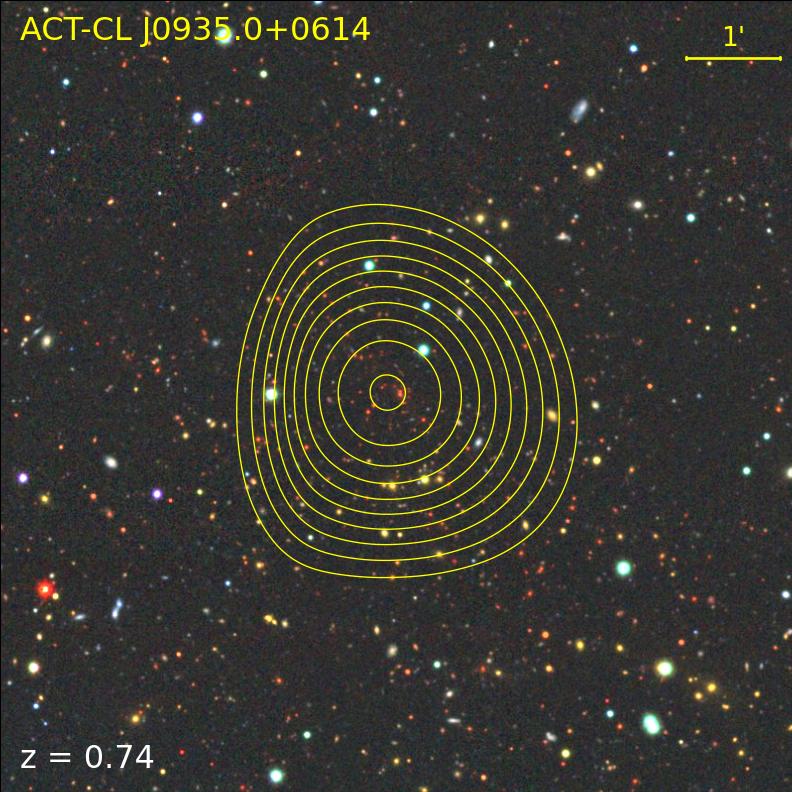}
\includegraphics[width=59mm]{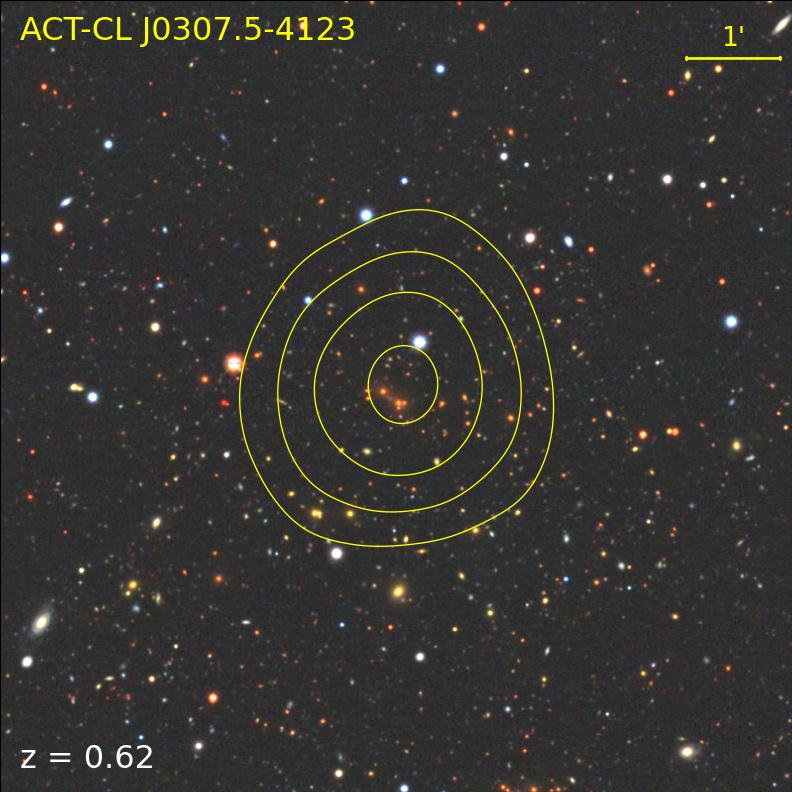}
\includegraphics[width=59mm]{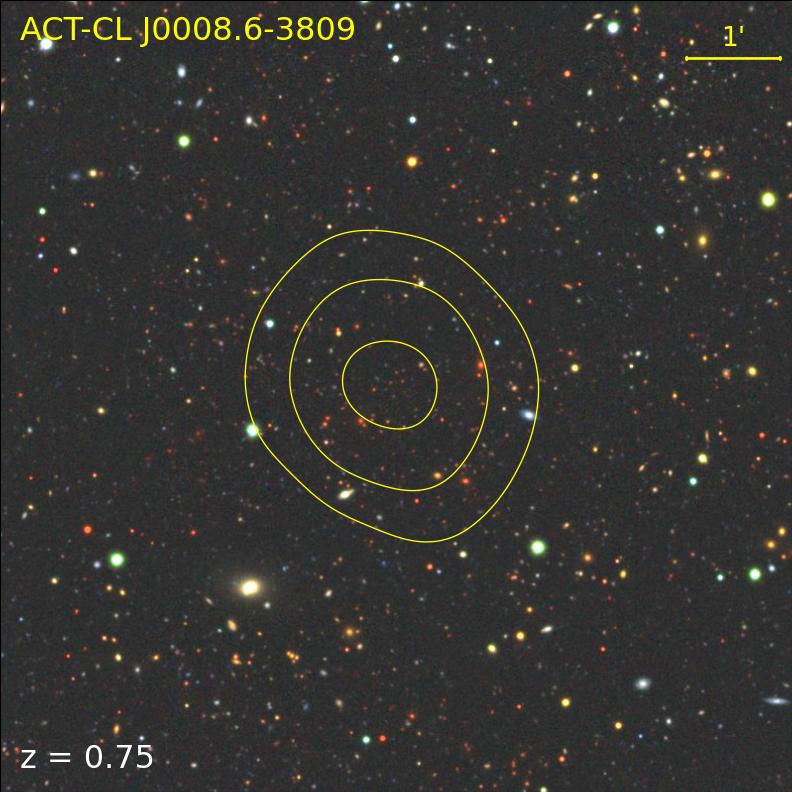}
\includegraphics[width=59mm]{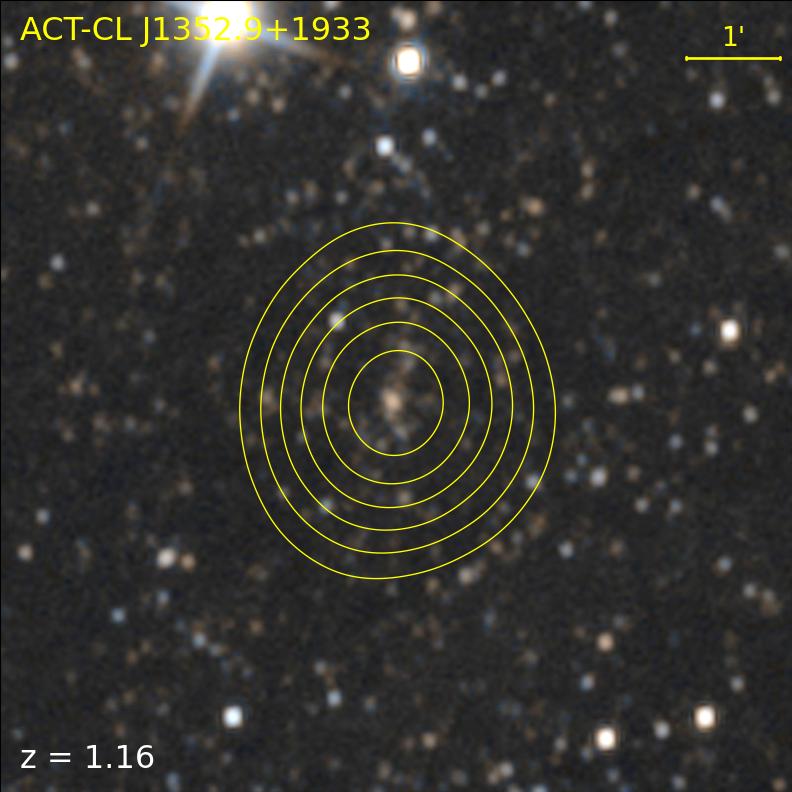}
\includegraphics[width=59mm]{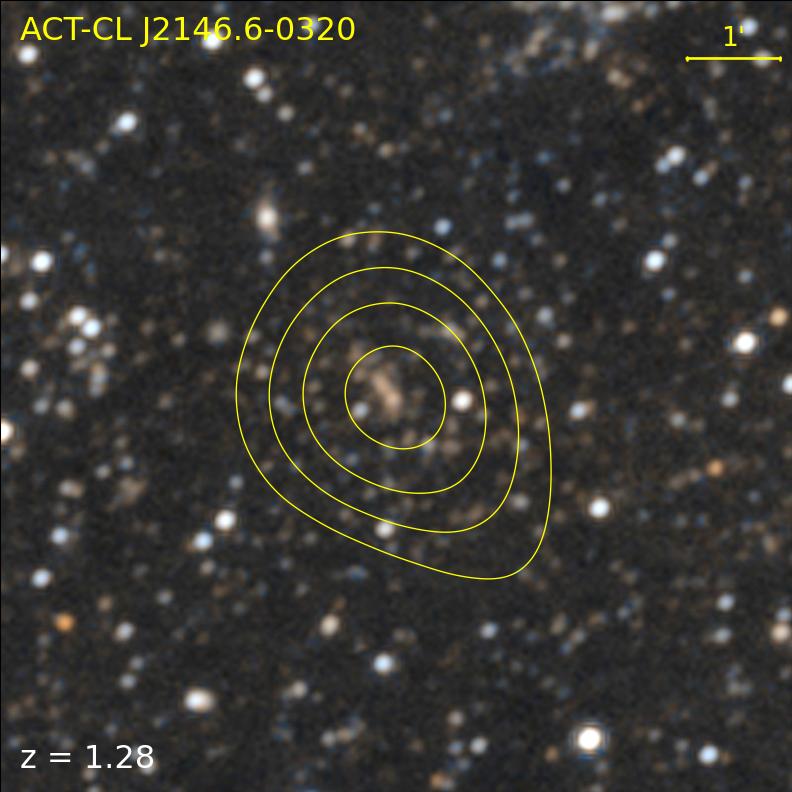}
\includegraphics[width=59mm]{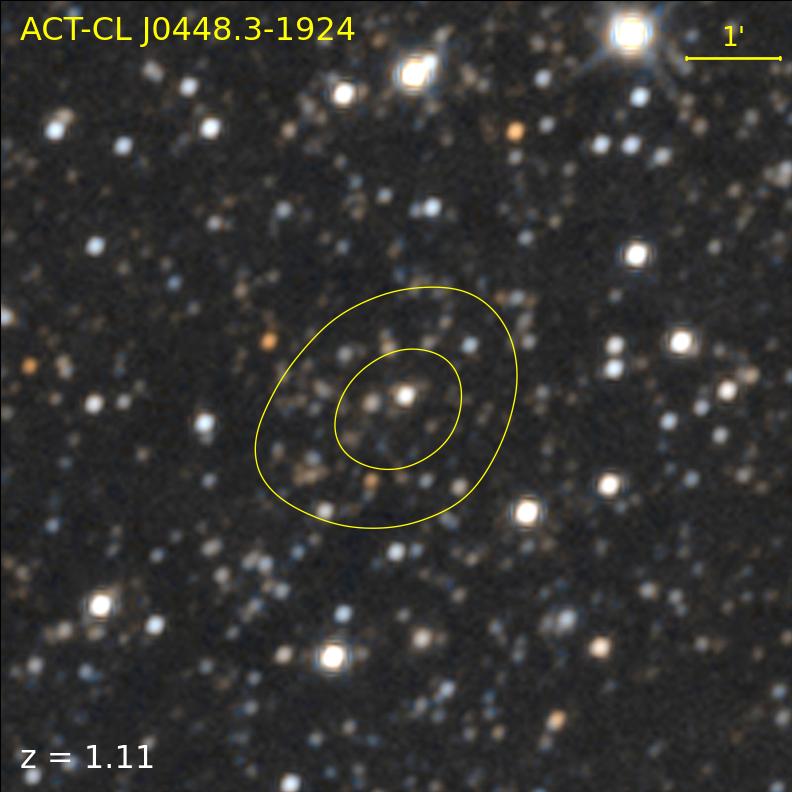}
\caption{Example optical/IR images of clusters in the ACT DR6 catalog. Objects in
columns from left to right have $\snrt > 20$, $10 < \snrt < 20$, and $5 < \snrt < 10$.
Objects in rows from top to bottom have $0 < z < 0.5$, $0.5 < z < 1$, $1 < z < 1.5$.
Images are taken from the DESI Legacy Imaging Surveys, with optical images (using $g$, $r$, $z$ data) shown
in the first
two rows, and WISE IR images (using $W1$, $W2$ data) shown in the bottom row. The contours mark regions of
constant signal-to-noise in the ACT filtered map, running from $3-25 \sigma$, with a
$2\sigma$ interval between each contour.}
\label{fig:opticalExamples}
\end{figure*}

Figure~\ref{fig:opticalExamples} shows optical/IR images of a selection of clusters
drawn from the catalog, arranged by $z$ and $\snrt$.

\subsection{Purity and Follow-up Completeness}
\label{sec:purity}

\begin{figure}
\includegraphics[width=\columnwidth]{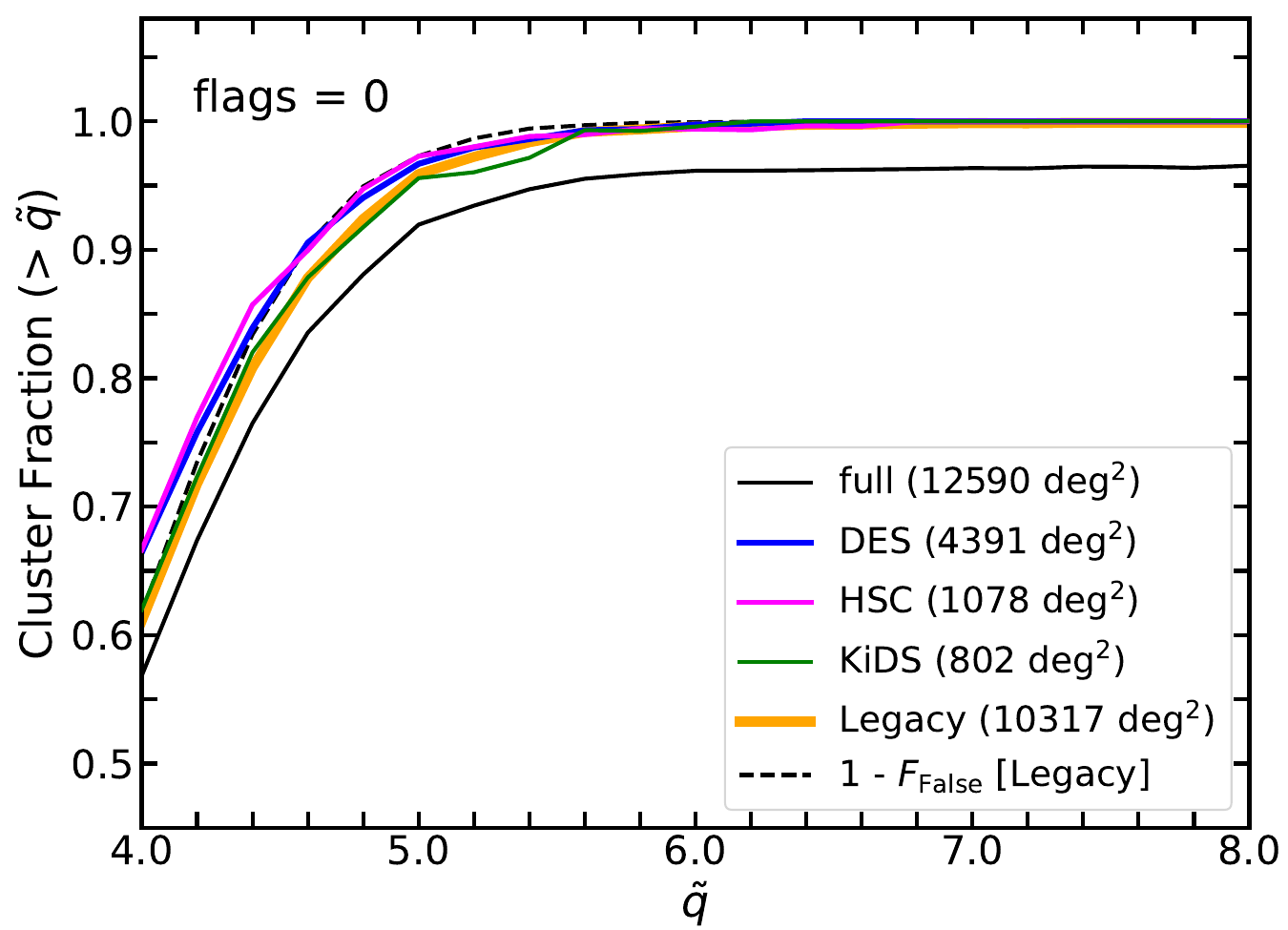}
\caption{
The fraction of DR6 cluster candidates with redshift measurements as a function of signal-to-noise
$\snrt$ in \texttt{flags = 0} areas of the cluster search region, within various
different footprints (solid colored lines). The dashed line (labeled $1-F_{\rm False}$)
shows the expected purity of the cluster sample within the \cosmo footprint, 
based on signal-free simulations (see Section~\ref{sec:signalFree}).
Redshift follow-up is $>99$\% complete for $\snrt > 5.5$ in the \cosmo footprint.
The full \texttt{flags = 0} search area, which includes regions at low Galactic
lattitude, is $>90$\% complete in terms of redshift follow-up (i.e., 90\% of candidates have been assigned redshifts) for $\snrt > 5.0$.
The slight excess above the $1-F_{\rm False}$ line at $\snrt \approx 4$ in the DES
and HSC regions is consistent within the expected shot noise, given the number
of cluster candidates.
}
\label{fig:purity}
\end{figure}

As in previous ACT work \citep[e.g.,][H21]{Menanteau_2010}, we assess follow-up
completeness from the fraction of cluster candidates with redshift estimates as a
function of $\snrt$ cut. Figure~\ref{fig:purity} shows this for the \texttt{flags = 0}
area in several footprints within the ACT cluster search area. We predict the 
purity of the sample as a function of $\snrt$ cut using signal-free simulations
(see Section~\ref{sec:signalFree})
as $1-F_{\rm False}$, where $F_{\rm False}$ is the fraction of false positives, 
shown by the dashed line in Figure~\ref{fig:purity}. This is expected to be the most
optimistic case, as the signal-free simulations do not contain any source of contamination
beyond that due to noise fluctuations in the map. Nevertheless, we see that the
$1-F_{\rm False}$ line is a reasonable match to the fraction of detected, optically
confirmed cluster candidates found in regions with deep optical data (DES, HSC; the small excess above the $1-F_{\rm False}$ line seen at $\snrt \approx 4$
may result from spurious cross-matching of some optically selected clusters with SZ cluster candidates that are noise fluctuations).
We find that the sample found within the \cosmoSearchArea{}\,deg$^2$ \cosmo footprint
is $>99$\% complete in terms of redshift follow-up for $\snrt > 5.5$.

\section{The ACT DR6 Cluster Catalog} 
\label{sec:catalog}

\subsection{Catalog Properties} 
\label{sec:catalogProperties}

The ACT DR6 cluster catalog consists of \totalConfirmed{} clusters detected at
$\snrt > 4$, with mass and redshift measurements. Of these, 
\totalFlagZeroConfirmed{} are located in the area with \texttt{flags = 0}.
These are drawn from a full list of \totalCandidates{} cluster candidates, though only \totalFlagZeroCandidates{} of these have \texttt{flags = 0}.
Table~\ref{tab:FITSTableColumns}
gives descriptions of the columns in the cluster catalog.

\begin{figure}
\includegraphics[width=\columnwidth]{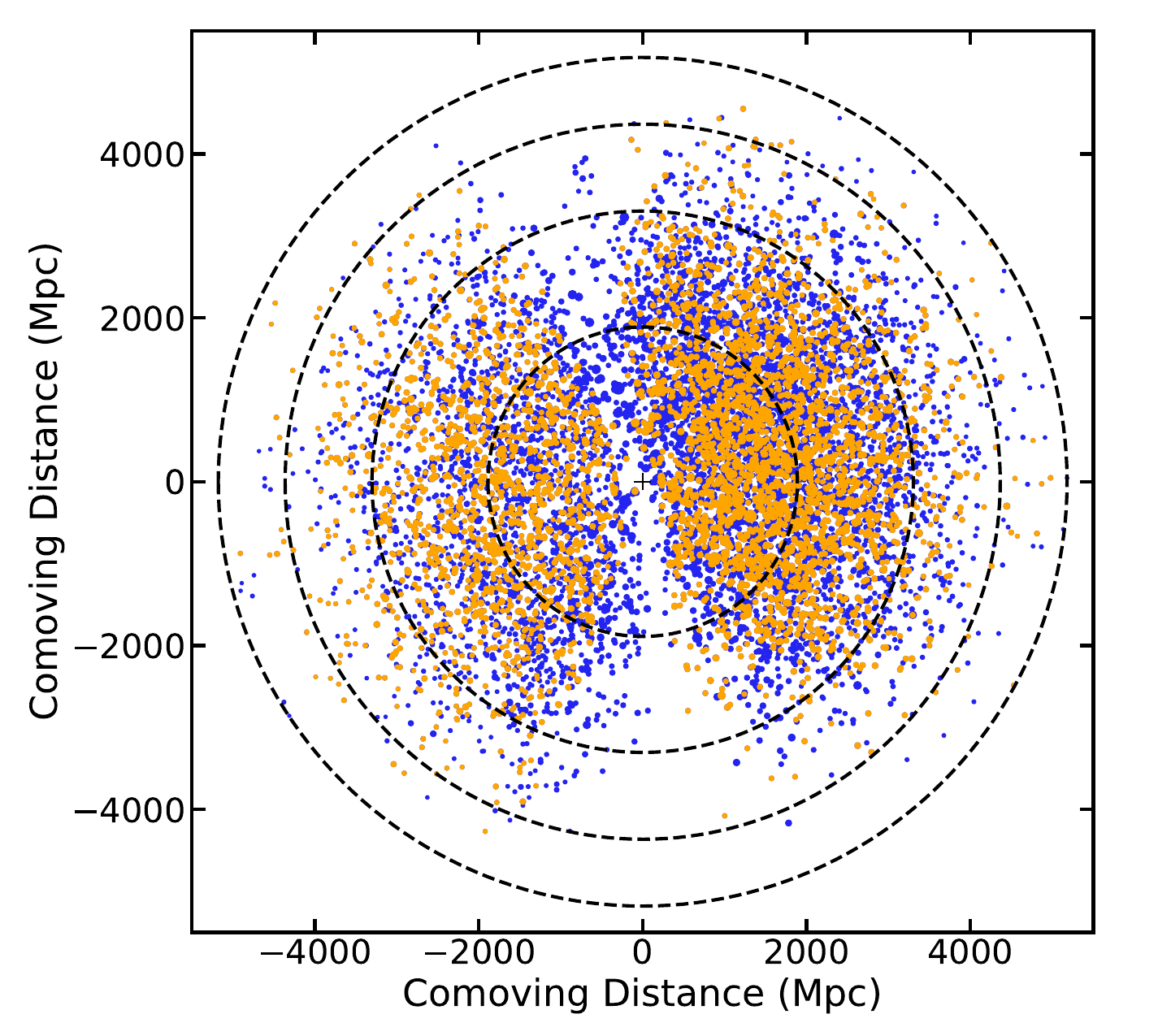}
\caption{Wedge plot showing the full ACT DR6 cluster catalog 
(\totalConfirmed{} blue points),
and the `Legacy sample' (\totalCosmoSample{} orange points), 
drawn in the equatorial plane. The size of each point scales with cluster
mass. RA is used as the angular coordinate, with 0\,deg RA pointing to the
right from the origin and increasing counterclockwise. The radial
coordinate is comoving distance in Mpc. The dashed circles mark the
distances equivalent to redshifts 0.5, 1.0, 1.5, and 2.0, 
starting from the observer's location at (0, 0).}
\label{fig:wedgePlot}
\end{figure}

We define a subset of \totalCosmoSample{} clusters with $\snrt > 5.5$ and 
\texttt{flags = 0} located within the \cosmoSearchArea{}\,deg$^2$ \cosmo 
footprint\footnote{The `Legacy' footprint was defined from visual inspection
of the confirmed cluster sample sky distribution, with reference to the dust mask.
It is an attempt to define a region with the maximum contiguous area (split by the
Galactic plane) that contains an almost complete sample of optically
confirmed $\snrt > 5.5$ clusters with redshifts (i.e., $>99\%$ of $\snrt > 5.5$
candidates within the `Legacy' footprint are confirmed clusters). Note that
the choice of the exact boundaries of this footprint are somewhat arbitrary.
Users of the ACT DR6 cluster search data products can easily provide their own alternative
mask to redefine this footprint if desired, see \url{https://nemo-sz.readthedocs.io/en/latest/dr6_tutorial.html} and \url{https://nemo-sz.readthedocs.io/en/latest/commands.html\#nemomask}.}
as the `Legacy sample',
which we suggest for use in cosmological analyses that require a complete,
statistical sample of clusters.
Figure~\ref{fig:wedgePlot} shows the distribution of the cluster sample in terms of
comoving distance in the celestial equatorial plane, with RA used as the angular
coordinate. The `Legacy sample', which is complete in terms of redshift follow-up,
probes out to comoving distance $\sim 5000$\,Mpc. 

\begin{figure*}
\includegraphics[width=\columnwidth]{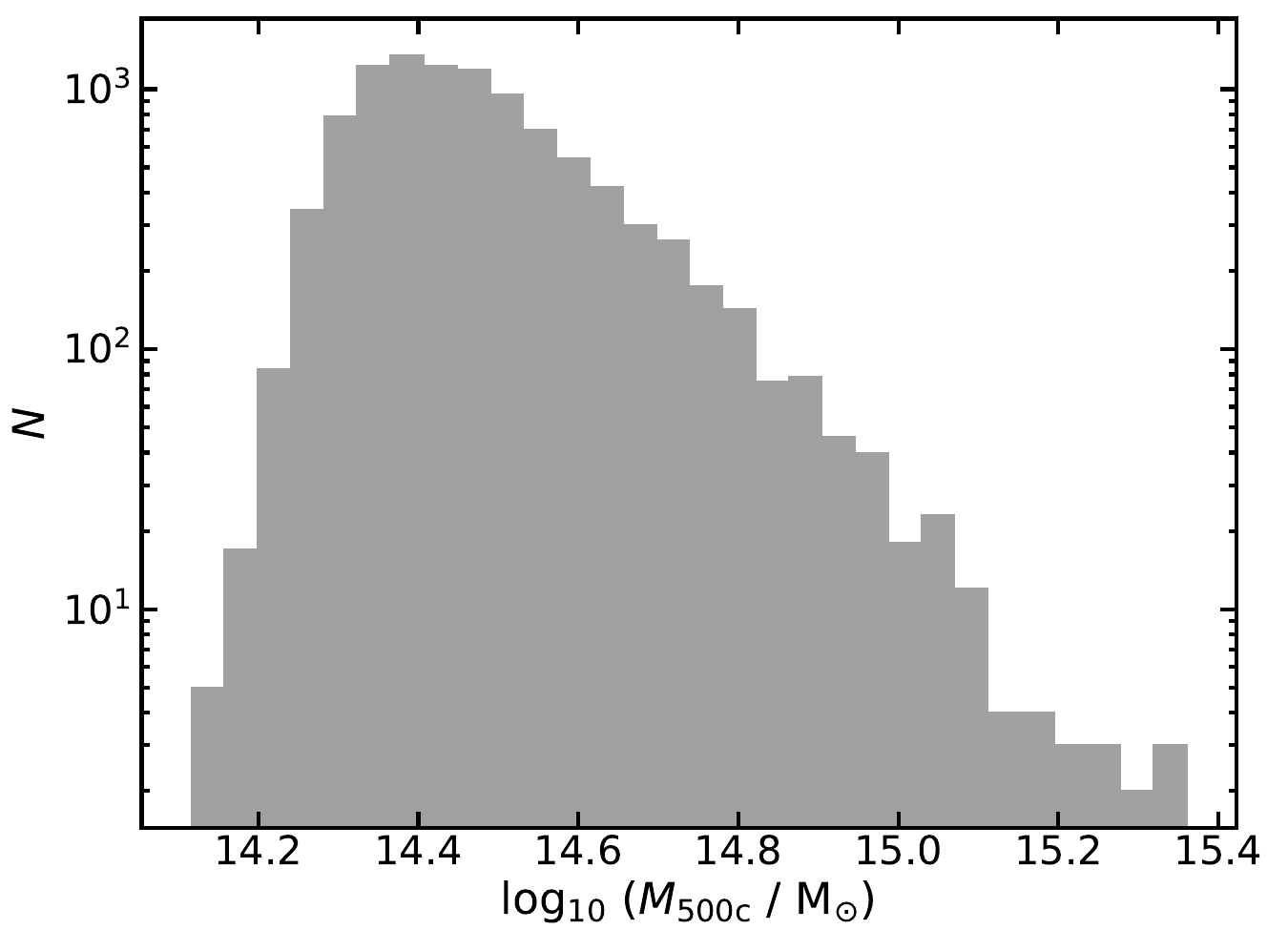}
\includegraphics[width=\columnwidth]{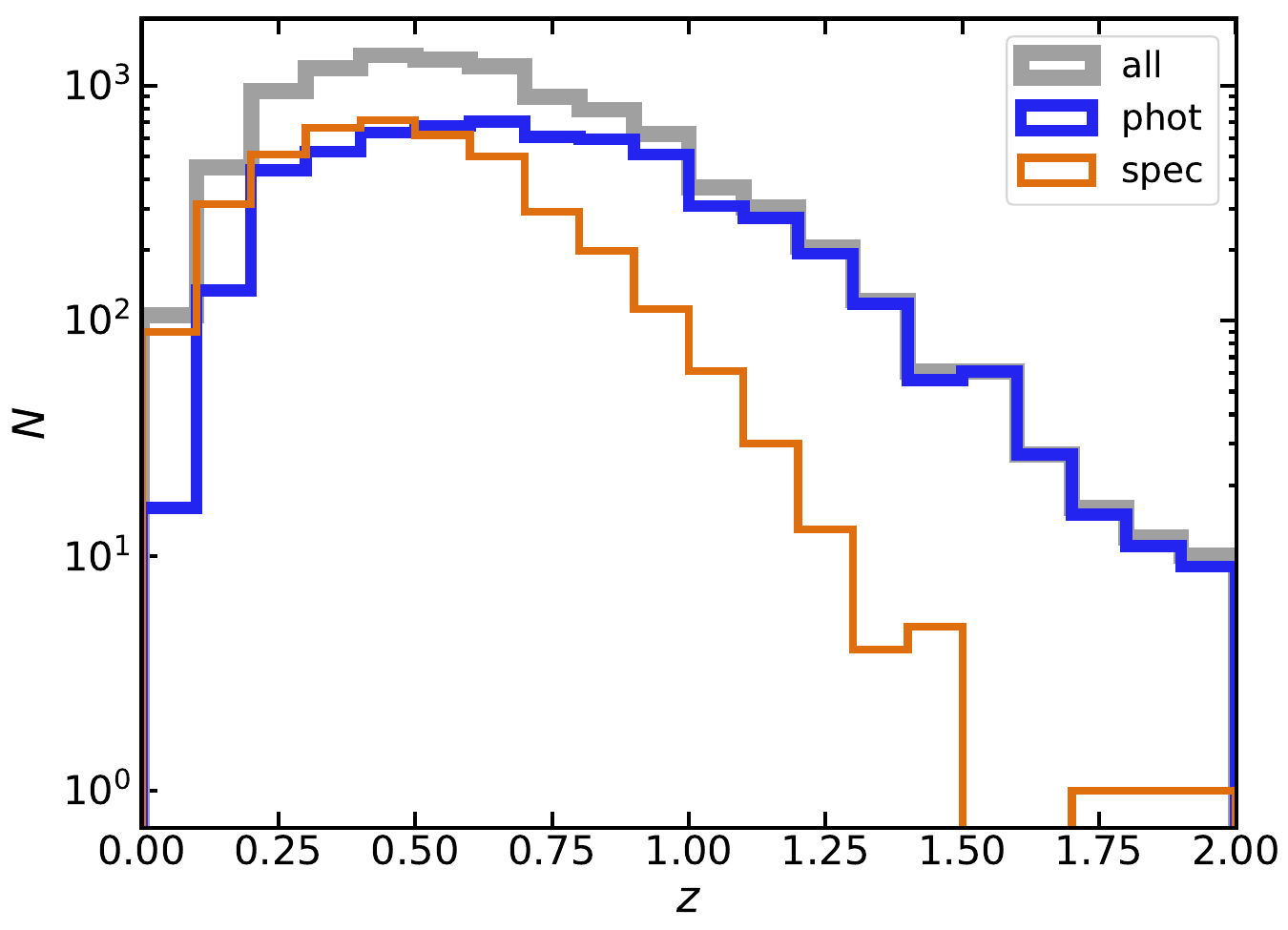}
\caption{Mass and redshift distributions for the ACT DR6 cluster
sample (see Section~\ref{sec:catalogProperties}). In the right panel,
objects with photometric redshifts are shown in blue, while those with
spectroscopic redshifts (\percentageSpecRedshifts{}\% of the total sample)
are shown in orange. The gray histogram shows the full sample.}
\label{fig:massRedshiftHists}
\end{figure*}

\begin{figure*}
\centering
\includegraphics[width=120mm]{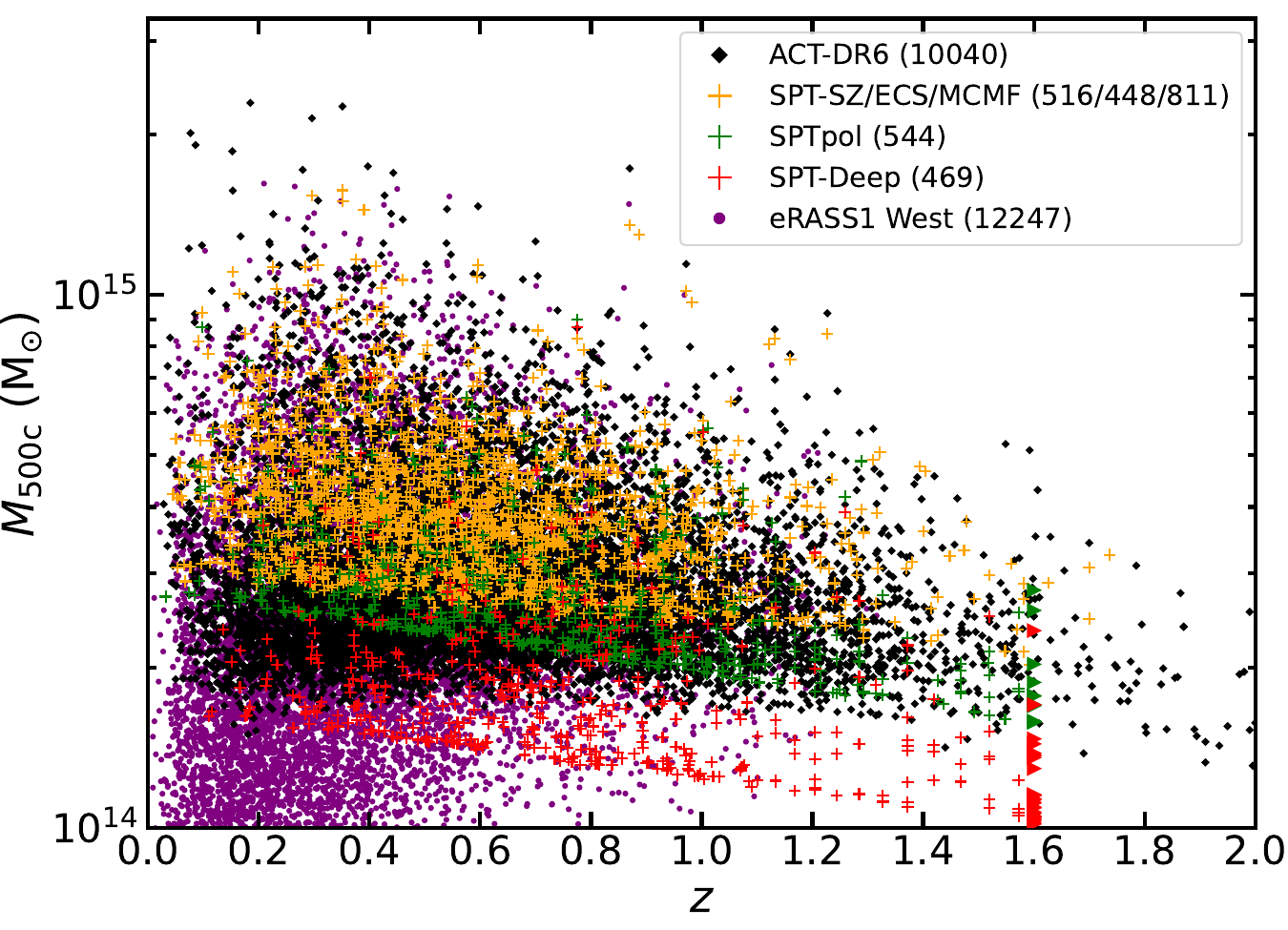}
\caption{The ACT DR6 cluster sample in the (mass, redshift) plane,
in comparison to recent SPT SZ cluster samples: SPT-SZ \citep{Bleem_2015}; SPT-ECS \citep{Bleem_2020}; SPT-MCMF \citep{KleinSPT_2024}; SPTpol \citep{Bleem_2024}; SPT-Deep \citep{Kornoelje_2025}; and the X-ray selected
eRASS1 catalog in the Western Galactic hemisphere \citep{Bulbul_2024}. 
Triangles indicate clusters with lower redshift limits in the SPTpol and SPT-Deep samples.
The number of clusters in each sample is given
in parentheses in the legend.
All of these surveys are on comparable mass scales (to within $<10$\% on average).}
\label{fig:massRedshiftPlane}
\end{figure*}

The mass and redshift distributions of the full cluster catalog are shown in
Figure~\ref{fig:massRedshiftHists}. The sample covers the mass range
$\minMass{} < M_{\rm 500c} / 10^{14}\,M_{\sun} < \maxMass{}$, 
with median $M_{\rm 500c} = \medianMass{} \times 10^{14}\,M_{\sun}$. 
The sample covers the redshift range $\minRedshift{} < z < \maxRedshift{}$, with
median $z = \medianRedshift{}$. It contains \totalSpecRedshifts{}
spectroscopic redshifts (\percentageSpecRedshifts\% of the sample). There are
\totalHighZ{} clusters at $z > 1$, of which \totalHighestZ{} have redshifts estimated to be at $z > 1.5$. 
Many of these have photometric redshifts from the \citet{Klein_2024} and
\citet{ThongkhamII_2024} catalogs, which used DESI Legacy Imaging Surveys data.

Figure~\ref{fig:massRedshiftPlane} shows the ACT DR6 cluster sample in comparison
to recent SPT surveys \citep{Bleem_2015, Bleem_2020, Bleem_2024, KleinSPT_2024, Kornoelje_2025}, and the first
release of the SRG/eROSITA cluster catalog in the Western Galactic hemisphere
\citep[eRASS1;][]{Bulbul_2024}. All of these catalogs have approximately
the same mass calibration (to within $< 10$\% on average).
Both SPT-Deep and eRASS1 are sensitive to lower mass
clusters than ACT DR6.
The ACT DR6 catalog is the largest SZ-selected cluster catalog produced to date.

\begin{figure}
\includegraphics[width=\columnwidth]{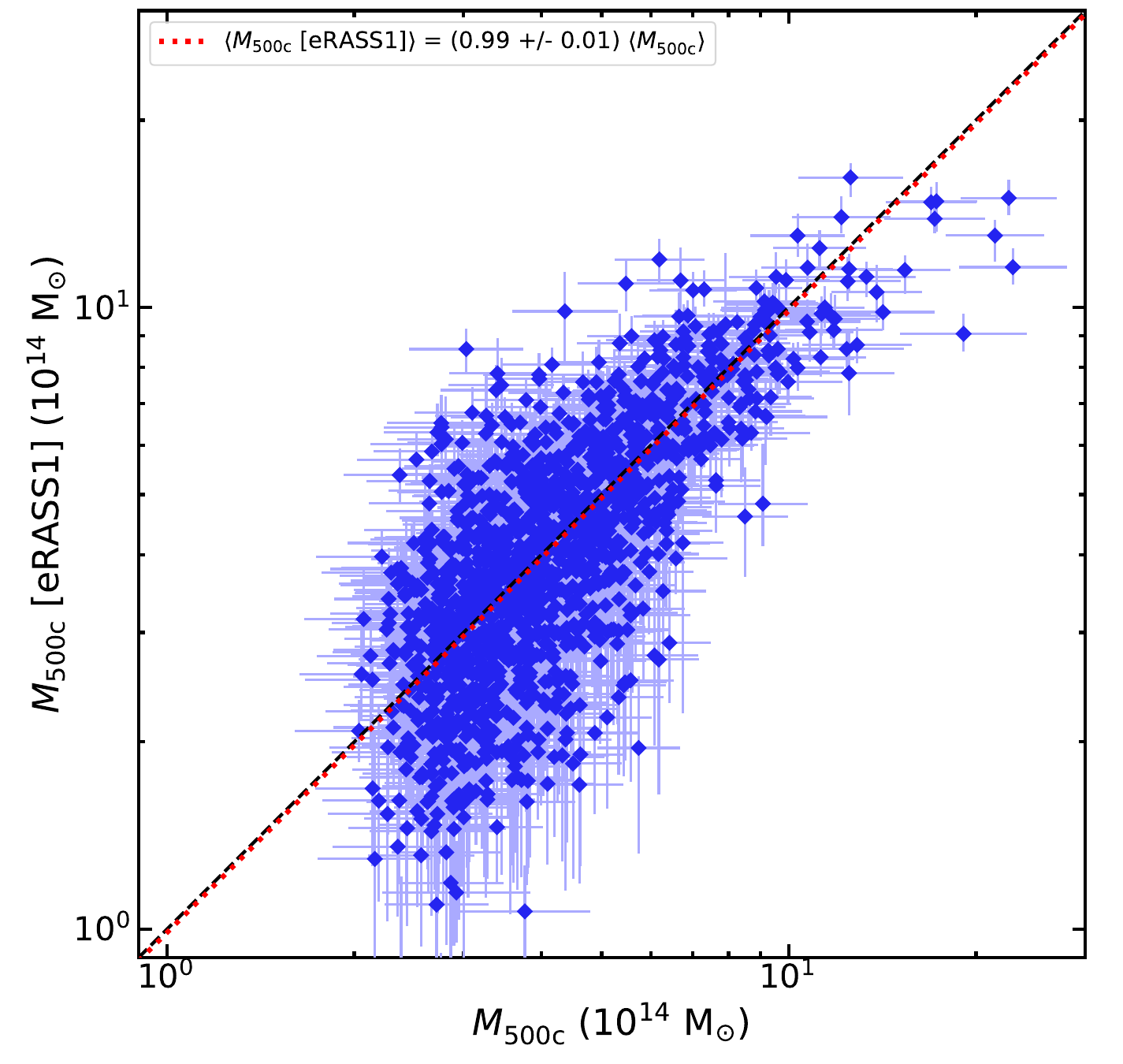}
\caption{Comparison between ACT DR6 mass estimates (this work, plotted along the horizontal
axis) and masses inferred from X-ray count rate measurements, calibrated using weak-lensing, as reported in the eRASS1 catalog
\citep{Bulbul_2024, Ghirardini_2024}. The large blue points are objects with $\snrt > 5$ in ACT DR6,
for which the unweighted mean ratio between the two cluster samples is calculated, 
shown by the dotted red line. The dashed black line shows the 1:1 correlation.}
\label{fig:eRASSMassComp}
\end{figure}

Inspection of Figure~\ref{fig:massRedshiftPlane} shows that the most massive ACT DR6 clusters have higher inferred masses than objects in the eRASS1 catalog. 
Since ACT and eRASS1 have both surveyed a large fraction of the sky in common, this might suggest a shift in the inferred masses between the two catalogs. Figure~\ref{fig:eRASSMassComp} shows a comparison of the mass estimates reported for clusters in common between the two samples. We find that the average masses of the samples are consistent; $\langle M_{\rm 500c} \left[\rm eRASS1 \right] \rangle = (0.99 \pm 0.01) \, \langle M_{\rm 500c} \rangle$,
where the uncertainty is the standard error on the mean. We note that the most massive clusters ($M_{\rm 500c} > 10^{15}\,M_{\sun}$) tend to have higher inferred masses in the ACT catalog. 
This could be due to physical reasons (e.g., mergers or line of sight projection effects, for example), 
but a detailed investigation is left for future work.

\subsection{Other Cluster Search Data Products} 
\label{sec:products}

We release a number of other data products related to the ACT DR6 cluster search:

\begin{enumerate}

\item{
\textit{Candidate list.}
The full list of \totalCandidates{} $\snrt > 4$ cluster candidates
detected by \textsc{Nemo}. This includes objects detected in regions
with high flag values, so this must be used with care.
}

\item{
\textit{Multiple systems catalog.}
We produce a catalog containing potential multiple systems, i.e., 
cluster pairs, triples, etc.\ that may be physically associated, using
the same method as described in H21 
(we link objects located within a projected distance of 10\,Mpc and within a line-of-sight velocity difference of 5000\,km\,s$^{-1}$).
One application of this kind of
catalog is stacking to detect bridges between clusters, as done
by \citet{Isopi_2024}.
The largest multiple systems that we find
contain four clusters, and there are three such systems which have spectroscopic redshifts for all components: 
ACT-CL\,J0857.8+0310 / J0857.9+0315 / J0859.5+0308 / J0901.5+0300 ($z = 0.20$);
ACT-CL\,J0908.8+1102 / J0909.2+1058 / J0909.3+1104 / J0911.9+1057 ($z = 0.17$);
and
ACT-CL\,J1030.3+1403 / J1030.5+1415 / J1031.3+1405 / J1032.8+1352 ($z = 0.32$).
}

\item{
\textit{Point source catalogs.}
These are extracted from the f090, f150, and f220 maps as a byproduct
of the \textsc{Nemo} cluster search (see Section~\ref{sec:multipass}),
and are used in constructing the flag mask (see below). 
The catalogs included here include amplitude ($\Delta T$), 
signal-to-noise, and position information.
A dedicated search for point sources and a report on their properties
will be presented in a future paper (Vargas et al., in prep.).
}

\item{
\textit{Filtered maps.}
Maps of $\snrt$ and $\yt$ ($\times 10^{-4}$) are provided in the standard
ACT \texttt{CAR} pixelization as FITS images.
}

\item{
\textit{Selection function information.}
This includes masks (such as the flag mask; see Figure~\ref{fig:flagMask}), noise tables and maps for various footprints that intersect with the ACT cluster search area (HSC, DES, KiDS, and the \cosmo footprint),
and tables that allow the filter mismatch function ($Q(\theta_{\rm 500c})$, see 
Section~\ref{sec:masses}) to be computed.
}

\item{
\textit{Cluster signal maps.}
These are FITS images in the standard ACT \texttt{CAR} pixelization
that allow the cluster thermal SZ signal to be subtracted from the
f090, f150, and f220 maps. Note that the estimate of cluster size is
quite noisy, and better results may be obtained by using more filter
scales than have been used for the ACT DR6 analysis. 
The UPP \citep{Arnaud_2010} is assumed. Tools are available in
\textsc{Nemo} to make these model maps with different parameters.
}

\end{enumerate}

Most of these data products can be reproduced by running 
\textsc{Nemo} on the public ACT DR6 maps.\footnote{See \url{https://nemo-sz.readthedocs.io/en/latest/dr6_tutorial.html}}

\section{Discussion} 
\label{sec:discussion}

\subsection{Comparison with the ACT DR5 cluster catalog}


We find that 508 clusters from the ACT DR5 catalog (H21) are not found in the
DR6 catalog. Differences in masking
account for 22 of these (i.e.,
4173/4195 of the DR5
clusters are located within the DR6 cluster search region, leaving 486 clusters from DR5 that are not found in the DR6 cluster catalog).
The difference in the signal-to-noise selection criterion between the DR5 and DR6 analyses accounts for a further 126 objects; in DR6, we require $\snrt > 4$ for inclusion in the catalog, whereas
for DR5 we included all clusters detected with ${\rm S/N} > 4$, regardless of the filter scale
at which the object was detected. 
So, 360/4173 (8.6\%) of
clusters with $\snrt > 4$ in the DR5 catalog that are within the DR6 cluster search area are missing in DR6.

All of the clusters previously reported in DR5 and
missing from the DR6 catalog were optically confirmed. Most are also found in other catalogs,
with 47\% of the missing clusters coming from the redMaPPer catalogs that provided the most
redshifts overall in DR5. They all have richness $\lambda > 20$, with median $\lambda = 40$
(for SDSS redMaPPer clusters) and $\lambda = 34$ (for DES redMaPPer clusters).
These red sequence richnesses can be boosted by projection, and as a result the halo mass and SZ signal for these systems has a significant skew to low values 
\citep{Cohn_2007, Cao_2025}.

Another explanation for the disappearance of these clusters below our
selection threshold is that they were boosted to higher $\snrt$ in the DR5 catalog by
noise fluctuations. Using \textsc{Nemo} to perform forced photometry in the DR6 $\snrt$
maps at the locations given in the DR5 catalog, we find that 178/354 (49\%) of the missing
clusters have $\snrt > 3$. The median difference in signal-to-noise between the catalogs for
the missing clusters is $\Delta \snrt = \snrt_{\rm DR5} - \snrt_{\rm DR6} =  1.5$.

Comparing the redshifts of clusters in common between the DR5 and DR6 catalogs,
the reported values have changed by $\Delta z > 0.1$ for 150 clusters, and by $\Delta z > 0.5$
for 23 clusters. In the latter case, this is due to projected systems (see
Section~\ref{sec:crossMatching}), where in DR5 we misidentified a lower redshift cluster
along the line of sight to the SZ source in some cases. Better optical/IR data and
new optical/IR cluster catalogs that were not available when H21 was published
have helped to identify these issues.

\subsection{Source Contamination and SZ Selection}
\label{sec:sourceContamination}
Incompleteness due to sources, either active galactic nuclei (AGN) or dusty star forming galaxies, filling in the SZ
decrement signal has long been a concern \citep{Lin_2009, Sayers_2013, Dicker_2021, Dicker_2024}, although studies to date
have not revealed it to be a significant issue \citep{Gupta_2017, Bleem_2024}.
To test this for the ACT DR6 sample, using \textsc{Nemo} we perform forced photometry in the $\snrt$ and
$\yt$ filtered maps at the locations of X-ray selected clusters from the eRASS1
catalog \citep{Bulbul_2024}, since we do not expect
X-ray selected clusters to be sensitive to this selection effect (although some X-ray clusters may have their luminosities boosted by contamination by X-ray
AGNs). 
We extract $\snrt$ and $\yt$ values at the positions of eRASS1 clusters in the 
valid ACT DR6 search area, and compute mass estimates using Equation~(\ref{eq:y0}).
Note that there are some eRASS1 clusters for which we adopt a different redshift
estimate in the ACT DR6 catalog compared to that listed in the eRASS1 catalog;
for estimating masses from forced photometry, we use the redshift as listed in
the eRASS1 catalog.
There are a total of 7043 eRASS1 clusters
within the ACT footprint, and the forced photometry procedure returns 1690 of these
with $\snrt > 4$. It is not surprising that most eRASS1 clusters are not detected
by ACT, as most have masses below the ACT detection limit (see
Figure~\ref{fig:massRedshiftPlane}). However, clusters that have large masses listed in
the eRASS1 catalog that are missed by ACT can be used to investigate SZ incompleteness
due to AGN contamination, under the assumption that the mass estimates provided by
\citet{Bulbul_2024} are reliable.

\begin{figure}
\includegraphics[width=\columnwidth]{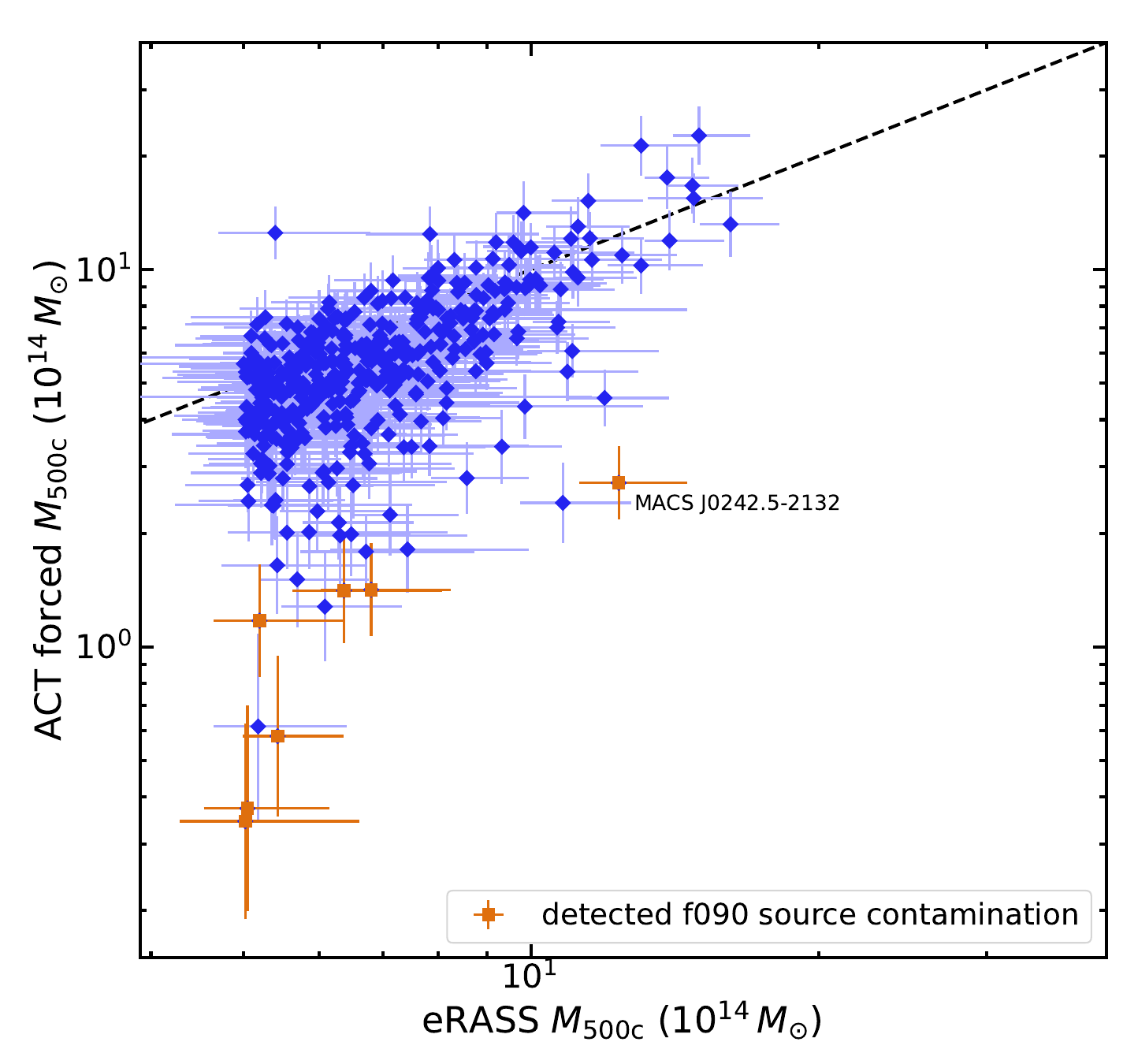}
\caption{Comparison of SZ mass estimates extracted via forced photometry (vertical
axis) with X-ray derived masses listed in the eRASS1 cluster catalog (horizontal
axis), for clusters with eRASS $M_{\rm 500c} > 5 \times 10^{14}\,M_{\sun}$ and at
$z > 0.2$ (approximately the $\snrt > 5$ 90\% mass completeness limit over the
ACT DR6 search area). The dashed line shows the 1:1 relation, and demonstrates that the
mass calibration used for both samples is comparable (see also Figure~\ref{fig:eRASSMassComp}). Objects that were
found to be located within 1.2$\arcmin$ of a f090 point source are shown as the
orange points. These have SZ mass estimates that are significantly lower than those
measured by eROSITA. We highlight the case of MACS J0242.5-2132
(1eRASS~J024235.9-213225), which is briefly discussed in Section~\ref{sec:sourceContamination}.
}
\label{fig:eRASSForced}
\end{figure}

As shown in Section~\ref{sec:completeness}, the 90\% mass completeness limit
is approximately $M_{\rm 500c} > 5 \times 10^{14}\,M_{\sun}$ for $z > 0.2$ over most
of the ACT cluster search area. There are 428 $z > 0.2$ eRASS1 clusters with eRASS
mass estimates above this limit. 
Figure~\ref{fig:eRASSForced} shows a comparison between the SZ masses extracted via
forced photometry, compared to the X-ray derived masses listed in the eRASS1 catalog.
We see overall agreement (most clusters scatter around the 1:1 line; see also Figure~\ref{fig:eRASSMassComp}), but there is
a tail of clusters with lower than expected SZ masses, given the eRASS1 mass.

We find $\snrt > 5$ for 92\% of the 428 clusters from the forced photometry procedure, 
while 32/428 clusters (7.5\%) have $\snrt < 5$. Of these, 14/33 have $3 < \snrt < 5$.
If we take the eRASS1 mass as `truth', then this indicates there is suppression
of the SZ signal in a small fraction of massive clusters. The most likely explanation
for this would be contamination due to radio AGN \citep[see also][]{Dicker_2021, Dicker_2024}, though miscentring of the X-ray cluster position with respect to the peak of the SZ signal may also play a role (since forced photometry relies on the X-ray positions in the eRASS1 catalog). 

In Figure~\ref{fig:eRASSForced},
we highlight clusters which are found within 1.2$\arcmin$ of a ${\rm S/N} >5$ f090 point source
(there are a further four clusters which are not shown as they have negative $\yt$ values
from the forced photometry). Note that sources are subtracted during the cluster
detection procedure (see Section~\ref{sec:nemo}), so while this process has resulted in
some SZ signal being recovered, it seems that this is not enough to match the mass
inferred from the X-ray data. MACS J0242.5-2132 (\citealt{Allen_2004}; also known as
1eRASS~J024235.9-213225, with $M_{\rm 500c} > 10^{15}\,M_{\sun}$ at $z = 0.314$) is one
example of a massive cluster that was missing in the ACT DR5 sample due to source contamination. This particular cluster hosts a compact radio source within the BCG with f090 flux density 65\,mJy and 1.4\,GHz flux density $\approx 1$\,Jy. It is recovered in the DR6 sample\footnote{Using \textsc{Nemo v0.9.0}; it was not recovered with $\snrt > 4$ using earlier versions of the code.} at $\snrt = 4.8$, and has an ACT mass
estimate that is 22\% of the eRASS1 mass estimate.

We found that 16/32 of the $\snrt < 5$ clusters have positions that are offset from the SZ peak S/N location from visual inspection of the ACT filtered map, and so miscentring is most likely to be responsible for the lower
than expected SZ signal for these systems. This includes one merging system that is visually resolved into two components by ACT, but which is not deblended in either the ACT or eRASS1 catalogs (ACT-CL\,J0810.3+1816 at $z = 0.42$, which is associated with 1eRASS J081025.1+181634).
Miscentring can be due to radio sources affecting the SZ peak position, or positional uncertainties in the eRASS1 catalog.

We note that some eRASS1 clusters are blended in the ACT DR6 catalog, and are likely to be mergers. There are three such ACT detections that have multiple eRASS1 clusters within $<1.2\arcmin{}$, which includes two mergers (both eRASS1 components at the same redshift; ACT-CL\,J0436.3$-$3317 at $z = 0.41$ and ACT-CL\,J0600.2$-$2007 at $z = 0.42$) and one possible projected system (ACT-CL\,J0547.1$-$5026, with eRASS1 clusters at $z = 0.68, 0.92$). Of these, only the eRASS1 clusters associated with ACT-CL\,J0600.2$-$2007 are part of the high-mass subsample with $M_{\rm 500c} > 5 \times 10^{14}\,M_{\sun}$, $z > 0.2$ considered above.

It is possible that some of the eRASS1 mass estimates may be biased high due to contamination by X-ray AGN, since X-ray count rate is the mass proxy used \citep[see][]{Bulbul_2024, Ghirardini_2024}. Objects with potentially high X-ray AGN contamination are flagged using the \texttt{PCONT} quantity in the eRASS1 catalogs, which gives an estimate of the probability that the X-ray source is a contaminant rather than a cluster. We find that 4/33 of the $\snrt < 5$ eRASS1 clusters have \texttt{PCONT}~$> 0.1$, and so their masses may be overestimated in the eRASS1 catalog.

After accounting for miscentring and X-ray AGN contamination, we conclude that radio source contamination directly results in 2--3\% of $M_{\rm 500c} > 5 \times 10^{14}\,M_{\sun}$, $z > 0.2$ eRASS1 clusters being missed by ACT.
The impact could be as high as 6\%, if we include the miscentered clusters, assuming that they are affected by radio sources that are not directly detected by ACT.
Obtaining a firmer estimate requires
taking into account the details of the eRASS1 cluster selection and mass estimates.
One should also note that the uncertainties on the SZ and X-ray signals considered here
are at the $> 10$\% level, and there may be other factors that cause the large
intrinsic scatter seen in Figure~\ref{fig:eRASSForced}.
A more detailed investigation, perhaps involving higher resolution SZ observations
\citep[e.g.,][]{Dicker_2021, Dicker_2024}, would help to clarify this issue.


\subsection{Extreme Clusters}
\label{sec:pinkElephants}
\begin{figure}
\includegraphics[width=\columnwidth]{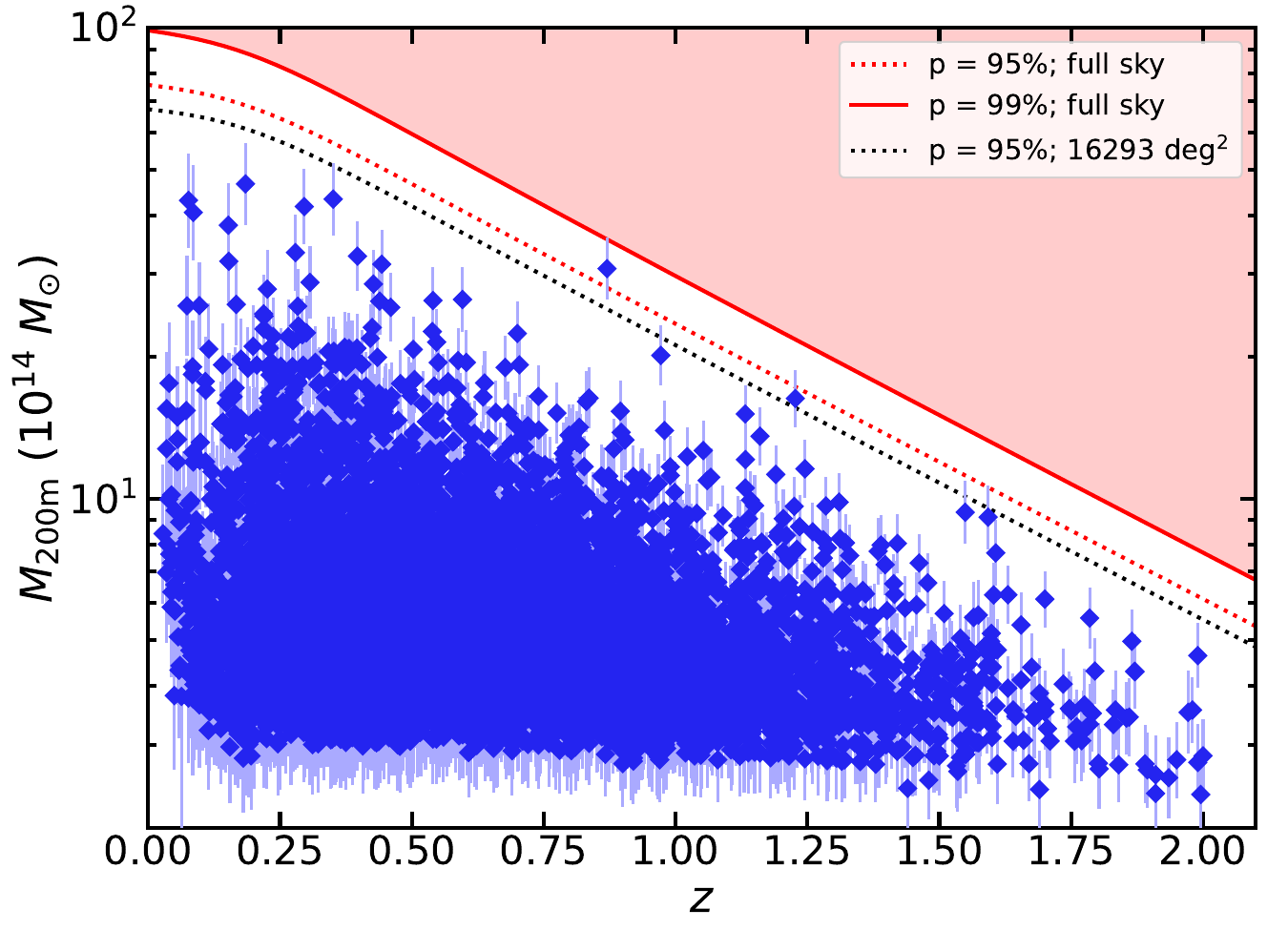}
\caption{Mass and redshift distribution for the ACT DR6 catalog in terms of
$M_{\rm 200m}$, compared to the flat $\Lambda$CDM exclusion curves of 
\citet{Mortonson_2011}, where the observation of even a single cluster above the shown lines
conflicts with $\Lambda$CDM at the confidence level indicated in
the figure legend. The two most extreme clusters are El Gordo
at $z = 0.87$ \citep{Menanteau_2012}, and ACT-CL\,\,J0329.2$-$2330
at $z = 1.23$ \citep{Bleem_2020, Sikhosana_2025}, which are both
undergoing mergers that may have boosted their SZ-inferred mass estimates.
(see Section~\ref{sec:pinkElephants}).}
\label{fig:mortonsonPlot}
\end{figure}

Thanks to the large sky area covered by ACT and the redshift independent SZ effect, 
we have now searched for massive clusters in a large fraction of the universe out to
$z = 2$. 
Here we revisit the question of whether any sufficiently massive clusters exist at early times in the universe
that can rule out the $\Lambda$CDM cosmological model with Gaussian initial conditions, given the expected time needed for such massive structures to grow \citep[e.g.,][]{Mortonson_2011,
HarrisonColes_2012,Sahlen2016}. 
We find that for the ACT DR6 catalog the answer is `no', as
shown by Figure~\ref{fig:mortonsonPlot}. However, there are two clusters that lie
over the 95\% line as calculated for the ACT DR6 footprint. These are
El Gordo \citep{Menanteau_2012}, a spectacular merging system that remains the most
extreme object in the ACT sample, despite the vastly increased sky area covered by
ACT DR6 compared to ACT DR1, and ACT-CL\,\,J0329.2-2330 at
$z = 1.23$ \citep[first reported by SPT;][]{Bleem_2020}. 
Like El Gordo, the latter system hosts a giant radio halo
\citep{Sikhosana_2025}, a turbulence-driven form of diffuse radio emission often seen in cluster mergers
\citep[see the review by][]{vanWeeren_2019}. Therefore it could be the
case that the masses inferred for these systems from the SZ signal are overestimated,
due to merger boosting as seen in hydrodynamical simulations 
\citep{Poole_2007, Wik_2008, Krause_2012}. These simulations show that such merger boosting is expected to be transient, lasting for $\sim 100$\,Myr.
For El Gordo, the SZ-derived mass in ACT DR6 is $\approx 30$\% higher than the
weak-lensing mass measured by \citet{Kim_2021}, although the two mass estimates
only differ at the $1.4\sigma$ level. Adopting the \citet{Kim_2021} weak-lensing mass for El Gordo would move it below the black dotted line 
($p = 95$\%, \maxSearchArea{}\,deg$^2$) in Figure~\ref{fig:mortonsonPlot}.


\section{Summary} 
\label{sec:summary}

This work presents the cluster catalogs and several of their properties derived from the ACT DR6 maps,
using data obtained over the lifetime of the project (2008--2022).
This includes a catalog of \totalConfirmed{} optically confirmed clusters with redshifts and mass
estimates, covering the redshift range $0 < z < 2$ 
(median $z = \medianRedshift$), with \totalHighZ{} clusters at $z > 1$,
largely made possible due to the overlap with recent optical/IR catalogs
exploiting the DESI Legacy Imaging Surveys \citep{Dey_2019},
including \citealt{ThongkhamI_2024} and \citealt{Klein_2024}.
It is the largest SZ-selected cluster catalog to date, being a factor of $>2$ larger than the ACT DR5 catalog (H21), and $>1.5$ times larger than the ACT-DR5 MCMF catalog \citep{Klein_2024}.

Using an SZ-signal--mass scaling relation that is consistent with
recent weak-lensing analyses \citep{Robertson_2024, Shirasaki_2024},
we find the confirmed clusters span the mass range
$\minMass < M_{\rm 500c} / 10^{14}\,M_{\sun} < \maxMass$,
with median $M_{\rm 500c} = \medianMass \times 10^{14}\,M_{\sun}$.
We find that the 90\% completeness limit at $\snrt = 5$
is $M_{\rm 500c} = 5 \times 10^{14}\,M_{\sun}$ over most
of the survey area. We find no evidence for a single cluster
massive enough to falsify $\Lambda$CDM; El Gordo
\citep{Menanteau_2012}, discovered early in the history
of the project, remains the most extreme cluster found by ACT, though we again note that the mass determined for El Gordo from weak lensing \citep{Kim_2021} is $\approx 30$\% lower than the ACT SZ-derived estimate.

\textsc{Nemo}, the cluster and source detection package used
to produce many of the results shown in this work, is
publicly available under a free software license, and its
documentation includes instructions on how to reproduce many of
the ACT DR6 cluster search data products. It should also
be capable of analyzing future maps from the Simons Observatory
Large Aperture Telescope, which is expected
to produce a catalog of $>30,000$ galaxy clusters \citep{SimonsObs_2019, SimonsObs_2025}.

\begin{acknowledgments}
We thank the referee for some suggestions which
helped to improve the quality of this paper.

Support for ACT was through the U.S.~National Science Foundation through awards AST-0408698, AST-0965625, and AST-1440226 for the ACT project, as well as awards PHY-0355328, PHY-0855887 and PHY-1214379. Funding was also provided by Princeton University, the University of Pennsylvania, and a Canada Foundation for Innovation (CFI) award to UBC. ACT operated in the Parque Astron\'omico Atacama in northern Chile under the auspices of the Agencia Nacional de Investigaci\'on y Desarrollo (ANID). The development of multichroic detectors and lenses was supported by NASA grants NNX13AE56G and NNX14AB58G. Detector research at NIST was supported by the NIST Innovations in Measurement Science program. Computing for ACT was performed using the Princeton Research Computing resources at Princeton University and the Niagara supercomputer at the SciNet HPC Consortium. SciNet is funded by the CFI under the auspices of Compute Canada, the Government of Ontario, the Ontario Research Fund–Research Excellence, and the University of Toronto. This research also used resources of the National Energy Research Scientific Computing Center (NERSC), a U.S. Department of Energy Office of Science User Facility located at Lawrence Berkeley National Laboratory, operated under Contract No. DE-AC02-05CH11231 using NERSC award HEP-ERCAPmp107 from 2021 to 2025. We thank the Republic of Chile for hosting ACT in the northern Atacama, and the local indigenous Licanantay communities whom we follow in observing and learning from the night sky.

We are grateful to the NASA LAMBDA archive for hosting our data. This work uses data from the Planck satellite, based on observations obtained with Planck (\url{http://www.esa.int/Planck}), an ESA science mission with instruments and contributions directly funded by ESA Member States, NASA, and Canada. 


MH and KM acknowledge support from the National Research Foundation of South Africa (grant nos. 97792, 137975, CPRR240513218388).
Computations were performed on Hippo at the University of KwaZulu-Natal, and at the Centre for High Performance Computing (project ASTR1534).
This work was performed in part at the Aspen Center for Physics, which is supported by a grant from the Simons Foundation (1161654, Troyer).
CS acknowledges support from the Agencia Nacional de Investigaci\'on y Desarrollo (ANID) through Basal project FB210003.
MM acknowledges support from NSF grants AST-2307727 and  AST-2153201 and NASA grant 21-ATP21-0145.
JD acknowledges support from NSF grant AST-2108126, from a Royal Society Wolfson Visiting Fellowship and from the Kavli Institute for Cosmology Cambridge and the Institute of Astronomy, Cambridge.
KMH is supported by NSF award 2206344.
AN acknowledges support from the Horizon 2020 ERC Starting Grant (project number 101163128).
ADH acknowledges support from the Sutton Family Chair
in Science, Christianity and Cultures, from the Faculty of Arts and
Science, University of Toronto, and from the Natural Sciences and
Engineering Research Council of Canada (NSERC) [RGPIN-2023-05014,
DGECR-2023- 00180].

The Legacy Surveys consist of three individual and complementary projects: the Dark Energy Camera Legacy Survey (DECaLS; Proposal ID \#2014B-0404; PIs: David Schlegel and Arjun Dey), the Beijing-Arizona Sky Survey (BASS; NOAO Prop. ID \#2015A-0801; PIs: Zhou Xu and Xiaohui Fan), and the Mayall z-band Legacy Survey (MzLS; Prop. ID \#2016A-0453; PI: Arjun Dey). DECaLS, BASS and MzLS together include data obtained, respectively, at the Blanco telescope, Cerro Tololo Inter-American Observatory, NSF’s NOIRLab; the Bok telescope, Steward Observatory, University of Arizona; and the Mayall telescope, Kitt Peak National Observatory, NOIRLab. Pipeline processing and analyses of the data were supported by NOIRLab and the Lawrence Berkeley National Laboratory (LBNL). The Legacy Surveys project is honored to be permitted to conduct astronomical research on Iolkam Du’ag (Kitt Peak), a mountain with particular significance to the Tohono O’odham Nation.

NOIRLab is operated by the Association of Universities for Research in Astronomy (AURA) under a cooperative agreement with the National Science Foundation. LBNL is managed by the Regents of the University of California under contract to the U.S. Department of Energy.

This project used data obtained with the Dark Energy Camera (DECam), which
was constructed by the Dark Energy Survey (DES) collaboration. Funding for
the DES Projects has been provided by the U.S. Department of Energy, the 
U.S. National Science Foundation, the Ministry of Science and Education of
Spain, the Science and Technology Facilities Council of the United
Kingdom, the Higher Education Funding Council for England, the National
Center for Supercomputing Applications at the University of Illinois at
Urbana-Champaign, the Kavli Institute of Cosmological Physics at the
University of Chicago, Center for Cosmology and Astro-Particle Physics at
the Ohio State University, the Mitchell Institute for Fundamental Physics
and Astronomy at Texas A\&M University, Financiadora de Estudos e Projetos,
Fundação Carlos Chagas Filho de Amparo, Financiadora de Estudos e
Projetos, Fundação Carlos Chagas Filho de Amparo à Pesquisa do Estado do
Rio de Janeiro, Conselho Nacional de Desenvolvimento Cient\'ifico e
Tecnol\'ogico and the Ministerio da Ciencia, Tecnologia e Inovacao, the
Deutsche Forschungsgemeinschaft and the Collaborating Institutions in the
Dark Energy Survey. The Collaborating Institutions are Argonne National
Laboratory, the University of California at Santa Cruz, the University of
Cambridge, Centro de Investigaciones Energ\'eticas, Medioambientales y
Tecnol\'ogicas-Madrid, the University of Chicago, University College London,
the DES-Brazil Consortium, the University of Edinburgh, the Eidgenossische
Technische Hochschule (ETH) Zurich, Fermi National Accelerator Laboratory,
the University of Illinois at Urbana-Champaign, the Institut de Ci\`encies
de l’Espai (IEEC/CSIC), the Institut de F\'isica d’Altes Energies, Lawrence
Berkeley National Laboratory, the Ludwig Maximilians Universität Munchen
and the associated Excellence Cluster Universe, the University of
Michigan, NSF’s NOIRLab, the University of Nottingham, the Ohio State
University, the University of Pennsylvania, the University of Portsmouth,
SLAC National Accelerator Laboratory, Stanford University, the University
of Sussex, and Texas A\&M University.


The Legacy Survey team makes use of data products from the Near-Earth Object Wide-field Infrared Survey Explorer (NEOWISE), which is a project of the Jet Propulsion Laboratory/California Institute of Technology. NEOWISE is funded by the National Aeronautics and Space Administration.

The Legacy Surveys imaging of the DESI footprint is supported by the Director, Office of Science, Office of High Energy Physics of the U.S. Department of Energy under Contract No. DE-AC02-05CH1123, by the National Energy Research Scientific Computing Center, a DOE Office of Science User Facility under the same contract; and by the U.S. National Science Foundation, Division of Astronomical Sciences under Contract No. AST-0950945 to NOAO.

The Hyper Suprime-Cam (HSC) collaboration includes the astronomical communities of Japan and Taiwan, and Princeton University. The HSC instrumentation and software were developed by the National Astronomical Observatory of Japan (NAOJ), the Kavli Institute for the Physics and Mathematics of the Universe (Kavli IPMU), the University of Tokyo, the High Energy Accelerator Research Organization (KEK), the Academia Sinica Institute for Astronomy and Astrophysics in Taiwan (ASIAA), and Princeton University. Funding was contributed by the FIRST program from the Japanese Cabinet Office, the Ministry of Education, Culture, Sports, Science and Technology (MEXT), the Japan Society for the Promotion of Science (JSPS), Japan Science and Technology Agency (JST), the Toray Science Foundation, NAOJ, Kavli IPMU, KEK, ASIAA, and Princeton University. 

This paper makes use of software developed for Vera C. Rubin Observatory. We thank the Rubin Observatory for making their code available as free software at \url{http://pipelines.lsst.io/}.

This paper is based on data collected at the Subaru Telescope and retrieved from the HSC data archive system, which is operated by the Subaru Telescope and Astronomy Data Center (ADC) at NAOJ. Data analysis was in part carried out with the cooperation of Center for Computational Astrophysics (CfCA), NAOJ. We are honored and grateful for the opportunity of observing the Universe from Maunakea, which has the cultural, historical and natural significance in Hawaii. 


This research has made use of the NASA/IPAC Extragalactic Database, which is funded by the National Aeronautics and Space Administration and operated by the California Institute of Technology.

\textit{Software:}
AstroPy \citep{Astropy_2013},
Core Cosmology Library \citep{Chisari_2019},
Pixell (\url{https://github.com/simonsobs/pixell/})

\end{acknowledgments}



\appendix

\renewcommand{\thetable}{A\arabic{table}}


\section{Contents of the Cluster Catalog}

\LongTables
\begin{deluxetable*}{lll}
\tablecaption{Description of the columns in the \texttt{FITS} Table format cluster catalog, 
available from LAMBDA (\url{https://lambda.gsfc.nasa.gov/product/act/actadv_dr6_szcluster_catalog_get.html}).
The Symbol column provides a mapping between column names and symbols used in the text and 
figures of this article.\label{tab:FITSTableColumns}}

\tablehead{
\colhead{Column}       &
\colhead{Symbol} &
\colhead{Description} \\
}
\startdata
\texttt{name} & \nodata & \begin{minipage}[t]{120mm}Cluster name in the format ACT-CL JHHMM.m$\pm$DDMM.\end{minipage}\\
\texttt{RADeg} & \nodata & \begin{minipage}[t]{120mm}Right Ascension in decimal degrees (J2000) of the SZ detection by ACT.\end{minipage}\\
\texttt{decDeg} & \nodata & \begin{minipage}[t]{120mm}Declination in decimal degrees (J2000) of the SZ detection by ACT.\end{minipage}\\
\texttt{SNR} & $q$ & \begin{minipage}[t]{120mm}Signal-to-noise ratio, optimized over all filter scales (in H21, the symbol SNR was used for this quantity).\end{minipage}\\
\texttt{y\_c} & $y_{0}$ & \begin{minipage}[t]{120mm}Central Comptonization parameter ($\times 10^{-4}$) measured using the optimal matched filter template (i.e., the one that maximizes detection S/N $q$). Note that \texttt{inferred\_y\_c} provides a lower scatter estimate of this quantity. Uncertainty column(s): \texttt{err\_y\_c}.\end{minipage}\\
\texttt{fixed\_SNR} & $\snrt$ & \begin{minipage}[t]{120mm}Signal-to-noise ratio at the reference 2.4\arcmin{} filter scale (in H21, the symbol SNR$_{2.4}$ was used for this quantity). This is the quantity used for estimating completeness.\end{minipage}\\
\texttt{fixed\_y\_c} & $\tilde{y}_0$ & \begin{minipage}[t]{120mm}Central Comptonization parameter ($\times 10^{-4}$) measured at the reference filter scale (2.4\arcmin{}). This is the SZ signal used for inferring mass estimates and estimating mass completeness limits (see Section~\ref{sec:masses}). Uncertainty column(s): \texttt{fixed\_err\_y\_c}.\end{minipage}\\
\texttt{template} & \nodata & \begin{minipage}[t]{120mm}Name of the matched filter template resulting in the highest S/N ($q$) detection of this cluster.\end{minipage}\\
\texttt{tileName} & \nodata & \begin{minipage}[t]{120mm}Name of the ACT map tile in which the cluster was found.\end{minipage}\\
\texttt{flags} & \nodata & \begin{minipage}[t]{120mm}Number of times a location on the sky was flagged, for whatever reason (see Section~\ref{sec:flagging}). Use \texttt{flags = 0} to select `clean' regions. Note that many regions are flagged due to the presence of bright stars (which make obtaining photometric redshifts difficult or impossible), and the flagging set-up used for ACT DR6 is conservative.\end{minipage}\\
\texttt{finderFlag} & \nodata & \begin{minipage}[t]{120mm}Number of times a location on the sky was flagged during cluster finding. A non-zero value for this quantity indicates that a point source was subtracted at this location. As the point source subtraction process is not perfect, objects with this flag should be treated with care (see Section~\ref{sec:flagging}).\end{minipage}\\
\texttt{starFlag} & \nodata & \begin{minipage}[t]{120mm}If non-zero, indicates an area of sky close to a bright star. While this should not affect detection of the SZ signal, it may make it difficult or impossible to obtain a redshift estimate for this object (see Section~\ref{sec:flagging}).\end{minipage}\\
\texttt{extObjFlag} & \nodata & \begin{minipage}[t]{120mm}If non-zero, indicates an area of sky associated with an extended foreground object, such as a galaxy or a nebula (see Section~\ref{sec:flagging}).\end{minipage}\\
\texttt{dustFlag} & \nodata & \begin{minipage}[t]{120mm}If non-zero, indicates an area of sky that is dusty, as traced by emission in the \textit{Planck} 353\,GHz map (see Section~\ref{sec:flagging}).\end{minipage}\\
\texttt{ringFlag} & \nodata & \begin{minipage}[t]{120mm}If non-zero, indicates an object flagged as a possible ring artifact. It may be that the object is in fact a cluster. If so, the SZ signal may have been artificially boosted by the presence of the ring, so use these objects with care (see Section~\ref{sec:flagging}).\end{minipage}\\
\texttt{redshift} & $z$ & \begin{minipage}[t]{120mm}Adopted redshift for the cluster. The uncertainty is only given for photometric redshifts. Uncertainty column(s): \texttt{redshiftErr}.\end{minipage}\\
\texttt{redshiftType} & \nodata & \begin{minipage}[t]{120mm}Redshift type (\texttt{spec} = spectroscopic, \texttt{phot} = photometric).\end{minipage}\\
\texttt{redshiftSource} & \nodata & \begin{minipage}[t]{120mm}Source of the adopted redshift.\end{minipage}\\
\texttt{M500c} & $M_{\rm 500c}$ & \begin{minipage}[t]{120mm}$M_{\rm 500}$ with respect to the critical density, in units of $10^{14}\,M_{\sun}$. This assumes a scaling relation normalization $10^{A_0} = 3.0 \times 10^{-5}$, which is consistent with recent weak-lensing measurements \citep{Robertson_2024, Shirasaki_2024}. Uncertainty column(s): \texttt{M500c\_errPlus, M500c\_errMinus}.\end{minipage}\\
\texttt{M200c} & $M_{\rm 200c}$ & \begin{minipage}[t]{120mm}$M_{\rm 200}$ with respect to the critical density, in units of $10^{14}\,M_{\sun}$, converted from $M_{\rm 500c}$ using the \citet{Bhattacharya_2013} c-M relation. Uncertainty column(s): \texttt{M200c\_errMinus, M200c\_errPlus}.\end{minipage}\\
\texttt{M200m} & $M_{\rm 200m}$ & \begin{minipage}[t]{120mm}$M_{\rm 200}$ with respect to the mean density, in units of $10^{14}\,M_{\sun}$, converted from $M_{\rm 500c}$ using the \citet{Bhattacharya_2013} c-M relation. Uncertainty column(s): \texttt{M200m\_errPlus, M200m\_errMinus}.\end{minipage}\\
\texttt{M500cUncorr} & $M_{\rm 500c}^{\rm Unc}$ & \begin{minipage}[t]{120mm}$M_{\rm 500c}$ in units of $10^{14}\,M_{\sun}$, uncorrected for bias due to the steepness of the cluster mass function. Uncertainty column(s): \texttt{M500cUncorr\_errPlus, M500cUncorr\_errMinus}.\end{minipage}\\
\texttt{A10\_M500c} & $M_{\rm 500c}^{\rm A10}$ & \begin{minipage}[t]{120mm}$M_{\rm 500c}^{\rm A10}$ in units of $10^{14}\,M_{\sun}$, assuming the \citet{Arnaud_2010} scaling relation (this was the default assumption in previous ACT data releases). Uncertainty column(s): \texttt{A10\_M500c\_errPlus, A10\_M500c\_errMinus}.\end{minipage}\\
\texttt{A10\_M200c} & \nodata & \begin{minipage}[t]{120mm}$M_{\rm 200}$ with respect to the critical density, in units of $10^{14}\,M_{\sun}$, converted from $M_{\rm 500c}^{\rm A10}$ using the \citet{Bhattacharya_2013} c-M relation. Uncertainty column(s): \texttt{A10\_M200c\_errMinus, A10\_M200c\_errPlus}.\end{minipage}\\
\texttt{A10\_M200m} & \nodata & \begin{minipage}[t]{120mm}$M_{\rm 200}$ with respect to the mean density, in units of $10^{14}\,M_{\sun}$, converted from $M_{\rm 500c^{\rm A10}}$ using the \citet{Bhattacharya_2013} c-M relation. Uncertainty column(s): \texttt{A10\_M200m\_errMinus, A10\_M200m\_errPlus}.\end{minipage}\\
\texttt{A10\_M500cUncorr} & \nodata & \begin{minipage}[t]{120mm}$M_{\rm 500c}^{\rm A10}$ in units of $10^{14}\,M_{\sun}$, assuming the \citet{Arnaud_2010} scaling relation (this was the default assumption in previous ACT data releases), but neglecting the Eddington bias correction. This is the mass estimate to use if you wish to compare with the \textit{Planck} PSZ2 catalog. Uncertainty column(s): \texttt{A10\_M500cUncorr\_errMinus, A10\_M500cUncorr\_errPlus}.\end{minipage}\\
\texttt{theta500Arcmin} & $\theta_{\rm 500c}$ & \begin{minipage}[t]{120mm}Inferred value for the angular size  $\theta_{\rm 500c}$ (arcmin), using the scaling relation model described in Section~\ref{sec:masses}. Uncertainty column(s): \texttt{theta500Arcmin\_err}.\end{minipage}\\
\texttt{Q} & $Q$ & \begin{minipage}[t]{120mm}Value of the filter mismatch function inferred for this cluster (see Section~\ref{sec:masses}). Uncertainty column(s): \texttt{Q\_err}.\end{minipage}\\
\texttt{inferred\_y\_c} & $y_{0}$ & \begin{minipage}[t]{120mm}Inferred value for the central Comptonization parameter $y_0$ ($\times 10^{-4}$), using the scaling relation model described in Section~\ref{sec:masses}. Uncertainty column(s): \texttt{inferred\_y\_c\_err}.\end{minipage}\\
\texttt{inferred\_Y500Arcmin2} & \nodata & \begin{minipage}[t]{120mm}Inferred value for the integrated Comptonization parameter $Y_{\rm 500c}$ (arcmin$^2$). Note that this is a model dependent quantity which depends on the assumed pressure profile and scaling relation parameters, and is not more physically meaningful than using $M_{\rm 500c}$ directly. Uncertainty column(s): \texttt{inferred\_Y500Arcmin2\_err}.\end{minipage}\\
\texttt{footprint\_DES} & \nodata & \begin{minipage}[t]{120mm}Flag indicating if the cluster falls within the Dark Energy Survey (DES) footprint.\end{minipage}\\
\texttt{footprint\_HSC} & \nodata & \begin{minipage}[t]{120mm}Flag indicating if the cluster falls within the Hyper Suprime-Cam Subaru Strategic Program (HSC) footprint.\end{minipage}\\
\texttt{footprint\_KiDS} & \nodata & \begin{minipage}[t]{120mm}Flag indicating if the cluster falls within the Kilo Degree Survey (KiDS) footprint.\end{minipage}\\
\texttt{footprint\_eROSITADe} & \nodata & \begin{minipage}[t]{120mm}Flag indicating if the cluster falls within the SRG/eROSITA All Sky Survey (eRASS1 Western Galactic hemisphere) footprint.\end{minipage}\\
\texttt{footprint\_Legacy} & \nodata & \begin{minipage}[t]{120mm}Flag indicating if the cluster falls within the ACT Legacy footprint (see Fig.~\ref{fig:dustOverlay}).\end{minipage}\\
\texttt{zCluster\_delta} & $\delta$ & \begin{minipage}[t]{120mm}Density contrast statistic measured at the \texttt{zCluster} photometric redshift. Uncertainty column(s): \texttt{zCluster\_errDelta}.\end{minipage}\\
\texttt{zCluster\_source} & \nodata & \begin{minipage}[t]{120mm}Photometry used for \texttt{zCluster} measurements. One of: DECaLS (DR10), KiDS (DR4), SDSS (DR16), PS1 (DR2).\end{minipage}\\
\texttt{RM} & \nodata & \begin{minipage}[t]{120mm}Flag indicating cross-match with a redMaPPer-detected cluster in the SDSS footprint \citep{Rykoff_2014}.\end{minipage}\\
\texttt{eRASS1CL} & \nodata & \begin{minipage}[t]{120mm}Flag indicating cross-match with an eRASS1 cluster \citep{Bulbul_2024}.\end{minipage}\\
\texttt{RMDESY6} & \nodata & \begin{minipage}[t]{120mm}Flag indicating cross-match with a redMaPPer-detected cluster in the DES Y6 footprint \citep{Rykoff_2016}.\end{minipage}\\
\texttt{WaZP} & \nodata & \begin{minipage}[t]{120mm}Flag indicating cross-match with a WaZP-detected cluster in the DES Y6 footprint \citep{Benoist_2025}.\end{minipage}\\
\texttt{CAMIRA} & \nodata & \begin{minipage}[t]{120mm}Flag indicating cross-match with a CAMIRA-detected cluster in the S23B catalog (\citealt{Oguri_2018}, Oguri et al. in prep.).\end{minipage}\\
\texttt{MaDCoWS2DR2} & \nodata & \begin{minipage}[t]{120mm}Flag indicating cross-match with a MaDCoWS DR2 cluster \citep{ThongkhamII_2024}.\end{minipage}\\
\texttt{opt\_RADeg} & \nodata & \begin{minipage}[t]{120mm}Alternative optically-determined Right Ascension in decimal degrees (J2000), from a heterogeneous collection of measurements (see \texttt{opt\_positionSource}).\end{minipage}\\
\texttt{opt\_decDeg} & \nodata & \begin{minipage}[t]{120mm}Alternative optically-determined Declination in decimal degrees (J2000), from a heterogeneous collection of measurements (see \texttt{opt\_positionSource}).\end{minipage}\\
\texttt{opt\_positionSource} & \nodata & \begin{minipage}[t]{120mm}Indicates the source of the alternative optically-determined cluster position. \texttt{CAMIRA} (position from the CAMIRA cluster finder; \citealt{Oguri_2018}, Oguri et al., in prep.), \texttt{MADCOWS2DR2} (position from MaDCoWS2 DR2; \citealt{ThongkhamII_2024}), \texttt{RM, RMDESY3, RMDESY6, RMDESY6ACT} (position from the redMaPPer cluster finder, in SDSS, DES Y3 or Y6; \citealt{Rykoff_2014, Rykoff_2016}), \texttt{Vis-BCG} (BCG position from visual inspection of available optical/IR imaging; this work and H21), \texttt{WaZP} (position from the WaZP cluster finder in DES Y6; \citealt{Benoist_2025}), \texttt{WH2015} (position from \citealt{WH_2015}), \texttt{WH2022} (position from \citealt{2022MNRAS.513.3946W}), \texttt{WH2024} (position from \citealt{2024ApJS..272...39W}). \end{minipage}\\
\texttt{notes} & \nodata & \begin{minipage}[t]{120mm}If present, at least one of: \texttt{AGN?} (central galaxy may have color or spectrum indicating it may host an AGN); \texttt{Lensing?} (cluster may show strong gravitational lensing features); \texttt{Merger?} (cluster may be a merger); \texttt{Star formation?} (a galaxy near the center may have blue colors which might indicate star formation if it is not a line-of-sight projection). These notes are not comprehensive and merely indicate some systems that were identified as potentially interesting during visual inspection of the available optical/IR imaging.\end{minipage}\\
\texttt{warnings} & \nodata & \begin{minipage}[t]{120mm}If present, a warning message related to the redshift measurement for this cluster (e.g., \texttt{Possible projected system}; see Section~\ref{sec:crossMatching}).\end{minipage}\\
\enddata
\end{deluxetable*}

\clearpage

\section{Optimization Bias in \textsc{Nemo} and Validation on End-to-End Simulations}
\label{ap:optBiasSims}


To accurately calculate the completeness at low values of $\snrt$, 
it is necessary to correct for the optimization bias \citep[e.g.,][]{Zub21} 
that results from the use of a matched filter for cluster finding, as in \textsc{Nemo}. 
This optimization bias arises due to the fact that clusters are identified 
with peaks in the filtered map, and positive noise fluctuations in the 
vicinity of a cluster drag the measured cluster location away from the true 
cluster location. This leads to a small positive bias in the recovered cluster 
SZ signal ($\yt$ in our case), which is naturally larger at lower 
signal-to-noise ($\snrt$) levels.

We find a model for the optimization
bias, and validate the completeness calculation described in 
Section~\ref{sec:completeness}, by running \textsc{Nemo} on 30 simulated 
realizations of the ACT DR6 maps.
We construct the simulations as follows.
First, we draw random samples in mass and $z$ from the \citet{Tinker_2008} HMF over the sky area observed by ACT.
SZ signals are assigned to each halo using an assumed scaling relation,
including log-normal intrinsic scatter. This produces an initial halo catalog,
referred to below as the `truth' catalog. 
The haloes in this catalog are then randomly placed within the ACT footprint.
Pressure profiles are pasted onto these halos following the \citet{Arnaud_2010}
profile, with angular size set according to the halo mass in the truth
catalog, and projected along the line-of-sight to produce a 2D Compton-$y$ map,
which is convolved with the appropriate ACT beam for a given frequency
channel. At a given frequency we multiply the Compton-y map by tSZ spectral 
distortion function and $T_{\rm CMB}$ to produce a map of temperature fluctuations
from the halos, to which we add simulated CMB (also convolved with the
appropriate beam). The final addition to these maps is simulated anisotropic
ACT noise, following the inverse variance maps for the real data at each 
frequency as produced by the map maker \citep[see][]{Naess_2020, Naess_2025}.
We scale up this white noise by a factor of 1.16 to match the median value
of the real DR6 $\yt$ noise map, since the simulations do not include all
possible sources of noise included in the real data.
We also add an additional $1/f$ noise component following
\citet{MacCrann_2024}, in order to approximate the impact of the atmosphere.

We process the simulated maps through \textsc{Nemo}, using the same settings as
used for the DR6 cluster search (see Section~\ref{sec:nemo}, except that we
do not find and subtract point sources, as these are not included in the simulated
maps used here). We cross-match
the \textsc{Nemo} output catalogs, which contain measurements of $\yt$ and $\snrt$
for detected clusters, with the `truth' catalog to assign redshifts.

We calculate the optimization bias by comparing the $\yt$ outputs recorded in the
\textsc{Nemo} output catalogs, to the $\yt$ values and $\snrt$ values extracted from the output maps
produced by \textsc{Nemo} at the known positions of the clusters as recorded in the truth catalog.
We refer to the latter as forced photometry measurements, and denote these with the subscript
FP in all that follows (i.e., $\yt$$_{\rm FP}$ and $\snrt_{\rm FP}$ refer to the forced
photometry values extracted at true cluster positions from the $\yt$ and $\snrt$ maps), and use them as proxies for unbiased measurements of the true values of these quantities. A similar approach was used by \citet{Zubeldia_2023}.

\begin{figure}
\centering
\includegraphics[width=90mm]{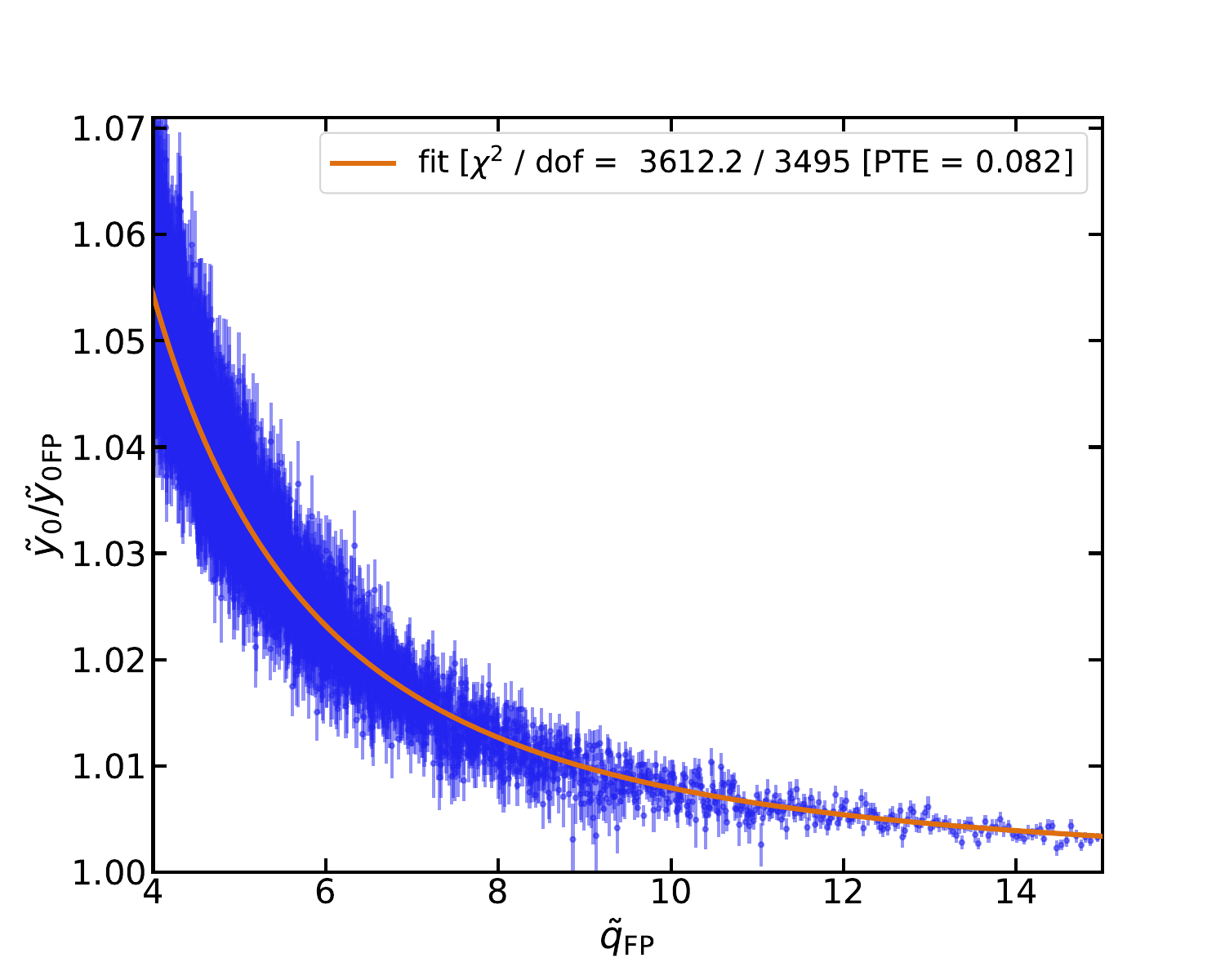}
\caption{Optimization bias on the extracted SZ signal ($\yt / \yt^{\rm FP}$) as
a function of signal-to-noise extracted by forced photometry at true cluster positions 
($\snr^{\rm FP}$), in a stack of 30 end-to-end simulation runs (the data are binned). 
The orange line is a fitted model of the form given in Equation~\ref{eq:optBiasModel}.}
\label{fig:optbias}
\end{figure}

We define the optimization bias to be the ratio $\yt/\yt$$_{\rm FP}$. In Figure~\ref{fig:optbias} we show this quantity as a function of $\snrt_{\rm FP}$.
We see that there is a clear trend for a larger deviation from 1 (no bias) as $\snrt_{\rm FP}$ decreases, with the bias reaching a value of 1.03 at $\snrt_{\rm FP} \approx 5$. We model the correction factor $C$ for this bias using
\begin{equation}
\label{eq:optBiasModel}
    C = 1 + \frac{1}{(\snrt_{\rm FP})^p},
\end{equation}
where $p = 2.1$ is found through fitting the data shown in Figure~\ref{fig:optbias}. 
This value is assumed in the calculations presented in Section~\ref{sec:completeness},
where we correct the true $\yt$ predicted from the scaling relation to $\yt^{\rm corr} = \yt (1 + \frac{1}{\snrt^p})$.
Note that Equation~\ref{eq:optBiasModel} diverges for very low values of $\snrt_{\rm FP}$; 
this is handled in the completeness calculations by setting $C = 1$ when $\snrt$ has a value
of 3 or more below the chosen $\snrt_{\rm cut}$ value.

\begin{figure*}
\includegraphics[width=90mm]{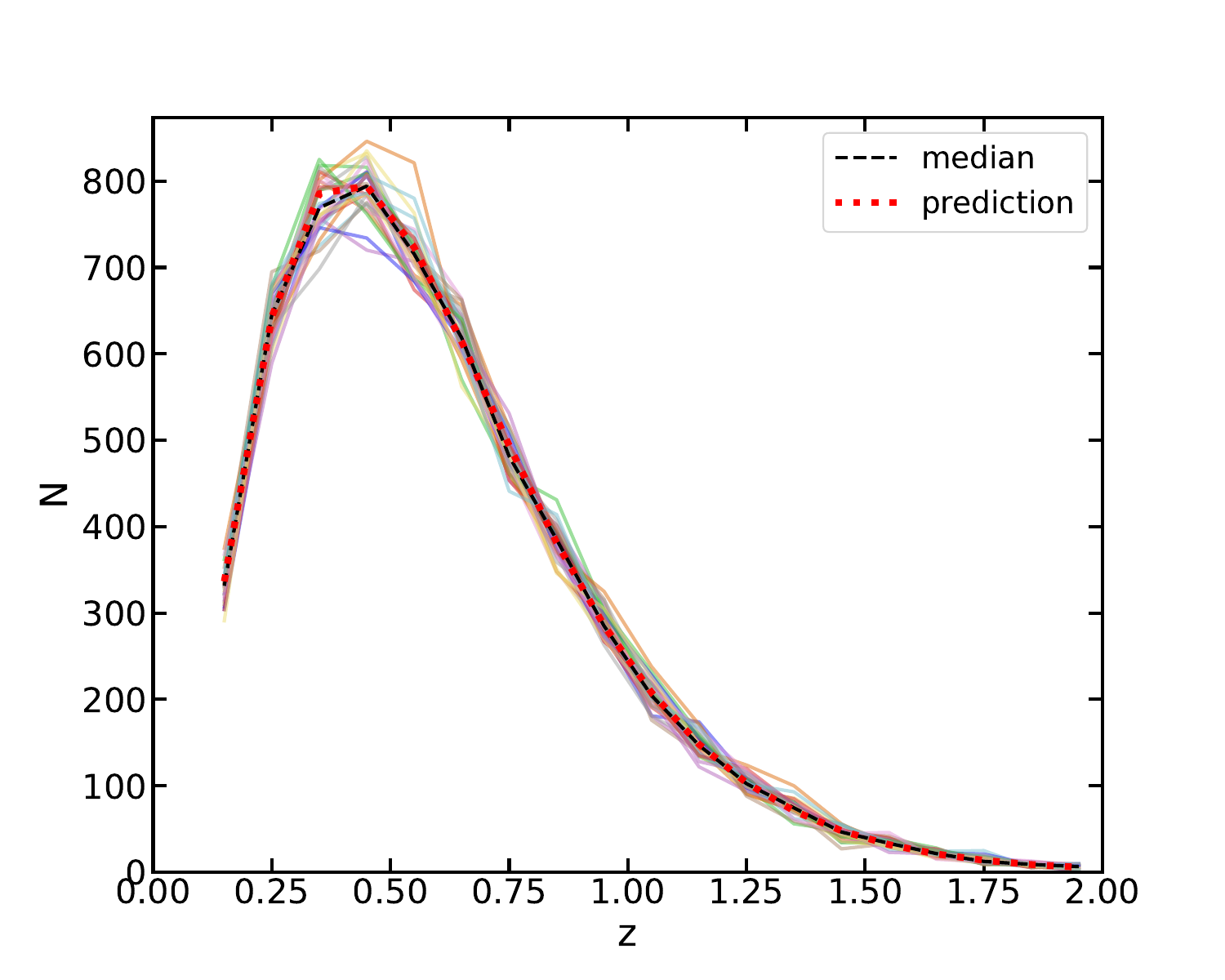}
\includegraphics[width=90mm]{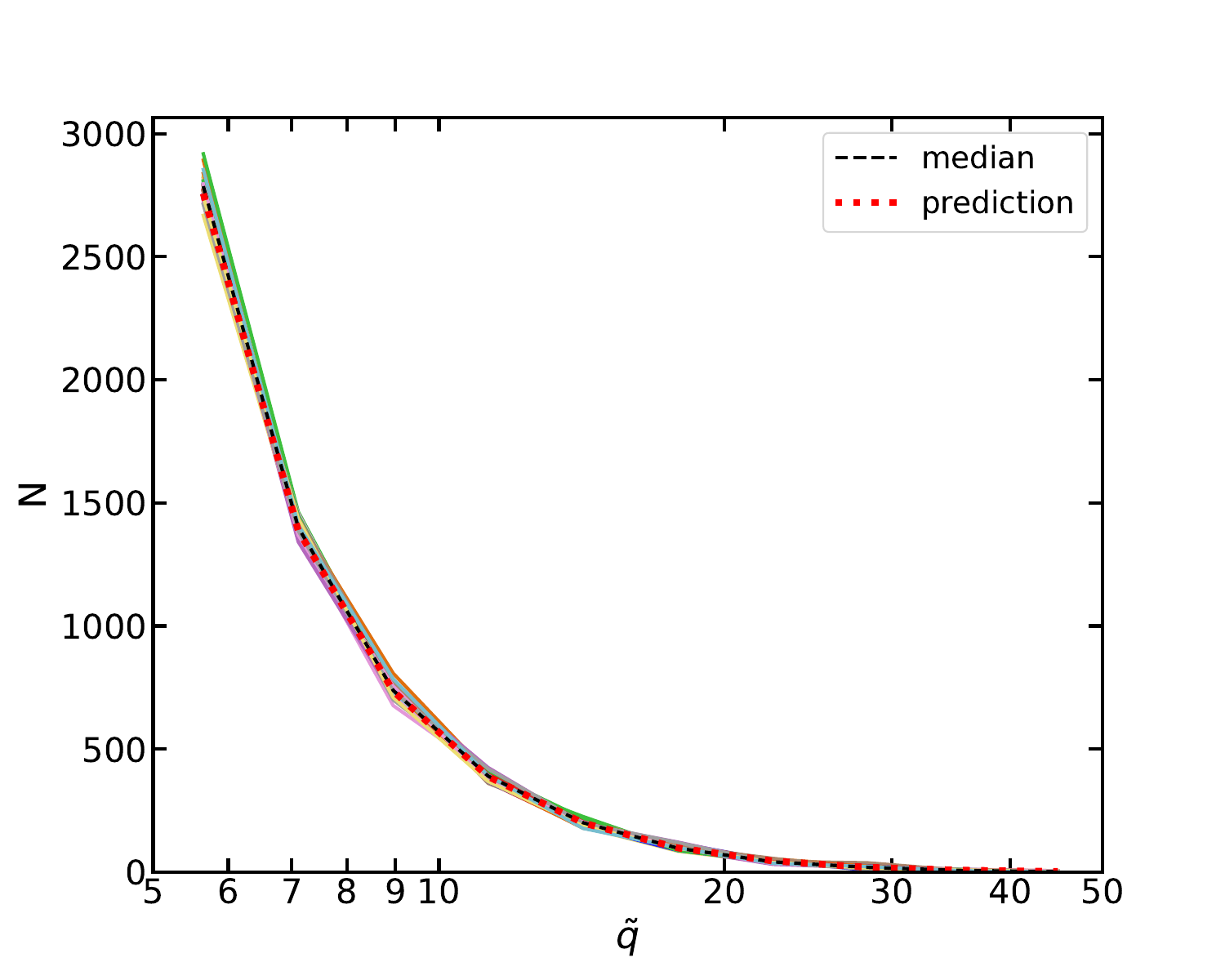}
\includegraphics[width=90mm]{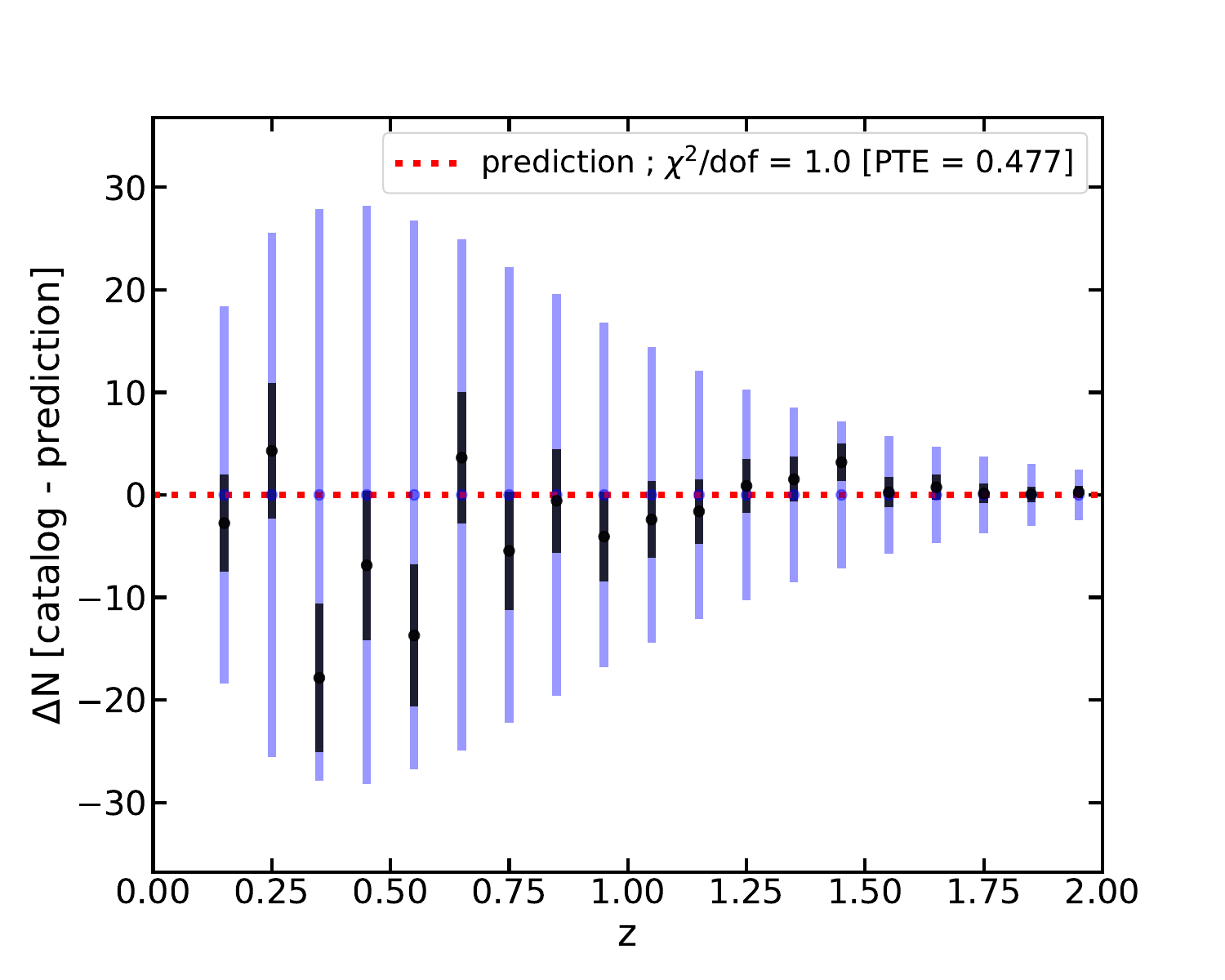}
\includegraphics[width=90mm]{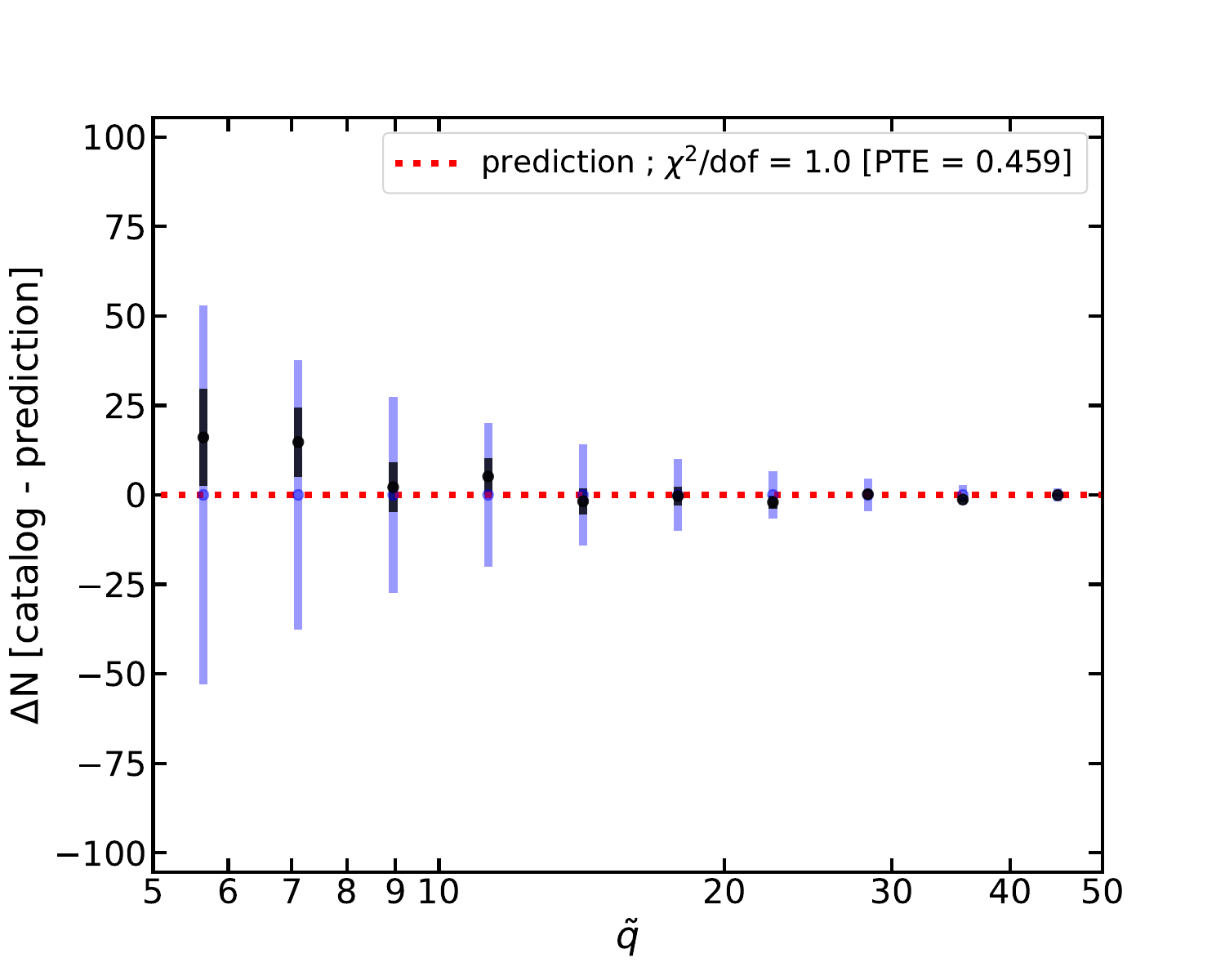}
\caption{Comparison of the number of $\snrt > 5$ clusters in the \cosmo footprint 
detected by \textsc{Nemo} with forward modeled predictions that use the optimization bias
model described in Appendix~\ref{ap:optBiasSims} and the completeness
calculation described in Section~\ref{sec:completeness}. 
The top row shows the number of clusters in bins of $z$ (left) and $\snrt$ (right). Different
colored lines represent each of the 30 simulation realizations. The dashed black line shows the median number of clusters detected in the simulated maps, while the dashed
red line shows the predicted number. The bottom row shows the difference (number of detected clusters minus the prediction), again in bins of $z$ (left) and $\snrt$ (right). The large blue error bars show the 1-$\sigma$ shot noise uncertainty on a single simulation realization,
while the black data points with error bars show the average over the 30 simulation
realizations. These show that the prediction is well within the accuracy needed to match
the number of clusters detected in the DR6-like simulated maps.}
\label{fig:NzNq}
\end{figure*}

We validate the completeness calculations and the optimization bias model
by comparing the number of clusters found by \textsc{Nemo} in the simulated
maps with predictions that forward model from the \citet{Tinker_2008} HMF,
using the noise distribution and filter mismatch function as calculated by
\textsc{Nemo}. These predictions assume the same cosmology and scaling
relation parameters as used to generate the simulated maps.
Figure~\ref{fig:NzNq} shows the results for all 30 simulation realizations, and their average, evaluated for $\snrt_{\rm cut} = 5$, within the \cosmo footprint. We find that we can
accurately predict the number of clusters detected by \textsc{Nemo} as a function of
$\snrt$ and $z$. On average, we detect a total of 5732 $\snrt > 5$ clusters with
\textsc{Nemo}, compared to 5766 predicted (i.e., within $<1$\%), within the \cosmo
footprint.

The \textsc{nemo-sim-kit}\footnote{\url{https://github.com/simonsobs/nemo-sim-kit}} package used to generate the simulated maps, 
run the \textsc{Nemo} cluster finder, fit the optimization bias model, 
and produce the validation plots shown here, will be made publicly available.


\bibliographystyle{yahapj}
\bibliography{ACTDR6Clusters, output_zSourceTable}

\section*{Affiliations}
\noindent
$^{1}$~INAF-Osservatorio Astronomico di Trieste, via G. B. Tiepolo 11, I-34143 Trieste, Italy\\
$^{2}$~Laborat\'orio Interinstitucional de e-Astronomia - LIneA, Av. Pastor Martin Luther King Jr, 126 Del Castilho, Nova Am\'erica Offices, Torre 3000/sala 817 CEP: 20765-000, Brazil\\
$^{3}$~Flatiron Institute, 162 5th Avenue, New York, NY 10010 USA\\
$^{4}$~Joseph Henry Laboratories of Physics, Jadwin Hall, Princeton University, Princeton, NJ, USA 08544\\
$^{5}$~Fermi National Accelerator Laboratory, P. O. Box 500, Batavia, IL 60510, USA\\
$^{6}$~Physik-Institut, University of Zürich, Winterthurerstrasse 190, CH-8057 Zürich, Switzerland\\
$^{7}$~Institute of Cosmology and Gravitation, University of Portsmouth, Portsmouth, PO1 3FX, UK\\
$^{8}$~Department of Astrophysical Sciences, Peyton Hall, Princeton University, Princeton, New Jersey 08544, USA\\
$^{9}$~Department of Astronomy, Cornell University, Ithaca, NY 14853, USA\\
$^{10}$~Sapienza University of Rome, Physics Department, Piazzale Aldo Moro 5, 00185 Rome, Italy\\
$^{11}$~University Observatory, LMU Faculty of Physics, Scheinerstr. 1, 81679 Munich, Germany\\
$^{12}$~Department of Physics, Madingley Road, Cambridge CB3 0HA, UK\\
$^{13}$~Kavli Institute for Cosmology Cambridge, Madingley Road, Cambridge CB3 0HA, UK\\
$^{14}$~Canadian Institute for Theoretical Astrophysics, University of Toronto, Toronto, ON, Canada M5S 3H8\\
$^{15}$~Department of Physics \& Astronomy, University College London, Gower Street, London, WC1E 6BT, UK\\
$^{16}$~School of Physics and Astronomy, Cardiff University, The Parade, Cardiff, Wales, UK CF24 3AA\\
$^{17}$~Institut de F\'{\i}sica d'Altes Energies (IFAE), The Barcelona Institute of Science and Technology, Campus UAB, 08193 Bellaterra (Barcelona) Spain\\
$^{18}$~Department of Physics and Astronomy, University of California, Riverside, CA 92521, USA\\
$^{19}$~Astronomy Unit, Department of Physics, University of Trieste, via Tiepolo 11, I-34131 Trieste, Italy\\
$^{20}$~Institute for Fundamental Physics of the Universe, Via Beirut 2, 34014 Trieste, Italy\\
$^{21}$~DAMTP, Centre for Mathematical Sciences, University of Cambridge, Wilberforce Road, Cambridge CB3 OWA, UK\\
$^{22}$~School of Mathematics and Physics, University of Queensland,  Brisbane, QLD 4072, Australia\\
$^{23}$~Department of Physics, IIT Hyderabad, Kandi, Telangana 502285, India\\
$^{24}$~Department of Physics and Astronomy, University of Pennsylvania, 209 South 33rd Street, Philadelphia, PA, USA 19104\\
$^{25}$~Max-Planck-Institut fur Astrophysik, Karl-Schwarzschild-Str. 1, 85748 Garching, Germany\\
$^{26}$~Department of Astrophysical Sciences, Peyton Hall, Princeton University, Princeton, NJ USA 08544\\
$^{27}$~Physics Division, Lawrence Berkeley National Laboratory, Berkeley, CA 94720, USA\\
$^{28}$~Department of Physics, University of California, Berkeley, CA, USA 94720\\
$^{29}$~Berkeley Center for Cosmological Physics, University of California, Berkeley, CA 94720, USA\\
$^{30}$~Department of Astronomy and Astrophysics, University of Chicago, Chicago, IL 60637, USA\\
$^{31}$~Kavli Institute for Cosmological Physics, University of Chicago, Chicago, IL 60637, USA\\
$^{32}$~Institut d'Estudis Espacials de Catalunya (IEEC), 08034 Barcelona, Spain\\
$^{33}$~Institute of Space Sciences (ICE, CSIC),  Campus UAB, Carrer de Can Magrans, s/n,  08193 Barcelona, Spain\\
$^{34}$~Department of Aeronautics \& Astronautics, Massachusetts Institute of Technology, 77 Mass. Avenue, Cambridge, MA 02139, USA\\
$^{35}$~Department of Physics, University of Chicago, Chicago, IL 60637, USA\\
$^{36}$~Center for Astrophysical Surveys, National Center for Supercomputing Applications, 1205 West Clark St., Urbana, IL 61801, USA\\
$^{37}$~Department of Astronomy, University of Illinois at Urbana-Champaign, 1002 W. Green Street, Urbana, IL 61801, USA\\
$^{38}$~Department of Physics and Astronomy, University of British Columbia, Vancouver, BC, Canada V6T 1Z4\\
$^{39}$~Department of Physics, Columbia University, New York, NY 10027, USA\\
$^{40}$~Wits Centre for Astrophysics, School of Physics, University of the Witwatersrand, Private Bag 3, 2050, Johannesburg, South Africa\\
$^{41}$~Astrophysics Research Centre, School of Mathematics, Statistics and Computer Science, University of KwaZulu-Natal, Durban 4001, South Africa\\
$^{42}$~David A. Dunlap Dept of Astronomy and Astrophysics, University of Toronto, 50 St George Street, Toronto ON, M5S 3H4, Canada\\
$^{43}$~Specola Vaticana (Vatican Observatory), V-00120 Vatican City State\\
$^{44}$~Santa Cruz Institute for Particle Physics, Santa Cruz, CA 95064, USA\\
$^{45}$~Center for Cosmology and Astro-Particle Physics, The Ohio State University, Columbus, OH 43210, USA\\
$^{46}$~Department of Physics, The Ohio State University, Columbus, OH 43210, USA\\
$^{47}$~NIST Quantum Sensors Group, 325 Broadway Mailcode 817.03, Boulder, CO, USA 80305\\
$^{48}$~Mitchell Institute for Fundamental Physics \& Astronomy and Department of Physics \& Astronomy, Texas A\&M University, College Station, Texas 77843, USA\\
$^{49}$~Department of Physics and Astronomy, Rutgers, The State University of New Jersey, Piscataway, NJ USA 08854-8019\\
$^{50}$~Center for Astrophysics $\vert$ Harvard \& Smithsonian, 60 Garden Street, Cambridge, MA 02138, USA\\
$^{51}$~Centre for Radio Astronomy Techniques and Technologies, Department of Physics and Electronics, Rhodes University, P.O. Box 94, Makhanda 6140, South Africa\\
$^{52}$~Department of Physics, Yale University, 217 Prospect St, New Haven, CT 06511\\
$^{53}$~Department of Physics and Astronomy, University of Pittsburgh, Pittsburgh, PA, USA 15260\\
$^{54}$~Institute of Astronomy and Astrophysics, Academia Sinica, Taipei 10617, Taiwan; Institute of Physics, National Yang Ming Chiao Tung University, Hsinchu 30010, Taiwan\\
$^{55}$~Institut de Fisica d'Altes Energies (IFAE), The Barcelona Institute of Science and Technology, Campus UAB, 08193 Bellaterra, Spain\\
$^{56}$~Kavli Institute for Particle Astrophysics \& Cosmology, P. O. Box 2450, Stanford University, Stanford, CA 94305, USA\\
$^{57}$~SLAC National Accelerator Laboratory, Menlo Park, CA 94025, USA\\
$^{58}$~Leiden Observatory, Leiden University, P.O. Box 9513, 2300 RA Leiden, The Netherlands\\
$^{59}$~George P. and Cynthia Woods Mitchell Institute for Fundamental Physics and Astronomy, and Department of Physics and Astronomy, Texas A\&M University, College Station, TX 77843,  USA\\
$^{60}$~Kavli Institute for Cosmological Physics, University of Chicago, 5640 S. Ellis Ave., Chicago, IL 60637, USA\\
$^{61}$~Department of Astronomy and Astrophysics, University of Chicago, 5640 S. Ellis Ave., Chicago, IL 60637, USA\\
$^{62}$~Enrico Fermi Institute, University of Chicago, Chicago, IL 60637, USA\\
$^{63}$~Universit\'e Grenoble Alpes, CNRS, LPSC-IN2P3, 38000 Grenoble, France\\
$^{64}$~Instituci\'o Catalana de Recerca i Estudis Avan\c{c}ats, E-08010 Barcelona, Spain\\
$^{65}$~Kobayashi-Maskawa Institute for the Origin of Particles and the Universe (KMI), Nagoya University, Furo-cho, Chikusa-ward, Nagoya, Aichi, 464-8602, Japan\\
$^{66}$~Institute for Advanced Research, Nagoya University, Furo-cho, Chikusa-ward, Nagoya, Aichi, 464-8601, Japan\\
$^{67}$~Kavli Institute for the Physics and Mathematics of the Universe (WPI), University of Tokyo, 5-1-5, Kashiwa-no-ha, Kashiwa, Chiba, 277-8583, Japan\\
$^{68}$~European Southern Observatory, Karl-Schwarzschild-Str. 2, D-85748, Garching, Germany\\
$^{69}$~Institute of Theoretical Astrophysics, University of Oslo, Norway\\
$^{70}$~Department of Physics, University of Milano - Bicocca, Piazza della Scienza, 3 - 20126, Milano (MI), Italy\\
$^{71}$~Jodrell Bank Centre for Astrophysics, Department of Physics and Astronomy, The University of Manchester, Manchester M13 9PL, UK\\
$^{72}$~Department of Physics, Cornell University, Ithaca, NY, USA 14853\\
$^{73}$~Centro de Tecnologia da Informa\c{c}\~ao Renato Archer, Campinas, SP, Brazil - 13069-901\\
$^{74}$~Observat\'orio Nacional, Rio de Janeiro, RJ, Brazil - 20921-400\\
$^{75}$~Center for Frontier Science, Chiba University, Chiba 263-8522, Japan\\
$^{76}$~Department of Physics, Graduate School of Science, Chiba University, Chiba 263-8522, Japan\\
$^{77}$~Department of Physics and Astronomy, Haverford College, Haverford, PA, USA 19041\\
$^{78}$~Hamburger Sternwarte, Universit\"{a}t Hamburg, Gojenbergsweg 112, 21029 Hamburg, Germany\\
$^{79}$~Centro de Investigaciones Energ\'eticas, Medioambientales y Tecnol\'ogicas (CIEMAT), Madrid, Spain\\
$^{80}$~Ruhr University Bochum, Faculty of Physics and Astronomy, Astronomical Institute, German Centre for Cosmological Lensing, 44780 Bochum, Germany\\
$^{81}$~Department of Physics, Stanford University, Stanford, CA 94305-4085, USA\\
$^{82}$~Kavli Institute for Particle Astrophysics and Cosmology, 382 Via Pueblo Mall Stanford, CA  94305-4060, USA\\
$^{83}$~SLAC National Accelerator Laboratory 2575 Sand Hill Road Menlo Park, California 94025, USA\\
$^{84}$~Department of Physics and Astronomy, Pevensey Building, University of Sussex, Brighton, BN1 9QH, UK\\
$^{85}$~Instituto de Astrofisica de Canarias, E-38205 La Laguna, Tenerife, Spain\\
$^{86}$~Universidad de La Laguna, Dpto. Astrofísica, E-38206 La Laguna, Tenerife, Spain\\
$^{87}$~Department of Physics, Northeastern University, Boston, MA 02115, USA\\
$^{88}$~Instituto de Física, Pontificia Universidad Católica de Valparaíso, Casilla 4059, Valparaíso, Chile\\
$^{89}$~Physics Department, Lancaster University, Lancaster, LA1 4YB, UK\\
$^{90}$~Computer Science and Mathematics Division, Oak Ridge National Laboratory, Oak Ridge, TN 37831\\
$^{91}$~Department of Physics, Duke University, Durham, NC, 27708, USA\\
$^{92}$~Department of Astronomy, University of California, Berkeley,  501 Campbell Hall, Berkeley, CA 94720, USA\\
$^{93}$~Lawrence Berkeley National Laboratory, 1 Cyclotron Road, Berkeley, CA 94720, USA\\
$^{94}$~Max Planck Institute for Extraterrestrial Physics, Giessenbachstrasse, 85748 Garching, Germany\\
$^{95}$~Universit\"ats-Sternwarte, Fakult\"at f\"ur Physik, Ludwig-Maximilians Universit\"at M\"unchen, Scheinerstr. 1, 81679 M\"unchen, Germany\\
$^{96}$~NASA/Goddard Space Flight Center, Greenbelt, MD, USA 20771\\
$^{97}$~Institute of Astronomy, Madingley Road, Cambridge CB3 0HA, UK\\

\end{document}